\documentclass[fleqn,usenatbib]{mnras}

\usepackage{newtxtext,newtxmath}

\usepackage[T1]{fontenc}

\DeclareRobustCommand{\VAN}[3]{#2}
\let\VANthebibliography\thebibliography
\def\thebibliography{\DeclareRobustCommand{\VAN}[3]{##3}\VANthebibliography}

\usepackage{graphicx}	
\usepackage{amsmath}	
\usepackage{hyperref}
\usepackage{xcolor}
\usepackage{soul}

\newcommand{\unit}[1]{\ensuremath{\, \mathrm{#1}}}

\title[Modelling Subhalo Density Profiles in TNG50]{Not All Subhaloes Are Created Equal: Modelling the Diversity of Subhalo Density Profiles in TNG50}

\author[F. M. Heinze, G. Despali and R. S. Klessen]{
Felix M. Heinze,$^{1}$
Giulia Despali$^{2,3}$
and Ralf S. Klessen$^{4,5}$
\\
$^{1}$Friedrich-Schiller-Universität Jena, Theoretisch-Physikalisches Institut, Fröbelstieg 1, D-07743 Jena, Germany\\
$^{2}$Alma Mater Studiorum - Università di Bologna, Dipartimento di Fisica e Astronomia "Augusto Righi", Via Gobetti 93/2, I-40129 Bologna, Italy\\
$^{3}$INAF-Osservatorio di Astrofisica e Scienza dello Spazio di Bologna, Via Gobetti 93/3, I-40129 Bologna, Italy \\
$^{4}$Universität Heidelberg, Zentrum für Astronomie, Institut für Theoretische Astrophysik, Albert-Ueberle-Straße 2, D-69120 Heidelberg, Germany \\
$^{5}$Universität Heidelberg, Interdisziplinäres Zentrum für Wissenschaftliches Rechnen, Im Neuenheimer Feld 205, D-69120 Heidelberg, Germany
}

\date{Accepted XXX. Received YYY; in original form ZZZ}

\pubyear{2023}

\begin{document}
\label{firstpage}
\pagerange{\pageref{firstpage}--\pageref{lastpage}}
\maketitle

\begin{abstract}
In this work, we analyse the density profiles of subhaloes with masses $M_\mathrm{sh} \geq 1.4 \times 10^8$ M$_\odot$ in the TNG50 simulation, with the aim of including baryonic effects. We evaluate the performance of frequently used models, such as the standard NFW, the Einasto, and a smoothly truncated version of the NFW profile. We find that these models do not perform well for the majority of subhaloes, with the NFW profile giving the worst fit in most cases. This is primarily due to mismatches in the inner and outer logarithmic slopes, which are significantly steeper for a large number of subhaloes in the presence of baryons. To address this issue, we propose new three-parameter models and show that they significantly improve the goodness of fit independently of the subhalo's specific properties. Our best-performing model is a modified version of the NFW profile with an inner log-slope of -2 and a variable truncation that is sharper and steeper than the slope transition in the standard NFW profile. Additionally, we investigate how both the parameter values of the best density profile model and the average density profiles vary with subhalo mass, $V_\mathrm{max}$, distance from the host halo centre, baryon content and infall time, and we also present explicit scaling relations for the mean parameters of the individual profiles. The newly proposed fit and the scaling relations are useful to predict the properties of realistic subhaloes in the mass range $10^8$ M$_\odot$ $\leq M_\mathrm{sh} \leq$ $10^{13}$ M$_\odot$ that can be influenced by the presence of baryons.
\end{abstract}




\section{Introduction}

The $\Lambda$ Cold Dark Matter ($\Lambda$CDM) model has so far been widely successful in predicting and explaining numerous observations on cosmological scales, such as the existence and properties of the cosmic microwave background (CMB), the expansion history of the Universe, the abundances of the chemical elements, as well as the formation and distribution of cosmic large-scale structure \citep{planck_main}. Another key prediction of the $\Lambda$CDM model is the presence of a large number of small dark matter subhaloes within larger host haloes \citep{halo_review}, which have first been studied in $N$-body simulations by \citet{moore1999, klypin1999}. The increase in available computational resources over the last few decades allowed for ever more precise theoretical predictions on galactic and subgalactic scales by using high-resolution simulations. On these small scales, a number of discrepancies between the theoretical model and observations have been identified in the past \citep{small_scale_problems}. A few common examples are the cusp-core problem \citep{cusp_core1, cusp_core2}, the missing satellites problem \citep{missing_satellites1, moore1999} and the too big to fail problem \citep{too_big_to_fail1}, which point out mismatches in the abundances, central densities and inner density profile shapes for subhaloes in simulations and dwarf galaxies in the Local Group. However, there are still a number of uncertainties in both the model predictions and the observations, which led to various attempts of resolving these small-scale issues, both within the $\Lambda$CDM framework \citep[for example by including baryonic feedback, see e.g.][]{feedback8, feedback7, feedback6, feedback3, feedback2, feedback1} and also by invoking other dark matter models, such as warm dark matter \citep[WDM, see e.g.][]{wdm_haloes1, wdm_haloes2, despali19,lovell2023} or self-interacting dark matter \citep[SIDM, see e.g.][]{sidm3, sidm4, sidm2019, sidm6,sidm5,sidm0, lovell2023}. Yet another idea to solve these problems is to modify the theory of gravity, which is done for example in Modified Newtonian Dynamics  \citep[MOND, see e.g.][]{mond5, mond4, mond3, mond2, mond1}. Up to the present day, these controversies and debates are still ongoing, and in order to find the most suitable model it is necessary to probe these subgalactic scales in ever more detail.

Investigating the properties of dwarf galaxies in the Local Group has for a long time been the primary method for probing these small scales observationally \citep{dwarf_galaxies1, dwarf_galaxies2}. More recently, other methods have been used, which even allow for studying the properties of small (sub-)haloes that do not contain enough luminous matter to be detectable via emission of radiation. In the Milky Way, these dark subhaloes can be detected by the perturbations they cause in stellar tidal streams \citep{stellar_streams1, stellar_streams3}, which led to the detection of a subhalo with a mass of $10^6$-$10^8$ M$_\odot$ in the GD-1 stellar stream by \citet{stellar_streams4}. Extragalactic subhaloes can be detected by the effect they have on the flux ratios of multiply-imaged quasars \citep{flux_ratios2, flux_ratios1} or by the perturbations they cause in strong galaxy-galaxy lens systems. Using this method, four subhalo detections have been claimed up to this day \citep{vegetti2010a, vegetti2010, vegetti2012, hezaveh2016}, with masses between $1.9 \times 10^8$ and $2.7 \times 10^{10}$ M$_\odot$, but many more are expected to follow from the upcoming Euclid and LSST surveys \citep{LSST_euclid} and high-resolution follow-up images. \citet{minor2021} and \citet{sengul2022} reanalysed some of the previously detected subhaloes and estimated that their concentrations are exceptionally high when modelled with NFW profiles, making them 2$\sigma$ outliers of the CDM model. The inferred concentration of the GD-1 stream perturber in \citet{stellar_streams4} is also higher than the average NFW concentration, pointing in the same direction.

In order to reliably test the $\Lambda$CDM model with these observations, accurate theoretical predictions for the properties of these subhaloes are needed, such as their mass function, their internal structure, their dynamical properties, as well as their structural evolution as they experience tidal stripping due to the presence of their host halo and tidal shocks due to close encounters with other subhaloes. Unfortunately, state-of-the-art simulations are limited in resolution and are therefore not able to resolve the entire possible mass range of subhaloes down to earth masses. The finite resolution also leads to several unwanted effects, such as artificial disruption \citep{artificial_disruption3, artificial_disruption2, artificial_disruption1}, which can severely limit the ability to obtain reliable data for the subhalo properties. 

\citet{moline2017, moline2023} recently investigated concentrations and other subhalo properties in great detail in $N$-body simulations for a wide range of different subhalo and host halo masses and how they evolve over cosmic time.  Furthermore, $N$-body simulations have shown that subhaloes generally exhibit much higher central densities and are on average more concentrated than isolated haloes of the same mass \citep{subhalo_conc1, subhalo_conc2}. 

The density profiles of dark matter subhaloes have also been studied extensively before. However, this has mostly been done in zoom-in simulations, which provide a high resolution but fewer objects for a proper statistical analysis. \citet{subhalo_profiles3} studied the evolution of the subhalo density profiles in idealised simulations as they experience tidal stripping and showed that the structural evolution solely depends on the initial subhalo concentration and the fraction of mass being stripped. \citet{subhalo_profiles1} used an approach similar to the one presented in this paper in order to investigate the density profiles of subhaloes in zoom-in simulations of the Local Group. They showed that the Einasto profile \citep{einasto} provides a better model than the commonly used Navarro-Frenk-White (NFW) profile \citep{nfw2, nfw1} and some modifications of it. They further demonstrated that the shape parameter of the Einasto profile strongly depends on the total subhalo mass, and is also being reduced by tidal stripping. Similar results have also been obtained before for isolated field haloes \citep{better_einasto2, better_einasto1, better_einasto3}. \citet{subhalo_profiles2} already pointed out that the NFW profile might not be an accurate model for the density profiles of subhaloes. \citet{springel2008} investigated the density profiles, concentrations, mass function and radial distribution of subhaloes in the dark-matter-only simulation Aquarius. They found that the density profiles of subhaloes show a similar behaviour to those of main haloes and that the Einasto profile provides a much better fit than the NFW or Moore profile \citep{moore_profile}, even with fixed shape parameter.

In this paper, we investigate the density profiles of subhaloes in the state-of-the-art cosmological simulation TNG50 of the IllustrisTNG project, which offers a good compromise between size and resolution. Recent studies have analysed the impact of baryons on the structural properties of (sub-)haloes as well as their mass function and found that they can have a significant impact, making haloes more spherical, reducing the abundance of lower mass subhaloes and leading to higher concentrations and different slopes in the inner regions of (sub-)halo density profiles \citep{baryon_effects1, baryon_effects2, baryon_effects3, baryon_effects4, baryon_effects5}. TNG50 includes the effect of baryons, together with many other physical model components. Our main goal is to provide a simple three-parameter analytical fit for the density profiles that can be used for a wide range of subhaloes with different properties. \\

The paper is structured as follows: In Section \ref{sec:methods}, we give a brief overview of the TNG50 simulation and its relevant technical details. We also discuss the selection of suitable subhaloes and the computation of their density profiles. In Section \ref{sec:analytical_models}, we present commonly used analytical models for describing the density profiles of dark matter haloes and subhaloes (the NFW profile, the Einasto profile and the truncated NFW profile) and further introduce new improved models. We compare the performances of all these models in Section \ref{sec:performace} for subhaloes with various different kinds of properties. In Section \ref{sec:parameter_relations}, we investigate how both the individual parameters of the best performing model and the average density profiles vary with different subhalo properties, including mass, V$_\mathrm{max}$, distance from the host halo centre, baryon fraction and infall time. Furthermore, we present simple scaling relations for how the model parameter values depend on the subhalo mass and V$_\mathrm{max}$, and discuss how baryons and tidal stripping affect the subhalo density profiles. Finally, our summary and conclusions are presented in Section \ref{sec:conclusions}.

\section{Methods}
\label{sec:methods}
\subsection{The TNG50 Simulation}
The subhaloes analysed in this paper are drawn from IllustrisTNG, a suite of large-volume cosmological gravo-magnetohydrodynamical simulations \citep{tng50data}.  All of these simulations were performed using the moving-mesh code \textsc{Arepo} \citep{arepo2010}, which calculates the gravitational forces using a Particle-Mesh-Tree method and solves the ideal magneto-hydrodynamic equations using a finite volume method on an adaptive mesh. Each simulation starts at a redshift of $z=127$ and finishes at the present day $z=0$, with cosmologically motivated initial conditions and a cosmology in agreement with Planck 2015, given by $\Omega_{\Lambda,0}=0.6911$, $\Omega_\mathrm{m,0}=0.3089$, $\Omega_\mathrm{b,0}=0.0486$, $\sigma_8=0.8159$, $n_\mathrm{s}=0.9667$ and $h=0.6774$ \citep{planck2015}. They also include a large number of physical processes such as primordial and metal-line cooling, heating by the extragalactic UV background, stochastic star formation, evolution of stellar populations, feedback from supernovae and AGB stars, as well as supermassive black hole formation and feedback. It has been shown that the TNG simulations produce results which are largely consistent with observations, beyond the regimes which were used to calibrate the model. For more information about the underlying model and its numerical details, see \citet{tng_model1} or \citet{tng_model2}.

In order to be able to reliably study the small subgalactic scales, we make use of the TNG50 simulation \citep{nelson2019, pillepich2019}, since it provides the highest resolution. TNG50-1 has a cubic simulation volume with a comoving side length of 51.7 Mpc, a dark matter mass resolution of $3.1 \times 10^5$ M$_\odot h^{-1}$, a baryonic mass resolution of $5.7 \times 10^4$ M$_\odot h^{-1}$ and a gravitational softening length of 0.288 kpc at $z=0$.

Haloes are identified using the friends-of-friends (FoF) group finder algorithm \citep{fof} with a linking length of 0.2. Subhaloes are subsequently identified using the \textsc{Subfind} algorithm \citep{subfind1, subfind2}. Additionally, merger trees have been created using \textsc{SubLink} \citep{sublink} and \textsc{LHaloTree} \citep{lhalotree}. The halo and subhalo finding algorithm has an impact on the on the total mass and radial extent of the structures \citep{subfind_comparison1, subfind_comparison2}. However, we will show in Section \ref{sec4} and \ref{sec:parameter_relations} that our newly proposed density profile model deviates significantly from the NFW profile in the inner regions, where the density is well constrained and where the effects on observations are the strongest.

\subsection{Subhalo Selection}
The TNG50-1 group catalogue of the snapshot at $z=0$ contains 5,688,113 \textsc{Subfind} groups in total. However, due to the limitations in resolution, not all of these objects are suitable for the analysis of the density profiles. Furthermore, some of the objects identified by the \textsc{Subfind} algorithm do not have a cosmological origin, in the sense that they have not formed due to the process of hierarchical structure formation. Instead, some of these objects might be fragments of already formed galaxies, produced by baryonic processes such as disk instabilities. These have been flagged in the simulation data and we exclude them from the analysis of the subhalo density profiles. In this work, we also exclude subhaloes with less than 300 dark matter particles. Below this threshold, the scatter in the parameter values of the analytical fit functions increases drastically, which we attribute to the limitations in resolution. Other authors use a higher number of particles as a threshold (e.g. 1000, see \citealt{subhalo_profiles1}), but this would only affect the first two mass bins of our sample. We also eliminate the main haloes as well as subhaloes which belong to a \textsc{Subfind} group that has a total mass of less than 10$^{10}$ M$_\odot$. After excluding all the objects which are not suitable for further analysis, 112,048 subhaloes remain in total. Their total masses lie in the range between 1.4 $\times$ 10$^8$ M$_\odot$ and 8.5 $\times$ 10$^{12}$ M$_\odot$, which we divide up into 17 mass bins that cover a large part of the subhalo mass range that can currently be probed observationally.

\subsection{Computation of the Density Profiles}
We computed the density profiles in spherical shells centered around the minimum gravitational potential. The corresponding 40 logarithmically spaced radial bins lie in the range between 10$^{-1}$ and 10$^3$ ckpc/$h$. We chose the lower limit such that the radial bins reach a bit below the gravitational softening length, which marks the limit below which the density profile can no longer be well resolved. The upper limit was determined by the largest subhalo. The fixed number of radial bins in this range allows for an easy computation of average density profiles, and it is large enough to guarantee that the smallest subhaloes with an extent of a few kpc are still well resolved. For computing the density value of each spherical shell we used all matter particles, including dark matter, gas, stars, and black holes. We further used the Poissonian error to estimate the uncertainties for the density values due to the finite number of particles that sample the density distribution. Figure \ref{fig:example_profiles} shows the density profiles of four representative subhaloes with different masses, together with some of the analytical models described in Section \ref{sec:analytical_models}.

\section{Analytical Density Profile Models}
\label{sec:analytical_models}
\subsection{Traditional Models}
\label{sec:common_models}
\subsubsection{The NFW Profile}
\label{sec:NFW}
\citet{nfw2, nfw1} demonstrated in their $N$-body simulations that haloes consisting of dark matter only have a universal density profile shape, which is independent of their mass and can be well described by the Navarro-Frenk-White (NFW) profile:
\begin{equation}
    \rho(r) = \frac{\rho_0}{\frac{r}{r_\mathrm{s}} \left( 1 + \frac{r}{r_\mathrm{s}} \right)^2}.
\label{eq:NFW}
\end{equation}
The profile has an inner logarithmic slope of -1 and an outer log-slope of -3, with the transition happening at the scale radius $r_\mathrm{s}$, which is the only free parameter besides the overall normalisation $\rho_0$. Therefore, the profile is completely characterised by the virial mass $M_{200}$ and the concentration $c_{200}=r_{200}/r_\mathrm{s}$ of the halo. The mass and concentration are not completely independent but correlated, with more massive haloes being less concentrated in general. Simulations have also shown that at fixed mass, the halo concentration is correlated with assembly time, which introduces a large halo-to-halo scatter due to the different mass assembly histories. This concentration-mass relation has been studied both for haloes \citep{c_m_rel1, c_m_rel2, c_m_rel3, better_einasto3, c_m_rel0} and for subhaloes \citep{subhalo_conc2, moline2017, moline2023}.

\subsubsection{The Truncated NFW Profile}
Since subhaloes can experience severe tidal truncation \citep{subhalo_trunc1, diemand2007}, it has been pointed out in the past that subhalo density profiles might not be well described by the NFW profile. One way to take the truncation into account is by using a smoothly truncated version of the NFW profile (tNFW) with a logarithmic slope of -5 well beyond the truncation radius $r_\mathrm{t}$:
\begin{equation}
    \rho(r) = \frac{\rho_0}{\frac{r}{r_\mathrm{s}} \left( 1 + \frac{r}{r_\mathrm{s}} \right)^2} \cdot \left( \frac{r_\mathrm{t}^2}{r^2 + r_\mathrm{t}^2} \right).
\label{eq:tNFW}
\end{equation}
This model has been used e.g. by \citet{minor2021} for the analysis of subhalo properties inferred from strong gravitational galaxy-galaxy lensing.

\subsubsection{The Einasto Profile}
The Einasto profile \citep{einasto} is another commonly used three-parameter model for describing the density profiles of dark matter haloes. Its functional form is given by:
\begin{equation}
    \rho(r) = \rho_{-2} \exp \left\{ \frac{-2}{\alpha} \left[ \left( \frac{r}{r_{-2}} \right) ^\alpha -1 \right]  \right\}.
\label{eq:einasto}
\end{equation}
$r_{-2}$ is the radius at which the log-slope is equal to -2, $\rho_{-2}$ is the corresponding density and $\alpha$ controls the shape of the profile. The Einasto profile is no longer a double power law but instead the logarithmic slope is a power law function of radius:
\begin{equation}
    \frac{d \log \rho}{d \log r} = -2 \left( \frac{r}{r_{-2}} \right) ^\alpha.
\label{eq:einasto_slope}
\end{equation}
It has been shown in the past that the Einasto profile generally provides a much better model for the density profile of both haloes and subhaloes \citep{better_einasto2, better_einasto1, subhalo_profiles2, better_einasto3}. \\

\begin{figure*}
	\includegraphics[width=\textwidth]{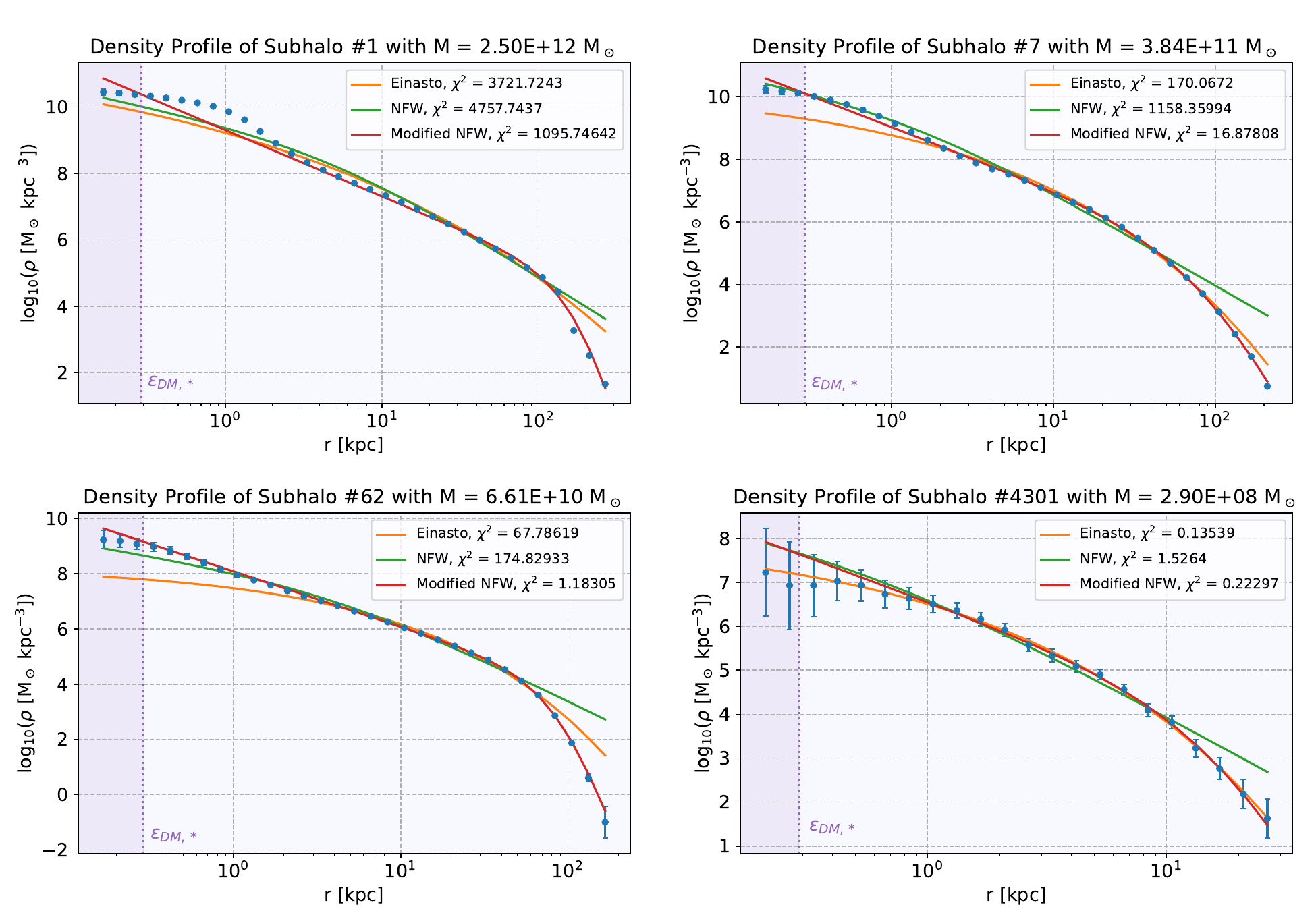}
    \caption{Four example density profiles of representative subhaloes with different masses, including the best fits for the NFW, the Einasto and the modified NFW profile. The softening length is highlighted in purple. Subhaloes with masses well above $10^{11}$ M$_{\sun}$ (upper left) tend to have a pronounced bump feature in the central regions which flattens towards the centre and forms a core. Subhaloes with masses in the range between $10^{10}$ and $10^{11}$ M$_{\sun}$ tend to have an inner log-slope of approximately -2 all the way to the centre and either a soft transition to a slightly steeper outer slope (upper right) or a sharper truncation (lower left). Subhaloes with masses below $10^9$ M$_{\sun}$ generally have a more regular shape, with a continuously changing logarithmic slope and can therefore often be better described by the Einasto profile.}
    \label{fig:example_profiles}
\end{figure*}

\subsection{Additional Models}
\label{sec:additional_models}
In Figure \ref{fig:example_profiles}, one can see that the previously mentioned density profile models do not generally provide a good fit for most of the subhaloes, especially the more massive ones. Furthermore, the best-fit parameter values of the NFW profile take extreme and unrealistic values. Because of that, we introduce some other fit functions that better capture the inner and outer logarithmic slopes as well as the truncation. 

\subsubsection{The Modified NFW Profile} \label{mfnw}
A generalisation of the NFW profile is the so-called $(\alpha, \beta, \gamma)$ model \citep[see e.g.][]{mnfw}, which has the following functional form:
\begin{equation}
    \rho(r) = \frac{\rho_0}{\left( \frac{r}{r_\mathrm{s}} \right)^\gamma \left[ 1 + \left( \frac{r}{r_\mathrm{s}} \right)^\alpha \right]^{(\beta - \gamma)/\alpha}}.
\label{eq:alpha_beta_gamma_model}
\end{equation}
Here $\gamma$ and $\beta$ set the inner and outer log-slope and $\alpha$ controls the sharpness of the transition from one slope to the other. At the scale radius $r_\mathrm{s}$, the log-slope is the average of the inner and outer slopes, which is $-(\gamma+\beta)/2$. For $(\alpha, \beta, \gamma)$ = (1, 3, 1) we get the NFW profile, but also other commonly used profiles can be obtained from it.

The $(\alpha, \beta, \gamma)$ model has five parameters in total. When looking at many of the computed density profiles, one can observe that profiles with a steeper outer log-slope generally also have a sharper transition from the inner to the outer slope. This motivates the following model with the functional form:
\begin{equation}
    \rho(r) = \frac{\rho_0}{\left( \frac{r}{r_\mathrm{s}} \right)^2 \left[ 1 + \left( \frac{r}{r_\mathrm{s}} \right)^\alpha \right]^6}.
\label{eq:modified_NFW}
\end{equation}

\begin{figure*}
\includegraphics[width=\columnwidth]{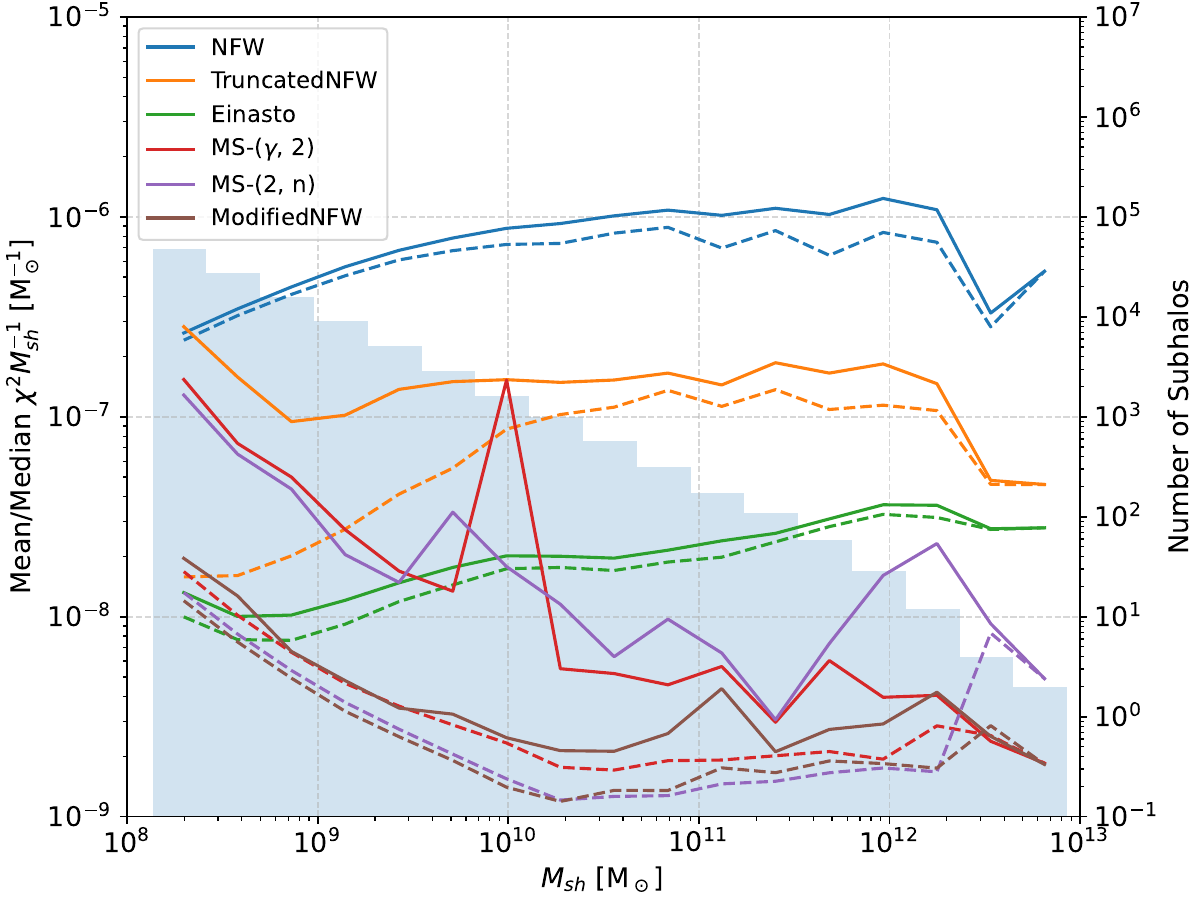}
\includegraphics[width=\columnwidth]{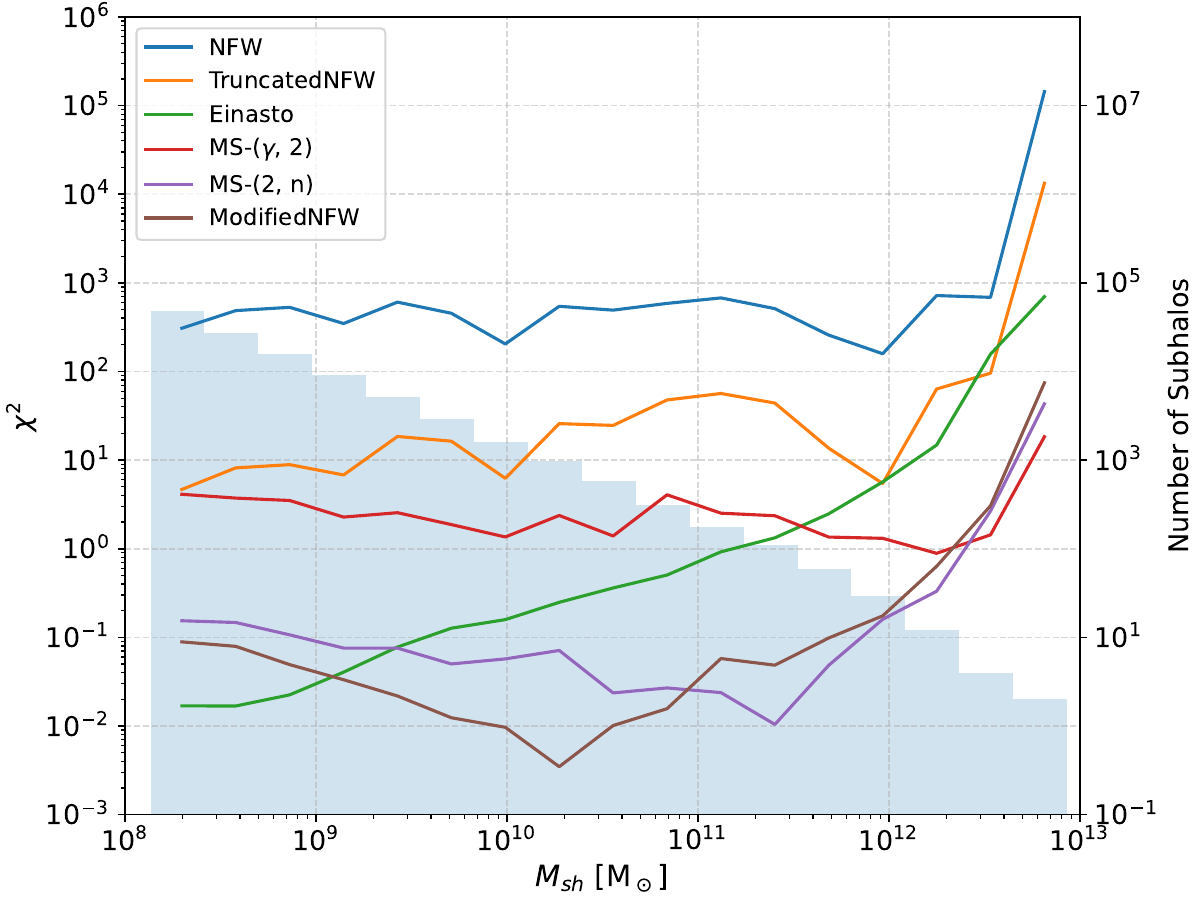}
\caption{Subhalo mass dependence of the density profile model goodness of fit values for the individual (left) and the average density profiles (right). The light blue histograms in the background indicate the number of subhaloes in each mass bin. For the individual subhaloes, the mean (solid line) and median (dashed line) values of the goodness of fit per unit mass have been computed for each mass bin. The scatter in $\chi^2$ has been omitted, since it is dominated by failing fits.}
\label{fig:performances_mass}
\end{figure*}

Here, $\alpha$ controls both the outer log-slope and the sharpness of the slope transition. It therefore determines the shape of the density profile. We set the inner log-slope to -2, and we determined the value of the other exponent by optimising the goodness of fit for the individual and average profiles. A value of 6 turns out to be the best choice for the subhaloes in our TNG50 sample. The scale radius $r_\mathrm{s}$ marks the radius at which the log-slope changes. In our new parametrisation it corresponds, in practice, to the truncation radius, and it could therefore be used to determine the extent of the inner part of the profile. As the log-slope beyond the truncation radius is very steep, most of the subhalo mass should be contained within $r_\mathrm{s}$.

Throughout the paper, we refer to the model in Equation (\ref{eq:modified_NFW}) as the \emph{modified NFW profile}.

\subsubsection{The Modified Schechter Profiles}
Instead of modelling the truncation with a double power law like we did in the modified NFW profile, one can use a simple power law profile with an exponential cutoff and an additional parameter $n$ that controls the sharpness of the truncation:
\begin{equation}
    \rho(r) = \rho_0 \left( \frac{r}{r_\mathrm{t}} \right)^{-\gamma} \cdot \exp \left[ - \left( \frac{r}{r_\mathrm{t}} \right)^n \right].
\label{eq:schechter}
\end{equation}
This looks very similar to the functional form of the Schechter luminosity function \citep{schechter1976} with the additional parameter $n$. Because of that, we will refer to it as the \emph{modified Schechter (MS) profile}. Here $\rho_0$ is again the overall normalisation factor, $r_\mathrm{t}$ is the truncation radius and $\gamma$ is the power law log-slope. From this four-parameter model we construct two three-parameter models by fixing one parameter and leaving the other parameters as free parameters.

In one case, we fix the inner log-slope to -2, similarly to the modified NFW profile and leave $n$ and the other parameters free. From now on, we will refer to this model as MS-(2, $n$).

For the other model we instead fix the sharpness of the truncation $n$ to 2 and leave $\gamma$ as a free parameter. From now on, we will refer to this model as MS-($\gamma$, 2). Our analysis of the parameters of the MS-(2, $n$) model in the appendix (see Figure \ref{fig:parameter_mass_scatter_tpl}, \ref{fig:parameter_vmax_scatter_tpl} and Table \ref{tab:scaling_rel}), suggests that the mean values of $n$ indeed lie very close to 2 for most of the mass bins. Allowing for more freedom to adjust the inner log-slope can lead to a better fit in the case of an additional bump in the central regions of the density profile, which is often present for subhaloes of higher mass (see Figure \ref{fig:example_profiles} in the upper left corner). Similar parametrisations with an exponential truncation for subhaloes have also been used in the past \citep[see e.g.][]{subhalo_trunc1, exp_truncation}. \\

Other frequently used radial profile models are: the Moore profile \citep{moore_profile}, the Hernquist profile \citep{hernquist_profile}, the Burkert profile \citep{burkert_profile} and the (Pseudo-)Jaffe profile \citep{jaffe_profile, corecusp_profile2}. While they have been used for modelling the density profiles of haloes and subhaloes in previous works, we do not consider them since the new models presented in this section provide a much better fit for our sample.

\section{Comparison of the Model Performances} \label{sec4}
\label{sec:performace}
\subsection{Fitting Process}
We obtain the optimal fit for each model and each density profile by using the non-linear least squares fitting algorithm of the \textsc{Python} library \textsc{SciPy}, together with the Poissonian density uncertainties. For calculating the goodness of fit, we use a reduced chi-squared statistic to include the influence of the uncertainties as well as the different numbers of radial bins and parameters:
\begin{equation}
    \chi^2 = \frac{1}{N-p} \cdot \sum_{i=1}^{N} \frac{(\log \rho_i - \log f_i)^2}{(\sigma_i/\rho_i)^2},
\label{eq:chi_squared}
\end{equation}
with the total number of nonzero data points $N$, the number of fitted parameters $p$, the difference between the log-values of the density $\rho_i$ and the model fit value $f_i$, and the uncertainties of the log-profile $\sigma_i/\rho_i$ at the data point $i$. We exclude the data points below the softening length as well as data points where the density is zero for both the fitting and the calculation of the goodness-of-fit.

We note that $\chi^2$ increases with subhalo mass, since lower-mass subhaloes exhibit larger relative density uncertainties due to the limitations in resolution. This makes it difficult to compare the goodness of fit for subhaloes with different masses. However, the comparison of different model performances within individual mass bins is not affected by this. We found that $\chi^2$ is roughly proportional to $M_\mathrm{sh}$. Because of that, we primarily look at $\chi^2$ per unit subhalo mass in order to have a quantity that is relatively independent of mass. Otherwise, high values of $\chi^2$ would only indicate where the most massive subhaloes can be found.

It should also be noted that it can make a significant difference whether one computes the fit for a linear density profile or a log-density profile. The difference in the obtained parameter values can be quite noticeable, especially for the models described in Section \ref{sec:common_models}, which do not perform well in most cases. However, for the models that provide an accurate fit, the difference is negligible. In order to be consistent, the further analysis will be done with fits for the linear density profiles for all the fit functions as they generally lead to more reliable results. Despite that, we still evaluate the goodness of fit by testing how well these linear fits can reproduce the overall shape of the log-density profile with the definition in Equation (\ref{eq:chi_squared}).

\begin{figure*}
\includegraphics[width=\columnwidth]{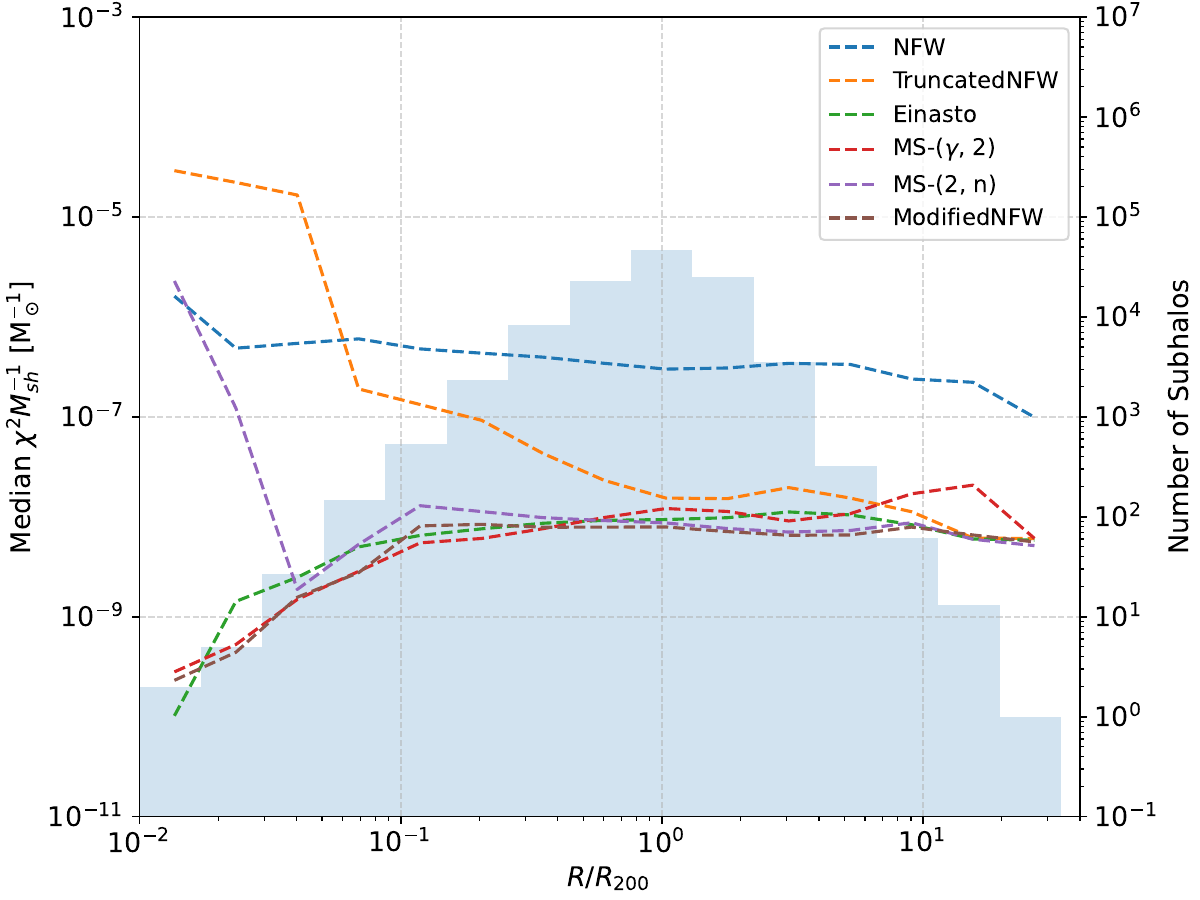}
\includegraphics[width=\columnwidth]{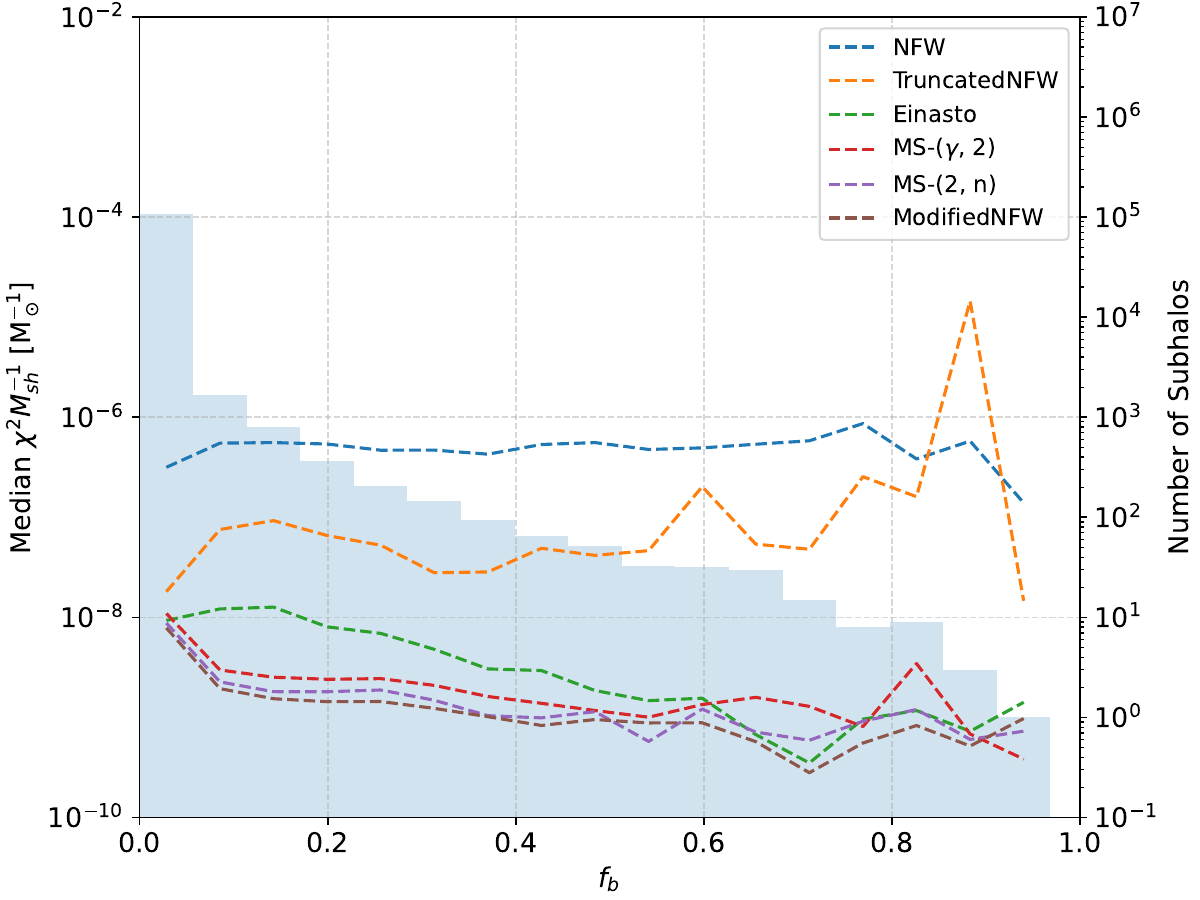}
\caption{Median values of the goodness of fit per unit subhalo mass for subhaloes with different distances from the host halo centre (on the left) and different baryon fractions (on the right). Additionally, the radial distribution of subhaloes and the distribution of baryon fractions is shown in the histograms.}
\label{fig:performances_distance_bf}
\end{figure*}

\subsection{Performance for Different Subhalo Masses}
\label{sec:model_performances}
In Figure \ref{fig:performances_mass}, we show the performances of the fits for both the individual and the average density profiles as a function of subhalo mass. The histogram indicates the number of subhaloes in each mass bin. 

The NFW profile performs the worst, followed by its smoothly truncated version. The reason for this is that most of the subhalo density profiles in TNG50 have an inner slope of -2 and not -1 (see Figure \ref{fig:example_profiles}). A slope of -2 however can only be found at the scale radius $r_\mathrm{s}$ for the NFW profile. Therefore, the parameters have to take extreme and unrealistic values, stretching the profile to a large extent in order to produce a reasonably good fit.

The Einasto profile has a better goodness of fit than the NFW profile. For lower-mass subhaloes with $M_\mathrm{sh} \sim 10^8$ M$_{\sun}$ it even shows a slightly better performance than the new models presented in Section \ref{sec:additional_models}, and it also provides the best fit for the average density profiles of the lowest-mass subhaloes. The reason for this can be seen in Figure \ref{fig:example_profiles}: the lower-mass subhaloes have a much more regular shape than the higher-mass ones, with no sharp truncation or central bump and a gradually changing logarithmic slope. Those are the properties reflected best by the Einasto profile. 

The modified NFW and Schechter profiles fit the individual density profiles equally well most of the time. However, the modified NFW profile is preferred in general, since the mean goodness of fit values for the MS-(2, $n$) and the MS-($\gamma$, 2) model are often a lot worse (solid lines in Figure \ref{fig:performances_mass}), which indicates that there are more outliers for which the fit fails. The modified NFW profile also provides a very good fit for most of the average density profiles. For subhaloes with masses above $10^{11}$ M$_\odot$ the MS-(2, $n$) works slightly better because the shape of the truncation can no longer be well-described by the modified NFW. The average profiles of subhaloes at the uppermost mass end are best described by the MS-($\gamma$, 2) profile, which is mostly due to the fact that it allows for variations of the inner logarithmic slope, and this can partially account for the central bumps that often occur for these subhaloes.  

Our results are compatible with those from \citet{subhalo_profiles1}, who looked at the density profiles of subhaloes with masses above $2 \times 10^8$ M$_{\sun} h^{-1}$, with 56 in the hydro and 66 in the dark-matter-only run. They computed the mean values of the goodness of fit for the entire sample of subhaloes, where most of them had masses in the range $10^{8.5}-10^{9.5} $ M$_\odot$, and they found that among their tested models, the Einasto profile provides the best fit for both the hydrodynamic and the dark-matter-only only run. They have not analysed the goodness of fit for individual mass bins and they also did not account for the density uncertainties.\\

\subsection{Performance for Different Distances from the host halo centre}
Figure \ref{fig:performances_distance_bf} (on the left) shows the goodness of fit as a function of the subhalo distance from the host centre (expressed as fraction of the host halo's virial radius R$_{200}$). The goodness of fit values are relatively independent of distance from the host halo centre. They only become better for subhaloes closer than 0.1 R$_{200}$ for most of the new models presented in Section \ref{sec:additional_models} as well as for the Einasto profile. The truncated NFW profile fit becomes worse with decreasing distance. We found similar trends for the $\chi^2$-value also when looking at individual mass bins. The radial distribution of subhaloes also looks similar for the individual mass bins, with the exception that more massive subhaloes are less likely to be found both closer to the centre and further out than the virial radius R$_{200}$. 

\subsection{Performance for Other Subhalo Properties}

We also checked how the goodness of fit varies with baryon fraction, triaxiality, group mass, concentration, infall time as well as $V_\mathrm{max}$ and $R_\mathrm{max}$, i.e. the peak circular velocity and the radius at which this velocity is reached. More details can be found in Appendix \ref{sec:appendix}.

\subsubsection*{\textbf{Baryon fraction}}
Figure \ref{fig:performances_distance_bf} (on the right) presents the goodness of fit as a function of the baryon fraction. It turns out that the fits become slightly better with higher baryon fractions, both for our new models and for the Einasto profile. The same is also true when looking at the majority of individual mass bins. For the truncated NFW profile, the fits become slightly worse with a higher baryon content. The typical baryon fractions that the most massive subhaloes have is around 10-15 per cent (see also Figure \ref{fig:mass_concentration_baryons}). \\

\begin{figure}
  \includegraphics[width=\columnwidth]{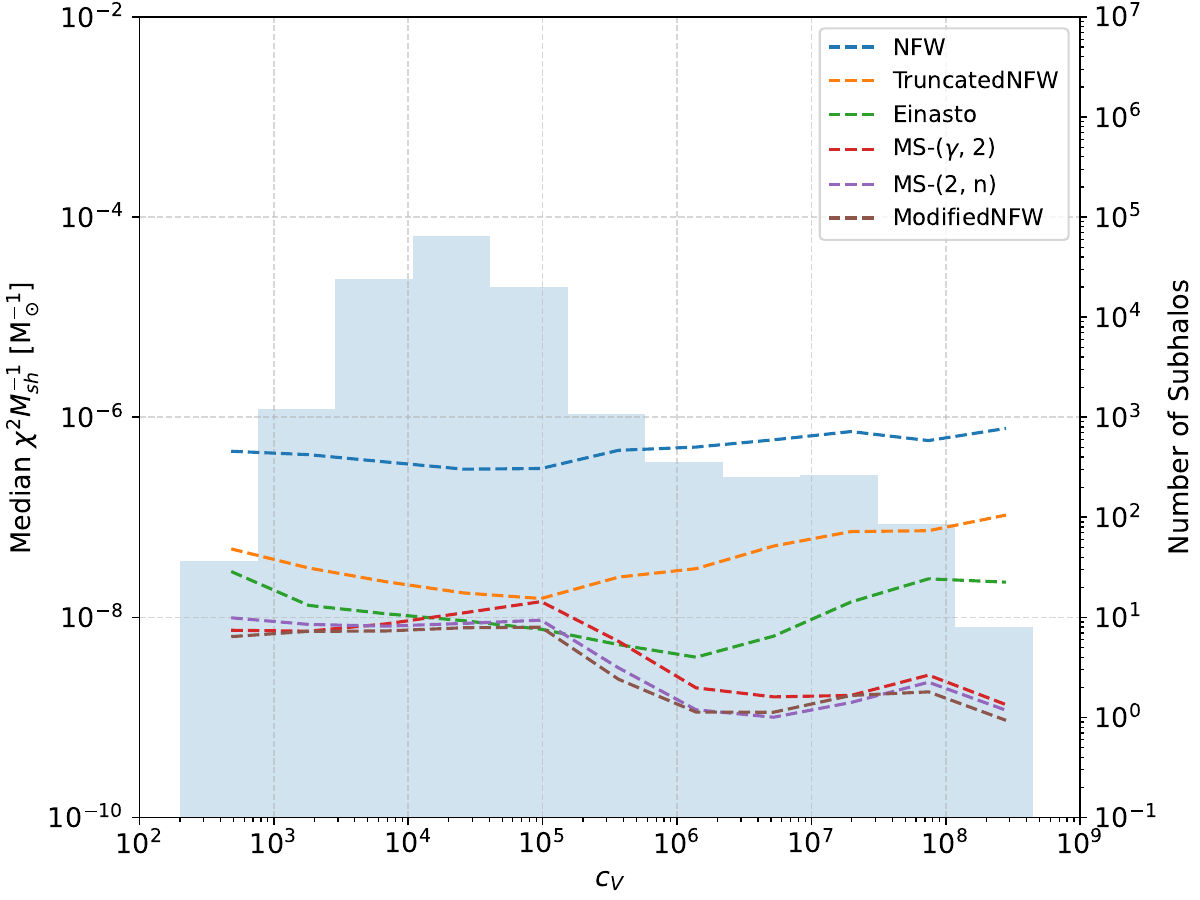}
  \caption{Median values of the goodness of fit per unit subhalo mass for subhaloes with different concentrations. Additionally, the distribution of concentrations is shown in the histogram.}
  \label{fig:performance_concentration_tot}
\end{figure}

\subsubsection*{\textbf{Triaxiality}}

 We define the subhalo triaxiality as \citep{triaxiality0}:
\begin{equation}
    T = \frac{a^2-b^2}{a^2-c^2},
\label{eq:triaxiality}
\end{equation}
where $a$, $b$ and $c$ are the major, intermediate and minor axes of a triaxial isodensity surface. We computed these values by diagonalising the shape tensor, which is also sometimes called inertia tensor \citep{triaxiality1, triaxiality2}: 
\begin{equation}
    S_{ij} = \frac{\sum_k m_k r_{k,i}r_{k,j}}{\sum_k m_k},
\label{eq:inertia_tensor}
\end{equation}
with the mass $m_k$ and the the $i$-th and $j$-th coordinate $r_{k,i}$ and $r_{k,j}$ of particle $k$. The axes $a$, $b$ and $c$ are the square roots of the eigenvalues of $S_{ij}$. Using these values to bin the data, no change of the goodness of fit with subhalo triaxiality has been found (see Figure \ref{fig:performance_triaxiality} in the Appendix \ref{sec:appendix}). 

\subsubsection*{\textbf{Concentration, V$_\mathrm{max}$ and R$_\mathrm{max}$}}

Isolated haloes which are well described by the NFW profile have a concentration defined as $c_{200}=r_{200}/r_\mathrm{s}$ with the virial radius $r_{200}$ and the scale radius $r_\mathrm{s}$. This definition is less suitable for subhaloes because the virial radius is not really well defined due to tidal stripping \citep{moline2017, halo_review}. The concentration of a subhalo can be described independently of a density profile by using the peak circular velocity $V_\mathrm{max}$, as:
\begin{equation}
    c_V = \frac{\bar{\rho}(< R_\mathrm{max})}{\rho_\mathrm{c}} = 2 \left( \frac{V_\mathrm{max}}{H R_\mathrm{max}} \right)^2,
\label{eq:subhalo_concentration}
\end{equation}
where $\bar{\rho}$ is the mean density within the radius $R_\mathrm{max}$ corresponding to $V_\mathrm{max}$ in units of the critical density $\rho_\mathrm{c}$ and $H$ is the Hubble parameter \citep{diemand2007, springel2008}. We thus use this definition for the concentration of a subhalo throughout this paper.

\begin{figure}
  \includegraphics[width=\columnwidth]{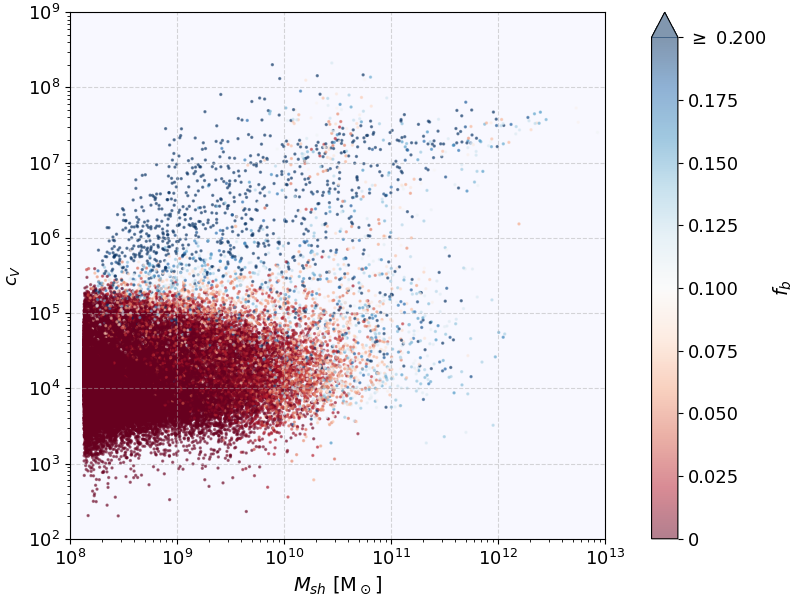}
  \caption{Subhalo mass vs. concentration with colour-coded baryon fraction for all of the subhaloes. A small population of highly concentrated, massive and baryon-rich subhaloes can be identified in the upper half of the diagram.}
  \vspace{-3pt}
  \label{fig:mass_concentration_baryons}
\end{figure}

For our new models, the goodness of fit per unit subhalo mass slightly decreases for $c_V>10^5$ (see Figure \ref{fig:performance_concentration_tot}). The reason for this can be inferred from Figure \ref{fig:performances_mass} and \ref{fig:mass_concentration_baryons}: the goodness of fit value per unit subhalo mass is a bit smaller for higher-mass subhaloes and very high concentrations occur primarily for more massive subhaloes. The performance of the new models also stays relatively constant for different values of $R_\mathrm{max}$ and $V_\mathrm{max}$, although it becomes slightly worse for values close to $R_\mathrm{max}=1$ kpc. More details can be found in the appendix in Figure \ref{fig:performace_vmax_rmax}.

Figure \ref{fig:mass_concentration_baryons} also shows that the subhaloes with $c_V>10^5$ generally also have high baryon fractions $f_\mathrm{b}>0.10$ and one can additionally show that they also have early infall times. This suggests that these high concentrations are not only facilitated by tidal stripping but also by baryonic processes, such as adiabatic contraction \citep{adiabatic_contraction1, adiabatic_contraction2}. These more concentrated subhaloes do not necessarily have an additional bump in the central regions, as the one in Figure \ref{fig:example_profiles} (upper left corner), but the majority of subhaloes with masses larger than $10^{11}$ M$_{\sun}$ has one, and it generally becomes a lot more pronounced for subhaloes with even higher masses and baryon contents.

\subsubsection*{\textbf{Infall time}}

In Figure \ref{fig:performance_infall_time} in the appendix, one can see how the goodness of fit value varies with infall time. The infall time has been obtained by using the \textsc{SubLink} merger tree in order to determine the snapshot at which both the subhalo and its host halo start sharing the same group number. It should be noted, though, that the subhalo could have also fallen into another halo at an even earlier time, which later became a subhalo of the host halo at $z=0$. For these sub-subhaloes the time at which severe tidal interactions happened is even earlier. We did not consider these cases specifically and only used the infall time into the main halo at $z=0$ as a first estimate of when tidal interactions became relevant for each subhalo. One can see that the goodness of fit does not depend much on the infall time for all models. We also found a similar behaviour for the individual mass bins. \\

\begin{figure*}
\includegraphics[width=\textwidth]{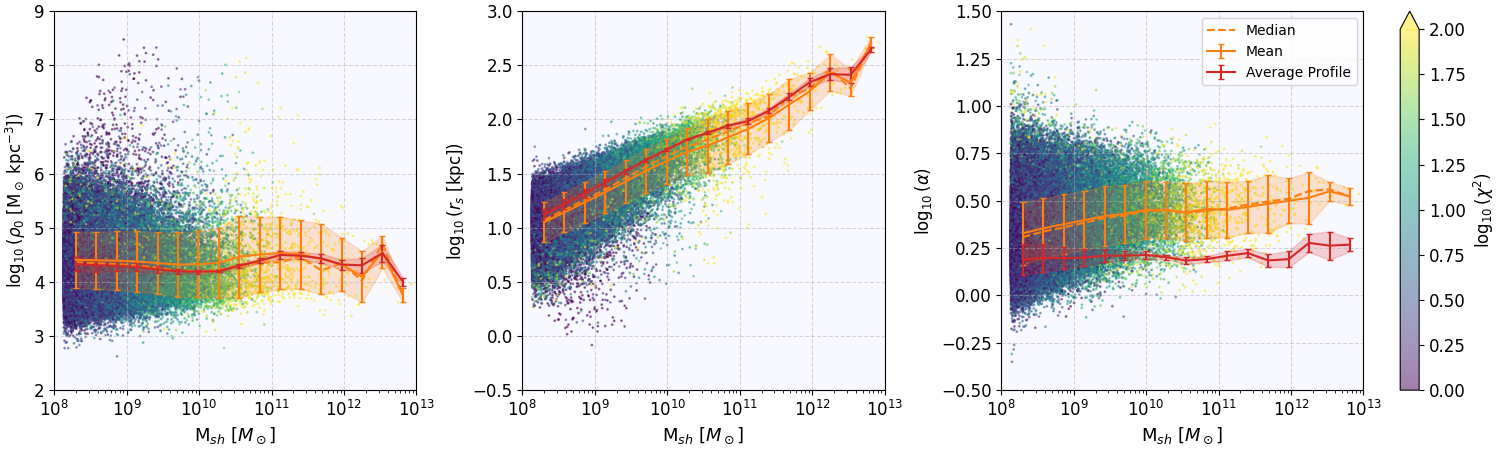}
\caption{Subhalo mass dependence of the parameters of the modified NFW profile from Equation (\ref{eq:modified_NFW}) with colour-coded goodness of fit. The solid and dashed orange lines show the mean and median values of the distribution for each mass bin, together with the standard deviation. The red solid line indicates the parameter values of the average density profiles for every mass bin, together with their uncertainties. The average density profiles can be found in Figure \ref{fig:average_profiles_mass}.}
\label{fig:parameter_mass_scatter}
\end{figure*}
\begin{figure*}
\includegraphics[width=\textwidth]{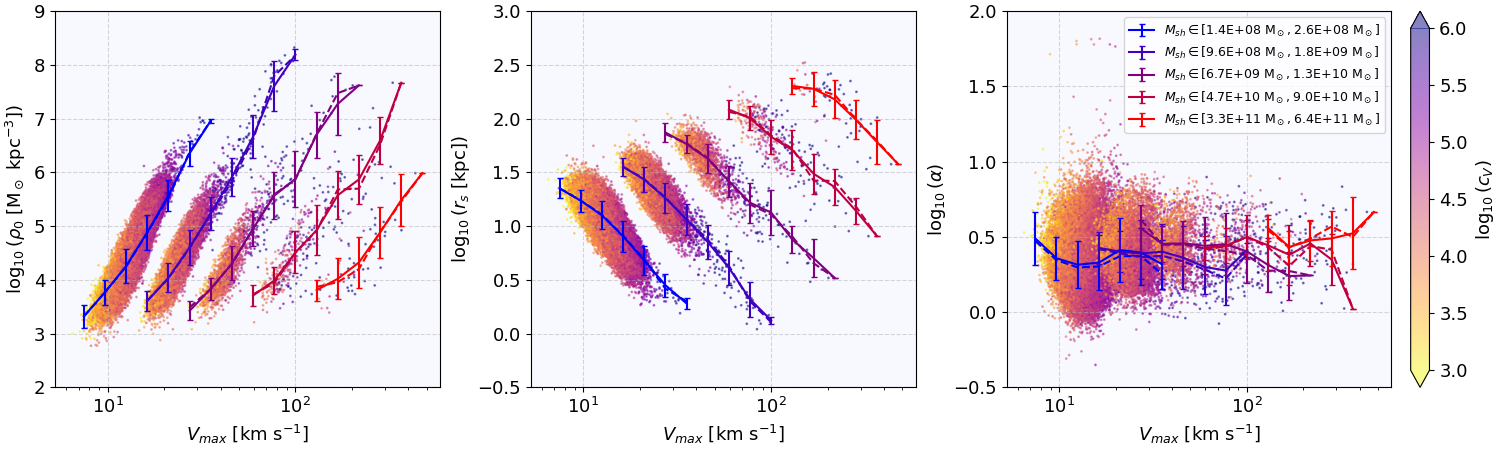}
\caption{Dependence of the modified NFW parameters on $V_\mathrm{max}$ for five different mass bins with colour-coded subhalo concentration. The solid and dashed lines indicate the mean and median of each distribution for several $V_\mathrm{max}$ bins, together with the standard deviations.}
\label{fig:parameter_vmax_scatter}
\end{figure*}

In summary, we were able to show that the new models proposed in Section \ref{sec:additional_models} perform much better than the NFW profile and its smoothly truncated version, both for the individual and for the average density profiles. The Einasto profile performs similarly well or even better for subhaloes with masses around $10^8$ M$_\odot$. The new models can provide a good fit, largely independent of concentration, baryon fraction, distance from the host halo centre, group mass and triaxiality. The model which performs best for most of the subhaloes is the modified NFW profile from Equation (\ref{eq:modified_NFW}). In some regimes, other models provide a better fit for the average density profiles, but the modified NFW profile still has a comparable goodness of fit. We will therefore focus the analysis in Section \ref{sec:parameter_relations} on the modified NFW profile and only mention the other models if they enable a better understanding of the density profiles in general. A more detailed analysis of the other models can be found in the Appendix \ref{sec:appendix}.

\section{Scaling relations for the Modified NFW Profile and average density profiles}
\label{sec:parameter_relations}
In this section, we discuss in more detail the new modified NFW profile introduced in Section \ref{mfnw}, which we find to be the best fit to the simulated subhaloes. We present the distribution of the fit parameters and derive scaling relations as a function of the subhalo mass $M_\mathrm{sh}$ and maximum circular velocity $V_\mathrm{max}$ that can be used to generate analytical profiles for a population of subhaloes. 

Finally, we discuss how the profile properties vary for individual subhaloes, depending on their distances from the host halo centre, their concentrations, their baryon fractions and their infall times, using both the distribution of the parameter values of the modified NFW profile and the average density profiles.

\subsection{Scaling relations}
Figure \ref{fig:parameter_mass_scatter} shows how the parameters ($\rho_0$,$r_\mathrm{s}$,$\alpha$) of the modified NFW profile (see Equation \ref{eq:modified_NFW}) depend on the subhalo mass. The parameter $r_\mathrm{s}$ correlates strongly with mass, which is not surprising since it describes the radial extent of the subhalo. $\alpha$ also increases with mass, with a mean value of 2 for subhaloes with masses of $\sim10^8$ M$_\odot$ up to a value of 3.5 for subhaloes with masses above $10^{12}$ M$_\odot$.

When looking at mass alone, the mean value of the normalisation $\rho_0$ stays relatively constant around $10^{4.5}$ M$_\odot$ kpc$^{-3}$, but with a very large scatter. Fitting the mean values of the other distributions with a power law, we obtain the following simple scaling relations:
\vspace{10pt}
\begin{equation}
    r_\mathrm{s} = (43.4 \pm 3.2) \ \mathrm{kpc} \left( \frac{M_{\mathrm{sh}}}{10^{10} \ \mathrm{M_\odot}} \right)^{0.351 \pm 0.016},
\label{eq:rs_mass_fit}
\end{equation}
\begin{equation}
    \alpha = (2.647 \pm 0.065) \ \left( \frac{M_{\mathrm{sh}}}{10^{10} \ \mathrm{M_\odot}} \right)^{0.0439 \pm 0.0047}.
\label{eq:alpha_mass_fit}
\end{equation}

\begin{figure}
\includegraphics[width=\columnwidth]{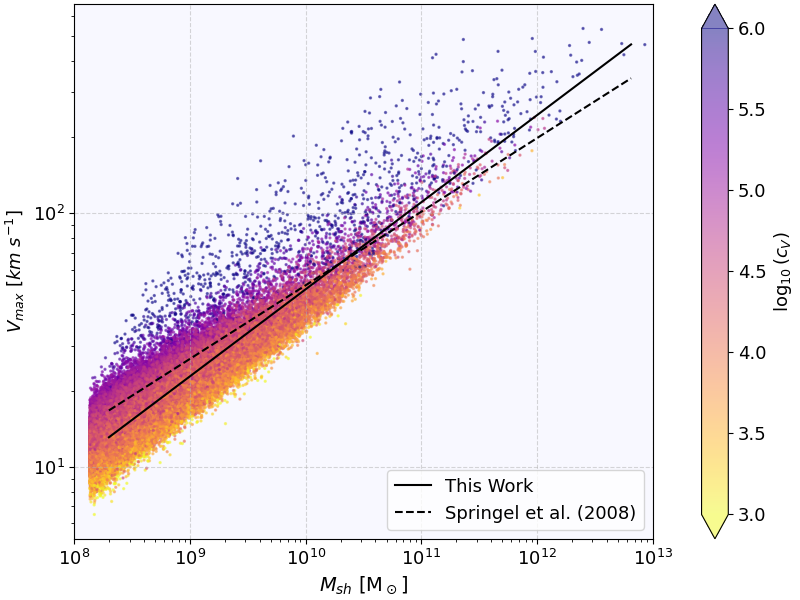}
\caption{$V_\mathrm{max}$ -- subhalo mass relation with colour-coded subhalo concentration. The fit to the mean values of each mass bin is represented by the solid black line. The dashed line indicates the result from \citet{springel2008}.}
\label{fig:vmax_m_relation}
\end{figure}

\begin{figure*}
\includegraphics[width=\columnwidth]{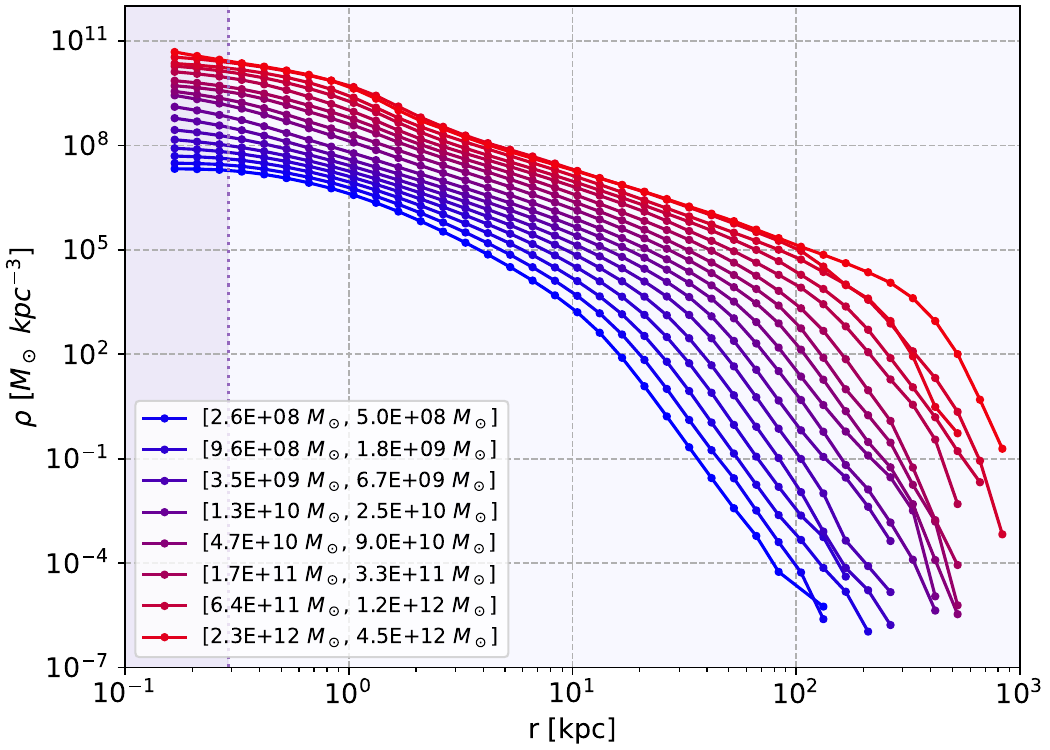}
\includegraphics[width=\columnwidth]{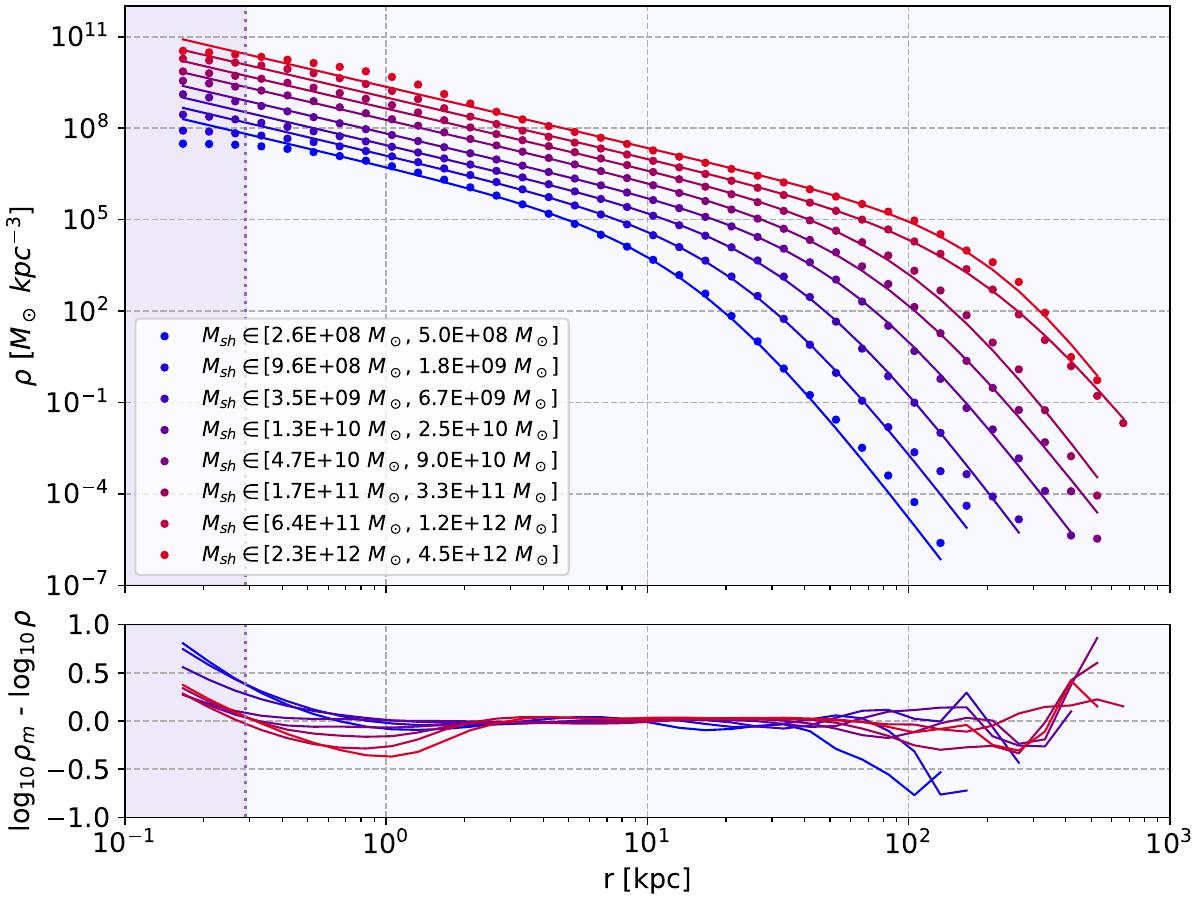}
\caption{The average density profiles for all the mass bins can be seen in the image on the left. On the right, the average density profiles for every second mass bin are plotted together with the best modified NFW profile fit. The residuals of the log-profiles are plotted below. The purple line indicates the softening length.}
\label{fig:average_profiles_mass}
\end{figure*}

Figure \ref{fig:parameter_mass_scatter} also shows the parameters obtained by fitting the average density profile of each mass bin with the Modified NFW (see Figure \ref{fig:average_profiles_mass}). While these values agree very well with the those of the individual profiles for the first two parameters, one can see that $\alpha$ is close to 1.5 for the average density profiles and therefore significantly differs from the mean values of the distribution. This is because within each mass bin, subhaloes of various different sizes go into the averaging process, which effectively creates a log-slope transition that looks quite different from the ones in most of the individual subhaloes.

We then obtain a more refined parametrisation by considering both the subhalo mass and $V_\mathrm{max}$. We obtain relations with much less scatter for both $\rho_0$ and $r_\mathrm{s}$, which can be seen in Figure \ref{fig:parameter_vmax_scatter}. One can see that for each mass bin $\rho_0$ increases with $V_\mathrm{max}$ while $r_\mathrm{s}$ decreases with $V_\mathrm{max}$, which is the expected behaviour for higher concentrations. The concentrations are colour-coded and follow the increase in $V_\mathrm{max}$. Within a single mass bin, the parameter $\alpha$ does not change much with $V_\mathrm{max}$. That is the reason why we only describe it in terms of the subhalo mass. Analytic relations that fit the other mean parameter values are given by:

\begin{equation}
    \rho_0 = (4.9 \pm 0.6) \mathrm{\frac{M_\odot} {kpc^{3}}} \left( \frac{M_{\mathrm{sh}}}{10^{13} \mathrm{M_\odot}} \right)^{-1.9 \pm 0.1} \left( \frac{V_{\mathrm{max}}}{10^2 \mathrm{km \ s^{-1}}} \right)^{5.2 \pm 0.1}
\label{eq:rho0_vmax_fit}
\end{equation}
\begin{equation}
    r_\mathrm{s} = (10.2 \pm 0.1) \mathrm{kpc} \left( \frac{M_{\mathrm{sh}}}{10^{10} \mathrm{M_\odot}} \right)^{0.9 \pm 0.1} \left( \frac{V_{\mathrm{max}}}{10^2 \mathrm{km \ s^{-1}}} \right)^{-1.6 \pm 0.1}.
\label{eq:rs_vmax_fit}
\end{equation}
Besides that, there is also a general relation between $V_\mathrm{max}$ and the subhalo mass. \citet{springel2008} found:
\begin{equation}
    V_\mathrm{max} = 10 \unit{km \ s^{-1}} \left( \frac{M_\mathrm{sh}}{3.37 \times 10^7 \unit{M_\odot}} \right)^{0.29},
\label{eq:c_vmax_relation_springel}
\end{equation}
which is close to the relation that we obtain:
\begin{equation}
    V_{\mathrm{max}} = 10 \ \mathrm{km \ s^{-1}} \left( \frac{M_{\mathrm{sh}}}{(9.0 \pm 1.7) \times 10^7 \ \mathrm{M_\odot}} \right)^{0.343 \pm 0.007}.
\label{eq:c_vmax_relation}
\end{equation}
The relation is depicted in Figure \ref{fig:vmax_m_relation}. One can see that there is a noticeable scatter due to different subhalo concentrations. $V_\mathrm{max}$ increases with $M_\mathrm{sh}$ according to Equation (\ref{eq:c_vmax_relation}) and within each mass bin $V_\mathrm{max}$ increases with concentration (which can also be seen in Figure \ref{fig:parameter_vmax_scatter}). This increase in concentration and $V_\mathrm{max}$ leads to an increase of $\rho_0$ and a decrease of $r_\mathrm{s}$ as described by the Equations (\ref{eq:rho0_vmax_fit}) and (\ref{eq:rs_vmax_fit}). If we plug Equation (\ref{eq:c_vmax_relation}) into Equation (\ref{eq:rs_vmax_fit}) we approximately obtain back Equation (\ref{eq:rs_mass_fit}). 

\begin{figure}
\includegraphics[width=\columnwidth]{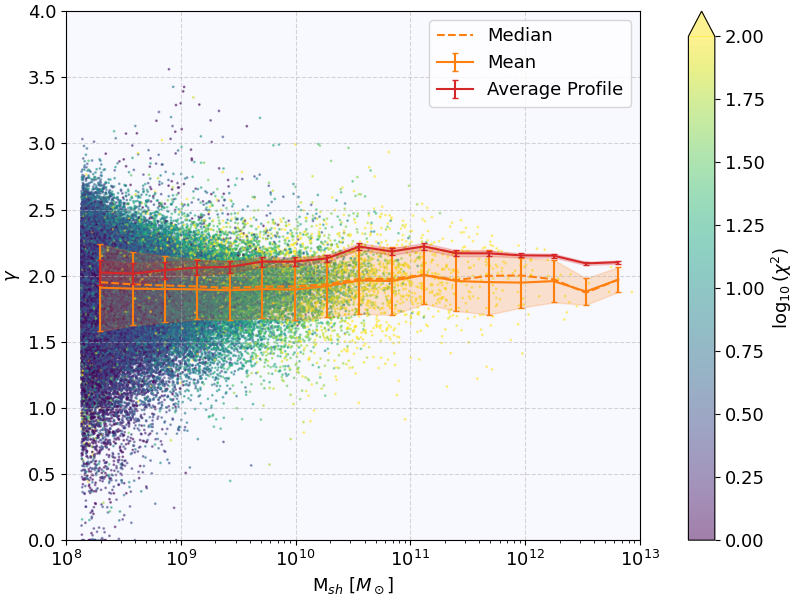}
\caption{Subhalo mass dependence of the $\gamma$-parameter of the MS-($\gamma$, 2) profile from Equation (\ref{eq:schechter}) with colour-coded goodness of fit as well as the mean, median and average density profile values for each mass bin.}
\label{fig:schechter_slope}
\end{figure}

In the Appendix \ref{sec:appendix}, we also included similar plots and analytical expressions for the scaling relations for the other density profile models. In Figure \ref{fig:schechter_slope}, we want to additionally highlight how the inner log-slope parameter $\gamma$ of the MS-($\gamma$, 2) profile depends on the subhalo mass. One can see that the mean values for the inner slope are close to -2 for all subhalo mass bins and the same applies to the parameter values of the average profiles, although with an offset to slightly higher values. The scatter becomes larger for lower-mass subhaloes since the behaviour of the log-slopes changes as discussed before.

\subsection{Average Density Profiles}
In Figure \ref{fig:average_profiles_mass}, we show the average density profiles for different mass bins, both with (right image) and without the modified NFW profile fits (left image). The average profiles have been computed by taking the average of the density values of each radial bin for all subhaloes in the same mass bin (including density values of zero). One can see that subhaloes with masses below $10^{10}$ M$_\odot$ have a gradually decreasing log-slope in the central regions, while above $10^{11}$ M$_\odot$ the central bump becomes more and more prominent. In both regimes the density profile has a central core, while for masses between $10^{10}$ M$_\odot$ and $10^{11}$ M$_\odot$ the log-slope stays constant all the way to the centre, resulting in a cusp. The modified NFW profile performs very well in almost all of the cases shown. The only regimes in which it strongly deviates from the true values are in the central regions of very massive subhaloes and the outer regions, where the standard deviations are very large. In the Appendix \ref{sec:appendix} in Figure \ref{fig:average_profile_example} one can see two examples of average density profiles for given mass bins, together with the best modified NFW profile fits as well as the combined density uncertainties and standard deviations.

\begin{figure*}
\includegraphics[width=0.95\columnwidth]{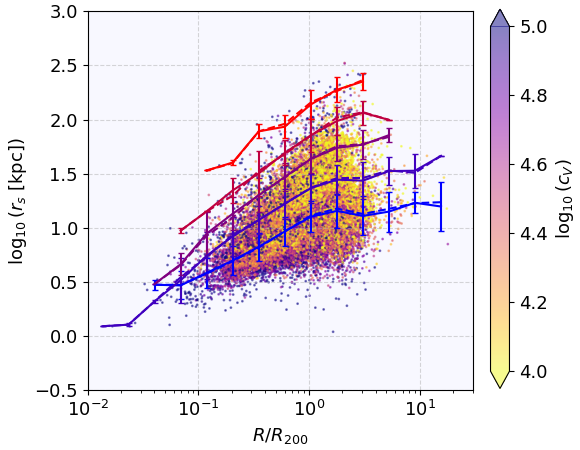}
\includegraphics[width=\columnwidth]{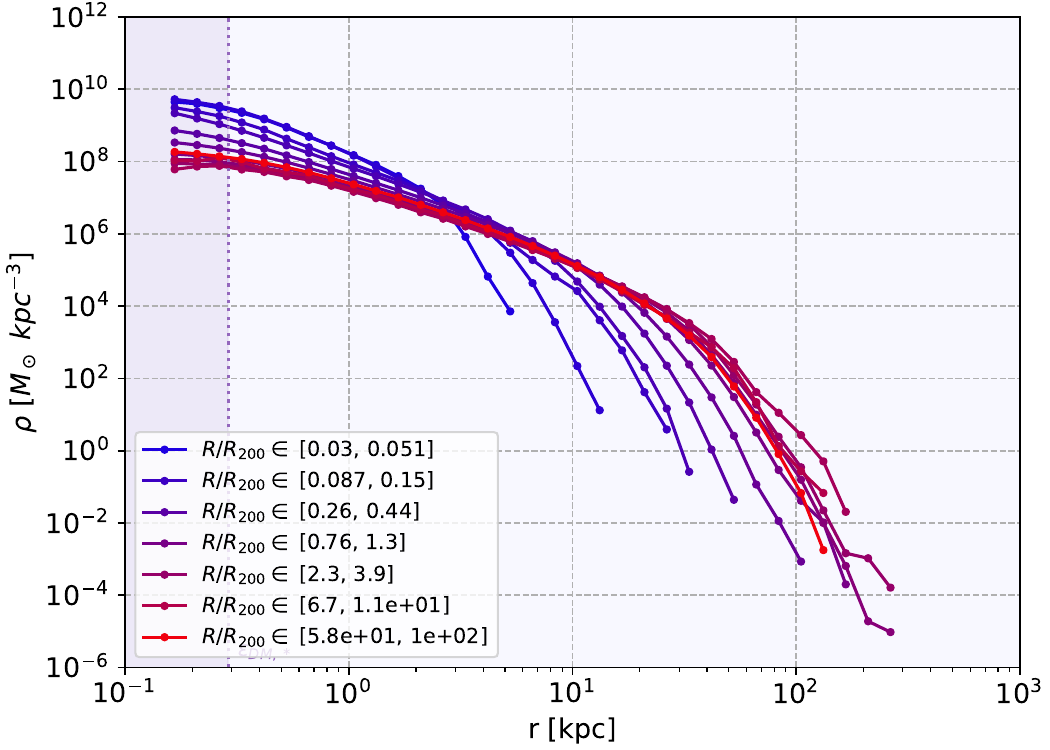}
\caption{\textit{Left:} dependence of the $r_\mathrm{s}$-parameter of the modified NFW profile on host halo centre distance for different mass bins with colour-coded concentration. The solid and dashed lines indicate the mean and median of each distribution for different distance bins, together with the standard deviations. A complete version of this plot can be found in Figure \ref{fig:parameter_distance_scatter_complete} in the Appendix \ref{sec:appendix}. \textit{Right:} average density profiles for different distances from the host halo centre for subhaloes with masses between $3.5 \times 10^9$ M$_\odot$ and $6.7 \times 10^9$ M$_\odot$. The purple region indicates the regime below the softening length.}
\label{fig:parameter_distance_scatter}
\end{figure*}

\begin{figure*}
\includegraphics[width=\columnwidth]{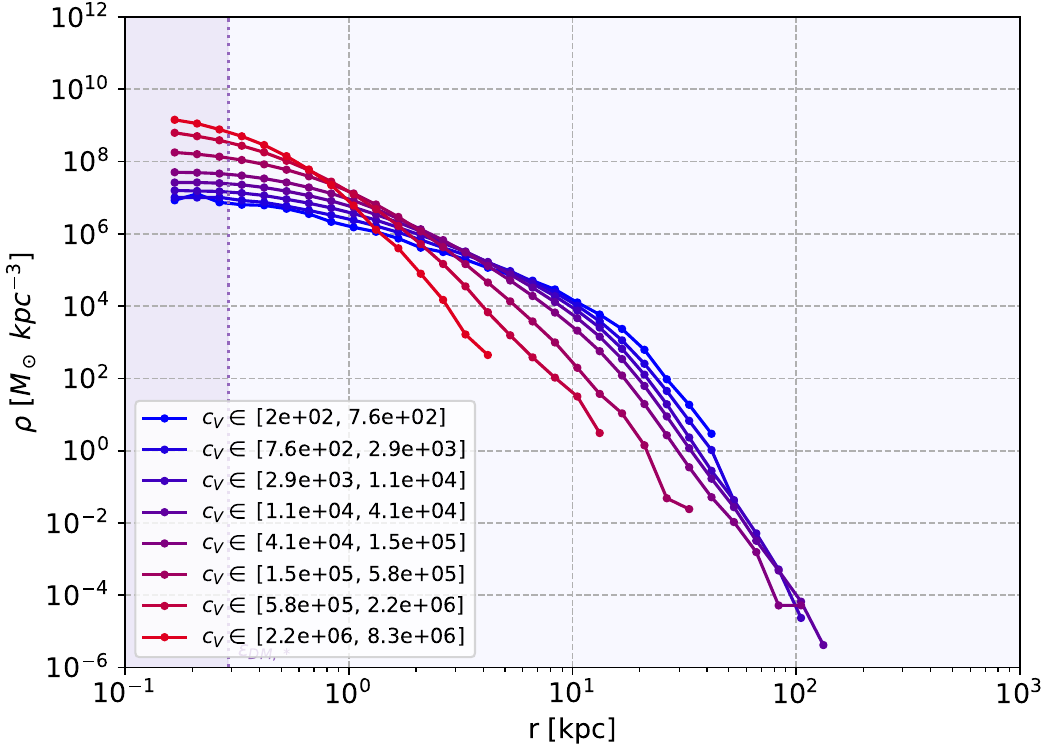}
\includegraphics[width=\columnwidth]{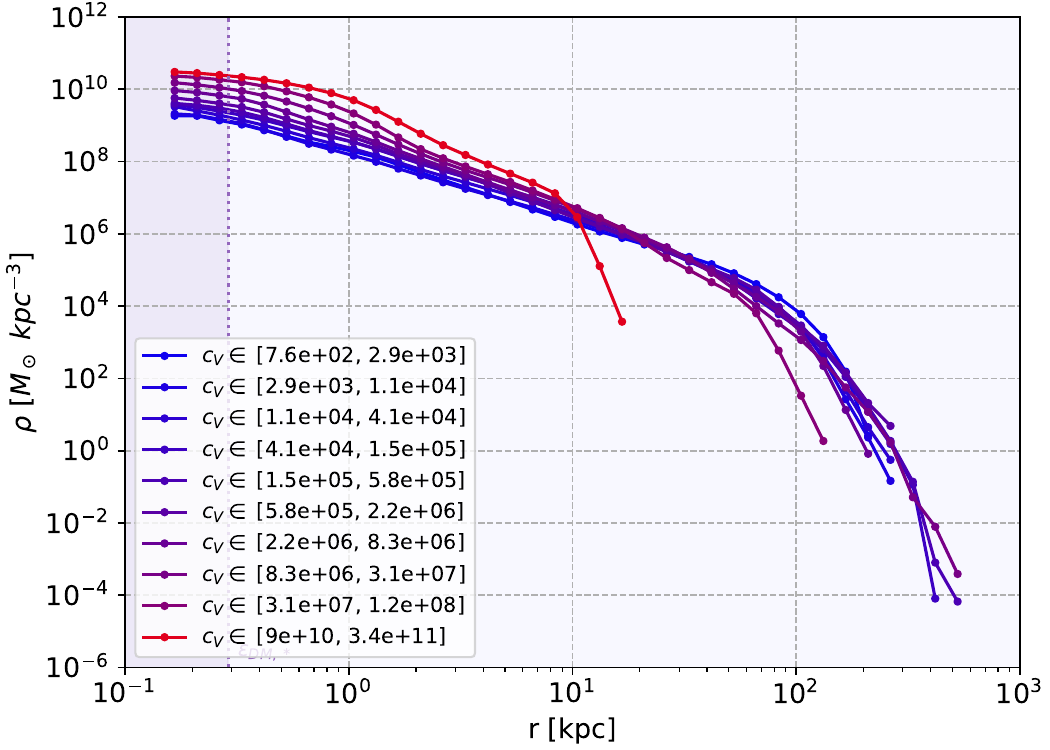}
\caption{The average density profiles for different subhalo concentration bins are shown. On the left for subhaloes with masses between $2.6 \times 10^8$ M$_\odot$ and $5.0 \times 10^8$ M$_\odot$ and on the right for subhaloes with masses between $1.7 \times 10^{11}$ M$_\odot$ and $3.3 \times 10^{11}$ M$_\odot$. The purple line indicates the softening length.}
\label{fig:average_profiles_concentration}
\end{figure*}

Within a single mass bin the profile shapes can be quite diverse. In the rest of this section, we investigate the deviations from these average profiles due to different subhalo distances from the host halo centre, different concentrations, baryon fractions and infall times (see Figures \ref{fig:parameter_distance_scatter}, \ref{fig:average_profiles_concentration}, \ref{fig:average_profiles_infalltime} and \ref{fig:average_profiles_baryons}).

\subsection{Dependence of the Parameters on Host Halo Distance}
The left panel in Figure \ref{fig:parameter_distance_scatter} shows how the $r_\mathrm{s}$-parameter of the modified NFW profile varies with distance to the host halo centre for different mass bins. One can see that below the virial radius $R_{200}$ the value of $r_\mathrm{s}$ decreases with distance in a very similar way for all subhalo mass bins by up to a factor of 10. Beyond the virial radius (i.e. for subhaloes associated with the FOF group, but located at $r>R_{200}$), the mean values of $r_\mathrm{s}$ stay more or less constant, which is consistent with the fact that these subhaloes are not yet subjected to tidal stripping. The colour coding also shows that the subhalo concentration increases towards the host halo centre. This behaviour is expected, since subhaloes closer to the centre generally also have earlier infall times and are therefore more tidally truncated. As $r_\mathrm{s}$ decreases, $\rho_0$ increases and $\alpha$ slightly increases as well for all mass bins in a similar way (see Figure \ref{fig:parameter_distance_scatter_complete} in the Appendix \ref{sec:appendix}).

The change in $r_\mathrm{s}$ can also be seen in the average profiles. The right panel in Figure \ref{fig:parameter_distance_scatter} shows the average density profiles for different distances from the host halo centre in the mass bin between $3.5 \times 10^9$ M$_\odot$ and $6.7 \times 10^9$ M$_\odot$. Inside the virial radius, the effect of tidal stripping clearly depends on distance. Besides that, the inner slope becomes slightly steeper.

\begin{figure*}
\includegraphics[width=\textwidth]{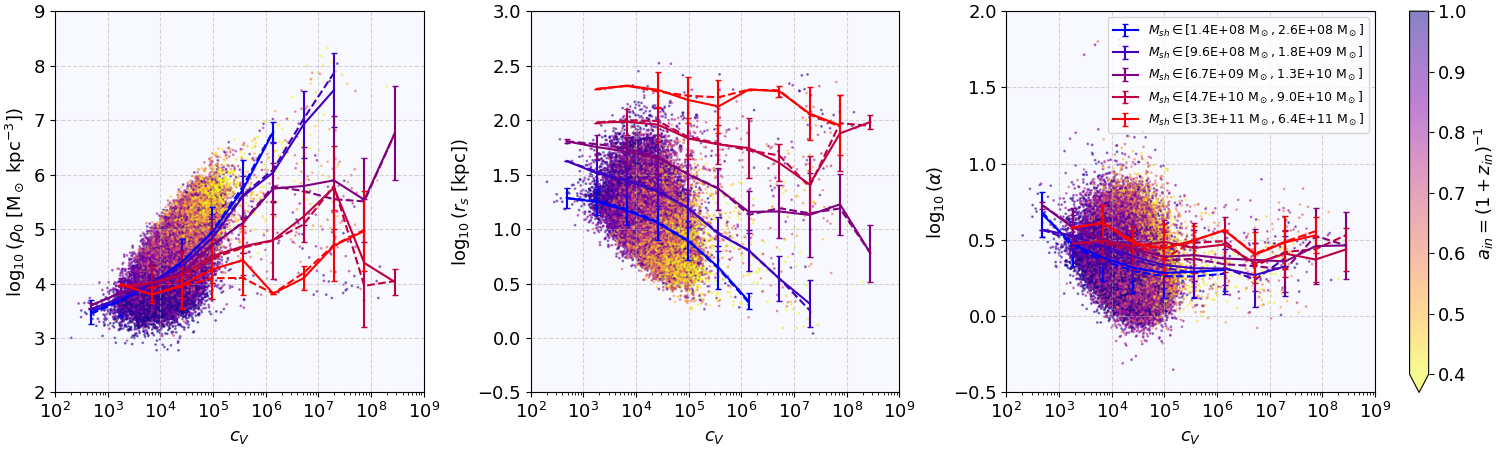}
\caption{Dependence of the modified NFW parameters on the subhalo concentration $c_V$ for five different mass bins with colour-coded infall time $a_\mathrm{in}$. The solid and dashed lines indicate the mean and median of each distribution for several $c_V$ bins, together with the standard deviations.}
\label{fig:parameter_concentration_scatter}
\end{figure*}

\subsection{Dependence of the Parameters on Subhalo Concentration}
In Figure \ref{fig:parameter_concentration_scatter} the relationships between the modified NFW parameters and the subhalo concentration are depicted for five different mass bins. The distributions are not as clearly separated as for $V_\mathrm{max}$, they also have a larger scatter and the mean values show more fluctuations. Nonetheless, one can see that for higher-mass subhaloes the parameter values change less with concentration than for lower-mass subhaloes. This can also be seen in the average density profiles for different concentrations in Figure \ref{fig:average_profiles_concentration}. For lower-mass subhaloes $r_\mathrm{s}$ becomes smaller and $\rho_0$ becomes larger for higher concentrations, as one would expect. However, for higher-mass subhaloes $\rho_0$ and $r_\mathrm{s}$ do not significantly vary with concentration and the increase in $c_V$ mainly comes from an increase in the strength and radius of the central bump, which is not parametrised by the model. Exceptions to this can be seen for the last two bins of exceptionally high concentrations, where there is an additional truncation besides the prominent central bump. However, concentrations above $10^9$ are extremely rare and only occur for four subhaloes in the sample. 

\subsection{The Effect of Tidal Interactions}
\label{sec:tides}
Once a subhalo gets sufficiently close to the centre of its host halo, the tidal forces become strong enough to remove matter from the outer regions: this starts to happen close to the virial radius $R_{200}$ (see Figure \ref{fig:parameter_distance_scatter}). The subhalo then continually loses orbital energy by dynamical friction and sinks closer towards the host halo centre. Matter gets removed beyond the tidal radius $r_\mathrm{tid}$, which is defined as the radius where the tidal force of the host halo is equal to the gravitational force of the subhalo. Besides tidal stripping in a slowly varying external potential, subhaloes can also experience tidal shocks if the external potential varies very rapidly, which happens either close to the main halo centre or during close encounters with other subhaloes \citep{halo_review}. This transfers orbital energy of the subhalo to internal energy of its particles, which can alter the subhalo's internal structure and make some of its particles unbound. \citet{artificial_disruption3} found that tidal shock heating due to close encounters with other subhaloes is negligible compared to the tidal effects of the host halo and that tidal shocks which significantly exceed the subhalo's binding energy generally do not lead to its complete disruption. 

\begin{figure*}
\includegraphics[width=\textwidth]{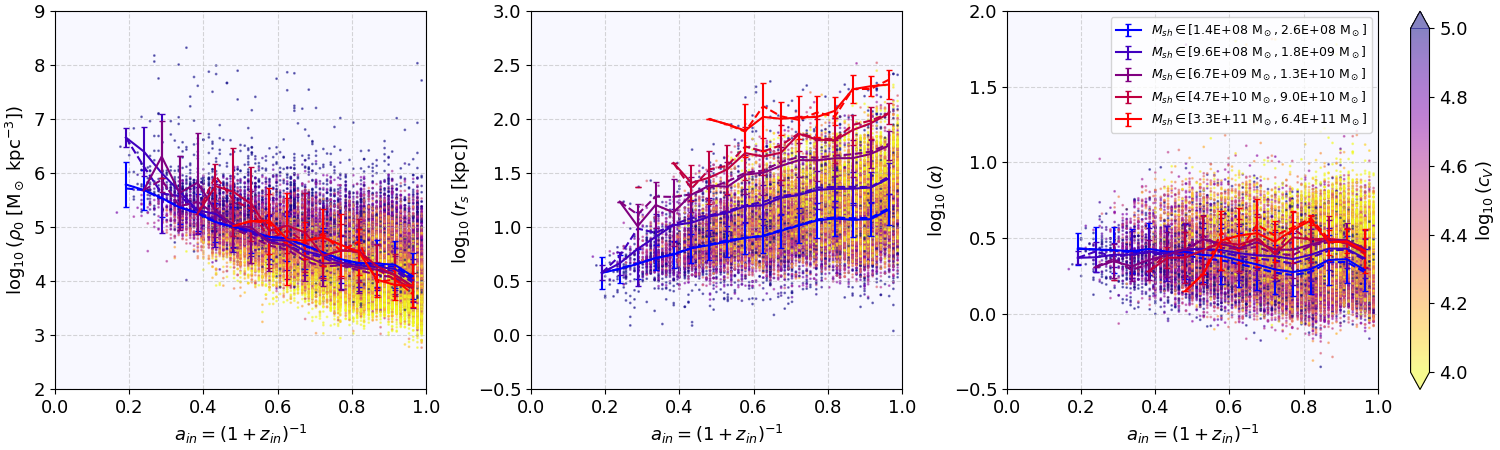}
\caption{Dependence of the modified NFW parameters on the infall time $a_\mathrm{in}$ for five different mass bins with colour-coded concentration $c_V$. The solid and dashed lines indicate the mean and median of each distribution for several $a_\mathrm{in}$ bins, together with the standard deviations.}
\label{fig:parameter_infalltime_scatter}
\end{figure*}

\begin{figure*}
\includegraphics[width=\columnwidth]{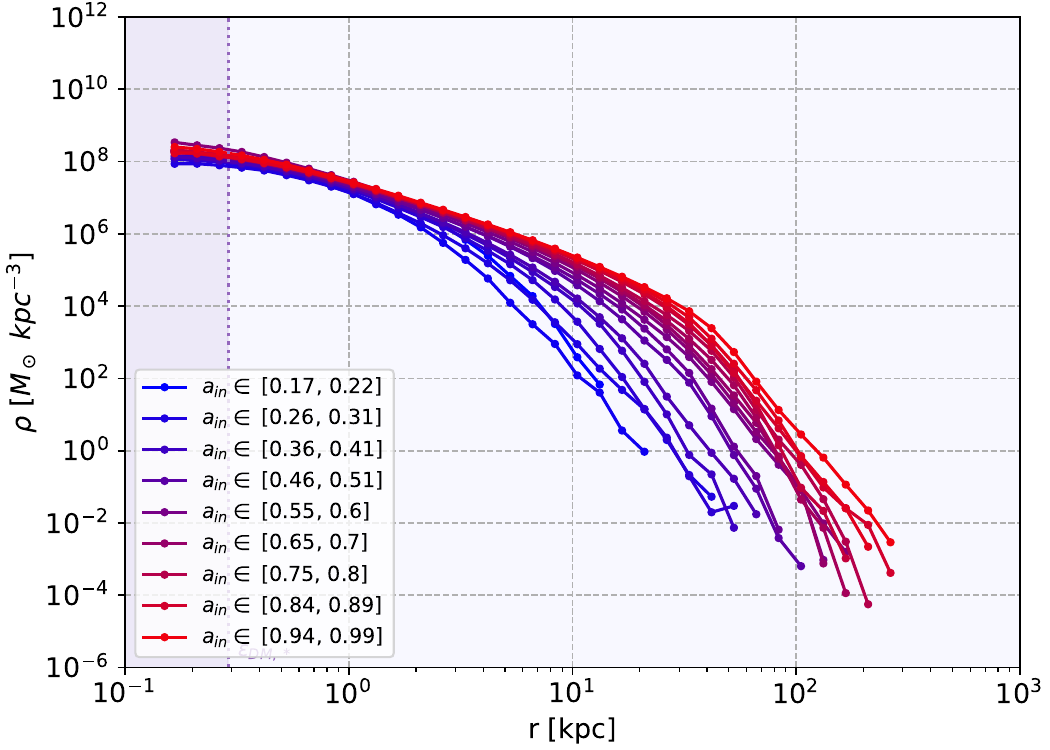}
\includegraphics[width=\columnwidth]{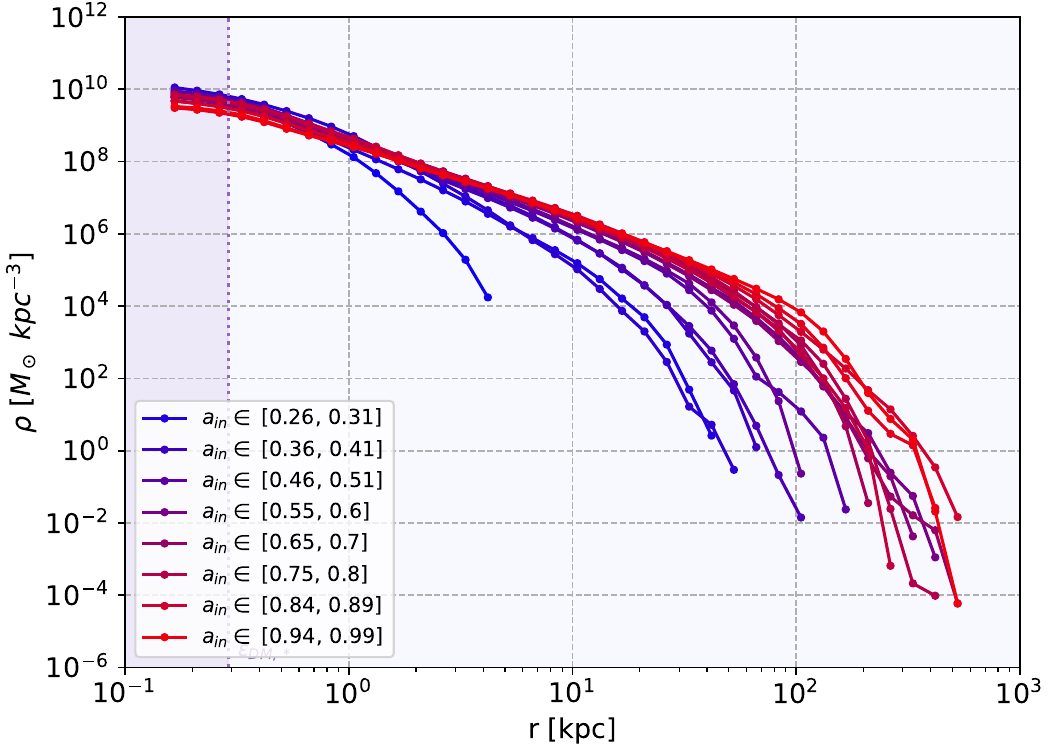}
\caption{The average density profiles for different subhalo infall time bins are shown. On the left for subhaloes with infall masses between $6.9 \times 10^9$ M$_\odot$ and $1.6 \times 10^{10}$ M$_\odot$ and on the right for subhaloes with masses between $1.9 \times 10^{11}$ M$_\odot$ and $4.3 \times 10^{11}$ M$_\odot$. The purple line indicates the softening length.}
\label{fig:average_profiles_infalltime}
\end{figure*}

\vspace{-10pt}
Because of the expansion of the dark matter particle orbits during tidal shocks, the inner density can be reduced, but the effect is usually not strong enough to produce a central core \citep{tidal_stripping}. Besides that, tidal effects mainly affect the outer regions of the subhalo, which can also be seen in our average density profiles for different infall times in Figure \ref{fig:average_profiles_infalltime}. However, the central density does not always decrease with earlier infall times and the shape parameter $\alpha$ stays relatively constant and depends primarily on the subhalo mass. Other studies, e.g. \citet{subhalo_trunc1} and \citet{subhalo_profiles3}, find a decrease of the central density and steeper truncations for individual subhalo density profiles as tidal stripping progresses. Figure \ref{fig:parameter_infalltime_scatter} shows that the parameters $\rho_0$ and $r_\mathrm{s}$ on average change in a very similar way for subhaloes with different masses and that the subhaloes become more concentrated for earlier infall times. There is, however, also a large number of subhaloes with recent infall times and a high concentration, which is not unexpected and reflects the different assembly histories. Subhaloes that formed earlier, when the Universe had a higher mean density, also have a higher concentration \citep{c_m_rel1}.

A more sophisticated analysis of the effects of tidal stripping on subhaloes with different properties at infall time would require following the evolution of each individual subhalo, which goes beyond the scope of this study.

\subsection{The Effect of Baryons}
Besides tidal truncation due to early infall times or small distances to the host halo centre, a high fraction of baryons is another indicator for a high subhalo concentration. We already saw in Figure \ref{fig:mass_concentration_baryons} that a high baryon fraction is generally present for subhaloes with $c_V > 10^6$. Baryons can cool, sink to the subhalo centre and gravitationally attract more mass, including dark matter. This leads to a significant density increase in the central regions, which could also lead to an increase in the inner log-slope. If the subhalo is massive and the gravitational potential well therefore deep enough, a large amount of baryons can accumulate in the centre and form stars. This is also the reason for the bump feature we see in subhaloes with masses larger than $10^{11}$ M$_\odot$ as these central bumps are mainly comprised of baryons. However, the dark matter distribution also gets reshaped due to gravitational interactions with the baryons, and thus the dark matter profile also shows an inner log-slope of -2 and can have a small central bump. For a comparison of the dark matter and baryon component of two subhalo density profiles, see Figure \ref{fig:density_profile_decomposition}. According to \citet{baryon_effects1}, feedback is the reason why the bump has a flat core in the centre. Tidal torques that the main galaxy disc exerts on subhaloes can disrupt them and additionally affect the shape of their density profiles \citep{baryon_effects4, baryon_effects5}. 

\begin{figure*}
\includegraphics[width=\textwidth]{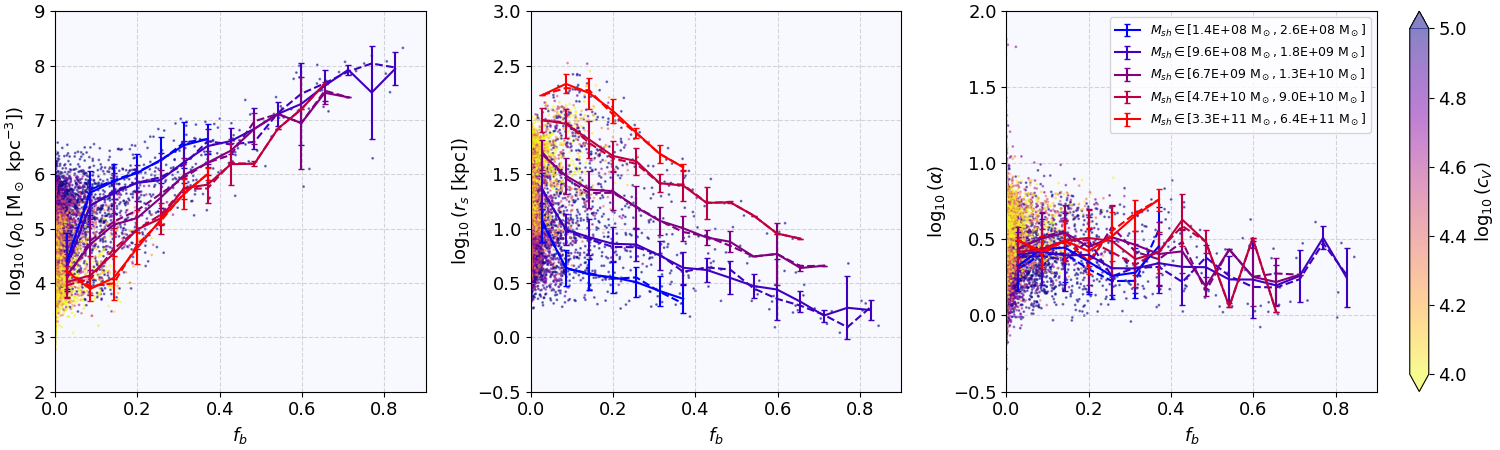}
\caption{Dependence of the modified NFW parameters on the subhalo baryon fraction $f_\mathrm{b}$ for five different mass bins with colour-coded concentration $c_V$. The solid and dashed lines indicate the mean and median of each distribution for several $c_V$ bins, together with the standard deviations.}
\label{fig:parameter_baryons_scatter}
\end{figure*}

\begin{figure*}
\includegraphics[width=\columnwidth]{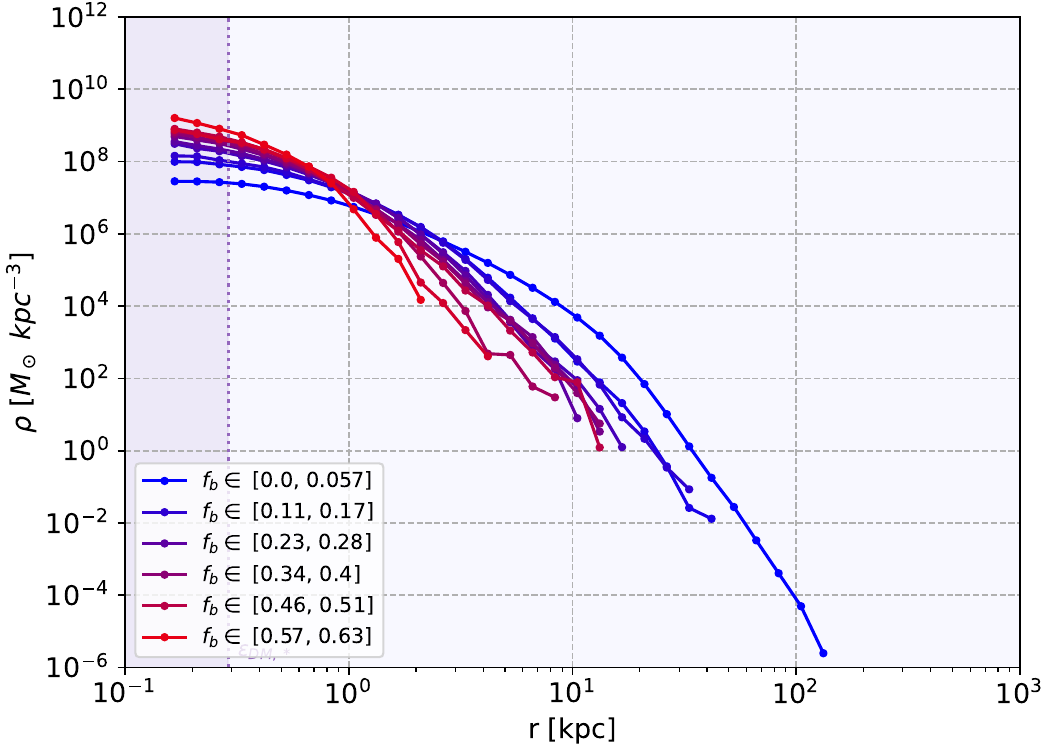}
\includegraphics[width=\columnwidth]{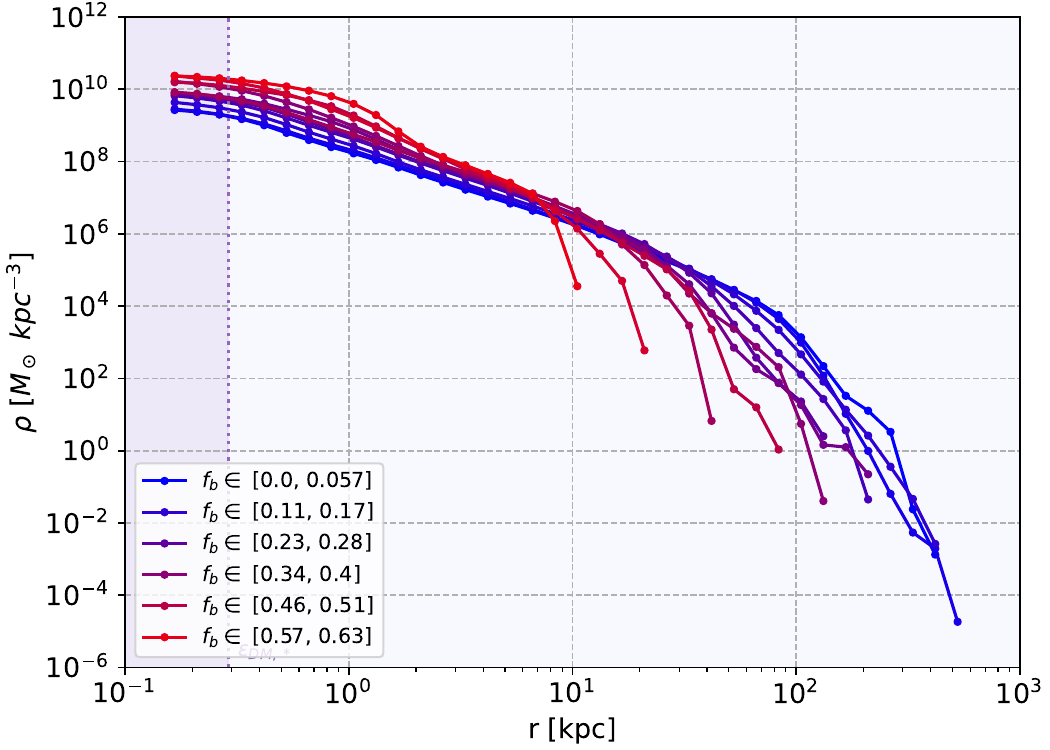}
\caption{The average density profiles for different subhalo baryon fraction bins are shown. On the left for subhaloes with masses between $2.6 \times 10^8$ M$_\odot$ and $5.0 \times 10^8$ M$_\odot$ and on the right for subhaloes with masses between $9.0 \times 10^{10}$ M$_\odot$ and $1.7 \times 10^{11}$ M$_\odot$. The purple line indicates the softening length.}
\label{fig:average_profiles_baryons}
\end{figure*}

In Figure \ref{fig:parameter_baryons_scatter}, the dependence of the modified NFW parameters on baryon fraction is shown for five different mass bins. We see that $r_\mathrm{s}$ decreases with increasing baryon fraction by up to a factor of 10 for very baryon-dominated subhaloes. Most of the subhaloes with a baryon fraction larger than 0.4 have a concentration higher than $10^6$. Analogously, $\rho_0$ increases with baryon fraction in a similar way for all mass bins, leading to a higher concentration.

Figure \ref{fig:average_profiles_baryons} shows the average density profiles for different baryon fractions for subhaloes with masses in the range between $2.6 \times 10^8$ M$_\odot$ and $5.0 \times 10^8$ M$_\odot$ (on the left) and for subhaloes in the mass bin between $9.0 \times 10^{10}$ M$_\odot$ and $1.7 \times 10^{11}$ M$_\odot$ (on the right). For less massive subhaloes $\rho_0$ increases and $r_\mathrm{s}$ decreases with baryon fraction, making the subhalo more concentrated, and additionally the inner log-slope becomes steeper. Here it should be noted, that tidal stripping removes mass from dark matter dominated outer regions, which also increases the ratio of baryon mass to total mass. For the more massive subhaloes, $r_\mathrm{s}$ also decreases with baryon fraction and the central bump becomes more pronounced.

Baryons can also have other effects, such as reducing the number of subhaloes by facilitating their disruption, especially for small subhalo masses and small distances from the host halo centre \citep{baryon_effects3, baryon_effects4}.

\section{Summary \& Conclusions}
\label{sec:conclusions}
In this paper, we have presented a detailed study of the density profiles of subhaloes with masses above $1.4 \times 10^8$ M$_\odot$ in the state-of-the-art cosmological simulation TNG50. As an improvement to many other simulations that have been used to study subhalo properties in previous works, TNG50 offers a good compromise between resolution and statistics, includes the effect of baryons and makes use of a large variety of other physical model components. The details of the outcomes presented in this paper could differ for simulations that use different ways to model the hydrodynamics, but overall, our main findings are mostly consistent with other studies.

We tested the performance of commonly used density profile models, such as the standard NFW, the Einasto and the truncated NFW profiles and further introduced new models, which show a much better performance on both the individual and the average density profiles. Of the models analysed, the best performing model is a modified version of the NFW profile, given by Equation (\ref{eq:modified_NFW}). The model parameter values follow simple scaling relations given by Equations (\ref{eq:rs_mass_fit}) - (\ref{eq:rs_vmax_fit}). For the lowest-mass subhaloes in the sample with masses around $10^8$ M$_\odot$, the Einasto profile shows a better performance compared to the new models. For the average density profiles of more massive subhaloes, the modified Schechter profiles, which model the truncation with an exponential factor, provide a slightly better fit. The goodness of fit of the new models is not significantly affected by other subhalo properties, including baryon fraction, distance from the host halo centre, triaxiality, concentration and infall time. This suggests that the new models are able to describe a large variety of subhaloes in the simulation. We have not considered models with more than three parameters in this study. It can be expected that the generalised NFW profile given by Equation (\ref{eq:alpha_beta_gamma_model}) provides an even better fit than the three-parameter model we proposed, since it allows for more variations in the inner log-slope and the exact shape of the truncation. An investigation of a model that additionally parametrises the strength and extent of the central bump that is present in the most massive subhaloes could provide further insights on the effects of baryons on the density profiles.

The other results of this study can be summarised as follows:
\begin{enumerate}
 \item The inner log-slope of most subhalo density profiles in TNG50 is close to -2, which is significantly larger than the results obtained in previous studies (e.g. \citet{springel2008} and \citet{subhalo_profiles1}). For subhaloes with masses in the range between $10^{10}$ M$_\odot$ and $10^{11}$ M$_\odot$, the inner log-slope stays relatively constant all the way to the centre, resulting in a central cusp. Subhaloes with masses above $10^{11}$ M$_\odot$ have an additional bump with a flat core in the centre that becomes more pronounced for more massive subhaloes with a higher baryon fraction. For subhalo masses smaller than $10^{10}$ M$_\odot$, the inner log-slopes continuously become smaller towards the centre, also resulting in a core. This core can be of physical origin and might be the result of the reduced amount of baryons, but limitations in the resolution could also have an impact. The inner log-slope of -2 is also present when looking only at the dark matter component of the density profiles, whereas the baryonic component falls of more strongly. The majority of matter in the central bump comes from the presence of baryons, but a small central bump can also manifest in the dark matter component due to the gravitational interactions with the baryons. \\
 \item The average density profiles for each mass bin have a nearly universal shape with a modified NFW shape parameter of $\alpha \approx 1.6$, while the mean shape parameters of the individual density profiles increase with mass according to Equation (\ref{eq:alpha_mass_fit}). The outer subhalo shape and extent might be affected by the subhalo finding algorithm. Nevertheless, the analysis of observations is predominantly influenced by the behavior in the inner regions. \\
 \item The model parameters of the modified NFW profile start to change below the virial radius of the host halo R$_{200}$. For smaller distances, the parameter $r_\mathrm{s}$ becomes smaller by up to a factor of 10 and the normalisation $\rho_0$ becomes larger by up to a factor $10^4$, resulting in a higher concentration for subhaloes in the same mass bin. This affects subhaloes of all masses in the same way. For subhaloes closer to the centre, the inner log-slope of the average density profiles can also increase. \\
 \item The parameters of the modified NFW profile vary more strongly with concentration for lower-mass subhaloes. For subhaloes with masses above $10^{11}$ M$_\odot$, the parameter values do not change significantly with $c_V$ and an increase in concentration is mostly associated with a more pronounced central bump feature. An additional tidal truncation can lead to exceptionally high concentrations with $c_V>10^9$, which are only seen in 4 subhaloes in the sample. \\
 \item Most subhalo concentrations lie in the range $10^3 \lesssim c_V \lesssim 10^5$. Subhaloes with $c_V > 10^5$ have in common that they generally have a much higher baryon fraction ($f_\mathrm{b}>0.10$). \\
 \item A larger baryon fraction is generally accompanied by a decrease in $r_\mathrm{s}$ by up to a factor of 10 and an increase in $\rho_0$, resulting in a much higher concentration. For subhaloes above $10^{11}$ M$_\odot$, the baryon fraction also correlates with the size of the central bump. \\
 \item Tidal interactions mainly affect the outer regions of the subhalo density profile, making the subhaloes more concentrated for earlier infall times. The mean shape parameter $\alpha$ stays relatively constant for different infall times and depends primarily on the subhalo mass. The average density profiles for different infall times within different infall mass bins show no consistent decrease in the central density with earlier infall times. This is different from the results of other studies on the tidal stripping of individual subhaloes (e.g. \citet{subhalo_trunc1} and \citet{subhalo_profiles3}), which find a decrease of the central density and steeper truncations as tidal stripping progresses. However, a more sophisticated analysis of how the density profiles change over time for different types of subhaloes requires following their evolution individually.
\end{enumerate}

The NFW and Einasto profiles, together with the concentration-mass relation, are routinely used to predict the internal structure of dark matter subhaloes and estimate their observable effects. In turn, the comparisons with observational data are fundamental to constrain the properties of dark matter. The new analytical expression for the best-fitting subhalo density profile model presented in this paper (Equation \ref{eq:modified_NFW}), as well as the scaling relations between the model parameters and subhalo properties (Equations \ref{eq:rs_mass_fit} - \ref{eq:rs_vmax_fit}) provide a more realistic description of subhaloes in the presence of baryonic physics and can thus improve our understanding of the detectability of substructures and their distribution, leading to more precise constraints on the dark matter distribution in galaxies. In particular, our results suggest that NFW profiles are too shallow to correctly represent the inner slope of hydrodynamical subhaloes and this can lead to a mismodelling of the subhalo concentrations and central densities.

\section*{Acknowledgements}
The authors acknowledge computing resources provided by the Ministry of Science, Research and the Arts (MWK) of the State of Baden-W\"{u}rttemberg through bwHPC and the German Science Foundation (DFG) through grants INST 35/1134-1 FUGG and 35/1597-1 FUGG. They also thank for data storage at SDS@hd funded through grants INST 35/1314-1 FUGG and INST 35/1503-1 FUGG.

The IllustrisTNG simulations were undertaken with compute time awarded by the Gauss Centre for Supercomputing (GCS) under GCS Large-Scale Projects GCS-ILLU and GCS-DWAR on the GCS share of the supercomputer Hazel Hen at the High Performance Computing centre Stuttgart (HLRS), as well as on the machines of the Max Planck Computing and Data Facility (MPCDF) in Garching, Germany.

GD was supported by a Gliese Fellowship.

RSK acknowledges financial support from the European Research Council via the ERC Synergy Grant ``ECOGAL'' (project ID 855130),  from the German Excellence Strategy via the Heidelberg Cluster of Excellence (EXC 2181 - 390900948) ``STRUCTURES'', and from the German Ministry for Economic Affairs and Climate Action in project ``MAINN'' (funding ID 50OO2206). 

\section*{Data Availability}

The data analysed in this work are available upon request to the corresponding author. The entire data of the IllustrisTNG simulations, including TNG50, are publicly available and accessible at \url{www.tng-project.org/data} \citep{tng50data}. This research made use of the public Python packages matplotlib \citep{matplotlib}, NumPy \citep{numpy} and SciPy \citep{jones_scipy:_2001}.

\newpage


\bibliographystyle{mnras}
\bibliography{mnras_template} 

\begin{thebibliography}{}
\makeatletter
\relax
\def\mn@urlcharsother{\let\do\@makeother \do\$\do\&\do\#\do\^\do\_\do\%\do\~}
\def\mn@doi{\begingroup\mn@urlcharsother \@ifnextchar [ {\mn@doi@} {\mn@doi@[]}}
\def\mn@doi@[#1]#2{\def\@tempa{#1}\ifx\@tempa\@empty \href {http://dx.doi.org/#2} {doi:#2}\else \href {http://dx.doi.org/#2} {#1}\fi \endgroup}
\def\mn@eprint#1#2{\mn@eprint@#1:#2::\@nil}
\def\mn@eprint@arXiv#1{\href {http://arxiv.org/abs/#1} {{\tt arXiv:#1}}}
\def\mn@eprint@dblp#1{\href {http://dblp.uni-trier.de/rec/bibtex/#1.xml} {dblp:#1}}
\def\mn@eprint@#1:#2:#3:#4\@nil{\def\@tempa {#1}\def\@tempb {#2}\def\@tempc {#3}\ifx \@tempc \@empty \let \@tempc \@tempb \let \@tempb \@tempa \fi \ifx \@tempb \@empty \def\@tempb {arXiv}\fi \@ifundefined {mn@eprint@\@tempb}{\@tempb:\@tempc}{\expandafter \expandafter \csname mn@eprint@\@tempb\endcsname \expandafter{\@tempc}}}

\bibitem[\protect\citeauthoryear{{Bailin} \& {Steinmetz}}{{Bailin} \& {Steinmetz}}{2005}]{triaxiality1}
{Bailin} J.,  {Steinmetz} M.,  2005, \mn@doi [\apj] {10.1086/430397}, \href {https://ui.adsabs.harvard.edu/abs/2005ApJ...627..647B} {627, 647}

\bibitem[\protect\citeauthoryear{{Behroozi} et~al.,}{{Behroozi} et~al.}{2015}]{subfind_comparison2}
{Behroozi} P.,  et~al., 2015, \mn@doi [\mnras] {10.1093/mnras/stv2046}, \href {https://ui.adsabs.harvard.edu/abs/2015MNRAS.454.3020B} {454, 3020}

\bibitem[\protect\citeauthoryear{{Blumenthal}, {Faber}, {Flores}  \& {Primack}}{{Blumenthal} et~al.}{1986}]{adiabatic_contraction1}
{Blumenthal} G.~R.,  {Faber} S.~M.,  {Flores} R.,   {Primack} J.~R.,  1986, \mn@doi [\apj] {10.1086/163867}, \href {https://ui.adsabs.harvard.edu/abs/1986ApJ...301...27B} {301, 27}

\bibitem[\protect\citeauthoryear{{Bode}, {Ostriker}  \& {Turok}}{{Bode} et~al.}{2001}]{wdm_haloes1}
{Bode} P.,  {Ostriker} J.~P.,   {Turok} N.,  2001, \mn@doi [\apj] {10.1086/321541}, \href {https://ui.adsabs.harvard.edu/abs/2001ApJ...556...93B} {556, 93}

\bibitem[\protect\citeauthoryear{{Bonaca}, {Hogg}, {Price-Whelan}  \& {Conroy}}{{Bonaca} et~al.}{2019}]{stellar_streams4}
{Bonaca} A.,  {Hogg} D.~W.,  {Price-Whelan} A.~M.,   {Conroy} C.,  2019, \mn@doi [\apj] {10.3847/1538-4357/ab2873}, \href {https://ui.adsabs.harvard.edu/abs/2019ApJ...880...38B} {880, 38}

\bibitem[\protect\citeauthoryear{{Boylan-Kolchin}, {Bullock}  \& {Kaplinghat}}{{Boylan-Kolchin} et~al.}{2011}]{too_big_to_fail1}
{Boylan-Kolchin} M.,  {Bullock} J.~S.,   {Kaplinghat} M.,  2011, \mn@doi [\mnras] {10.1111/j.1745-3933.2011.01074.x}, \href {https://ui.adsabs.harvard.edu/abs/2011MNRAS.415L..40B} {415, L40}

\bibitem[\protect\citeauthoryear{{Brooks} \& {Zolotov}}{{Brooks} \& {Zolotov}}{2014}]{feedback6}
{Brooks} A.~M.,  {Zolotov} A.,  2014, \mn@doi [The Astrophysical Journal] {10.1088/0004-637X/786/2/87}, \href {https://ui.adsabs.harvard.edu/abs/2014ApJ...786...87B} {786, 87 pp.}

\bibitem[\protect\citeauthoryear{{Bullock} \& {Boylan-Kolchin}}{{Bullock} \& {Boylan-Kolchin}}{2017}]{small_scale_problems}
{Bullock} J.~S.,  {Boylan-Kolchin} M.,  2017, \mn@doi [\araa] {10.1146/annurev-astro-091916-055313}, \href {https://ui.adsabs.harvard.edu/abs/2017ARA&A..55..343B} {55, 343}

\bibitem[\protect\citeauthoryear{{Bullock}, {Kolatt}, {Sigad}, {Somerville}, {Kravtsov}, {Klypin}, {Primack}  \& {Dekel}}{{Bullock} et~al.}{2001}]{subhalo_conc2}
{Bullock} J.~S.,  {Kolatt} T.~S.,  {Sigad} Y.,  {Somerville} R.~S.,  {Kravtsov} A.~V.,  {Klypin} A.~A.,  {Primack} J.~R.,   {Dekel} A.,  2001, \mn@doi [\mnras] {10.1046/j.1365-8711.2001.04068.x}, \href {https://ui.adsabs.harvard.edu/abs/2001MNRAS.321..559B} {321, 559}

\bibitem[\protect\citeauthoryear{{Burkert}}{{Burkert}}{1995}]{burkert_profile}
{Burkert} A.,  1995, \mn@doi [\apjl] {10.1086/309560}, \href {https://ui.adsabs.harvard.edu/abs/1995ApJ...447L..25B} {447, L25}

\bibitem[\protect\citeauthoryear{{Carlberg}}{{Carlberg}}{2012}]{stellar_streams1}
{Carlberg} R.~G.,  2012, \mn@doi [\apj] {10.1088/0004-637X/748/1/20}, \href {https://ui.adsabs.harvard.edu/abs/2012ApJ...748...20C} {748, 20}

\bibitem[\protect\citeauthoryear{{Collett}}{{Collett}}{2015}]{LSST_euclid}
{Collett} T.~E.,  2015, \mn@doi [\apj] {10.1088/0004-637X/811/1/20}, \href {https://ui.adsabs.harvard.edu/abs/2015ApJ...811...20C} {811, 20}

\bibitem[\protect\citeauthoryear{{Despali} \& {Vegetti}}{{Despali} \& {Vegetti}}{2017}]{baryon_effects3}
{Despali} G.,  {Vegetti} S.,  2017, \mn@doi [\mnras] {10.1093/mnras/stx966}, \href {https://ui.adsabs.harvard.edu/abs/2017MNRAS.469.1997D} {469, 1997}

\bibitem[\protect\citeauthoryear{{Despali}, {Sparre}, {Vegetti}, {Vogelsberger}, {Zavala}  \& {Marinacci}}{{Despali} et~al.}{2019}]{sidm2019}
{Despali} G.,  {Sparre} M.,  {Vegetti} S.,  {Vogelsberger} M.,  {Zavala} J.,   {Marinacci} F.,  2019, \mn@doi [\mnras] {10.1093/mnras/stz273}, \href {https://ui.adsabs.harvard.edu/abs/2019MNRAS.484.4563D} {484, 4563}

\bibitem[\protect\citeauthoryear{{Despali}, {Lovell}, {Vegetti}, {Crain}  \& {Oppenheimer}}{{Despali} et~al.}{2020}]{despali19}
{Despali} G.,  {Lovell} M.,  {Vegetti} S.,  {Crain} R.~A.,   {Oppenheimer} B.~D.,  2020, \mn@doi [\mnras] {10.1093/mnras/stz3068}, \href {https://ui.adsabs.harvard.edu/abs/2020MNRAS.491.1295D} {491, 1295}

\bibitem[\protect\citeauthoryear{{Despali}, {Walls}, {Vegetti}, {Sparre}, {Vogelsberger}  \& {Zavala}}{{Despali} et~al.}{2022}]{sidm6}
{Despali} G.,  {Walls} L.~G.,  {Vegetti} S.,  {Sparre} M.,  {Vogelsberger} M.,   {Zavala} J.,  2022, \mn@doi [\mnras] {10.1093/mnras/stac2521}, \href {https://ui.adsabs.harvard.edu/abs/2022MNRAS.516.4543D} {516, 4543}

\bibitem[\protect\citeauthoryear{{Di Cintio}, {Knebe}, {Libeskind}, {Yepes}, {Gottl{\"o}ber}  \& {Hoffman}}{{Di Cintio} et~al.}{2011}]{subhalo_profiles2}
{Di Cintio} A.,  {Knebe} A.,  {Libeskind} N.~I.,  {Yepes} G.,  {Gottl{\"o}ber} S.,   {Hoffman} Y.,  2011, \mn@doi [\mnras] {10.1111/j.1745-3933.2011.01123.x}, \href {https://ui.adsabs.harvard.edu/abs/2011MNRAS.417L..74D} {417, L74}

\bibitem[\protect\citeauthoryear{{Di Cintio}, {Knebe}, {Libeskind}, {Brook}, {Yepes}, {Gottl{\"o}ber}  \& {Hoffman}}{{Di Cintio} et~al.}{2013}]{subhalo_profiles1}
{Di Cintio} A.,  {Knebe} A.,  {Libeskind} N.~I.,  {Brook} C.,  {Yepes} G.,  {Gottl{\"o}ber} S.,   {Hoffman} Y.,  2013, \mn@doi [\mnras] {10.1093/mnras/stt240}, \href {https://ui.adsabs.harvard.edu/abs/2013MNRAS.431.1220D} {431, 1220}

\bibitem[\protect\citeauthoryear{{Diemand}, {Kuhlen}  \& {Madau}}{{Diemand} et~al.}{2007}]{diemand2007}
{Diemand} J.,  {Kuhlen} M.,   {Madau} P.,  2007, \mn@doi [\apj] {10.1086/520573}, \href {https://ui.adsabs.harvard.edu/abs/2007ApJ...667..859D} {667, 859}

\bibitem[\protect\citeauthoryear{{Diemer} \& {Kravtsov}}{{Diemer} \& {Kravtsov}}{2015}]{c_m_rel0}
{Diemer} B.,  {Kravtsov} A.~V.,  2015, \mn@doi [\apj] {10.1088/0004-637X/799/1/108}, \href {https://ui.adsabs.harvard.edu/abs/2015ApJ...799..108D} {799, 108}

\bibitem[\protect\citeauthoryear{{Dolag}, {Borgani}, {Murante}  \& {Springel}}{{Dolag} et~al.}{2009}]{subfind2}
{Dolag} K.,  {Borgani} S.,  {Murante} G.,   {Springel} V.,  2009, \mn@doi [\mnras] {10.1111/j.1365-2966.2009.15034.x}, \href {https://ui.adsabs.harvard.edu/abs/2009MNRAS.399..497D} {399, 497}

\bibitem[\protect\citeauthoryear{{Dutton} \& {Macci{\`o}}}{{Dutton} \& {Macci{\`o}}}{2014}]{better_einasto3}
{Dutton} A.~A.,  {Macci{\`o}} A.~V.,  2014, \mn@doi [\mnras] {10.1093/mnras/stu742}, \href {https://ui.adsabs.harvard.edu/abs/2014MNRAS.441.3359D} {441, 3359}

\bibitem[\protect\citeauthoryear{{Einasto}}{{Einasto}}{1965}]{einasto}
{Einasto} J.,  1965, Trudy Astrofizicheskogo Instituta Alma-Ata, \href {https://ui.adsabs.harvard.edu/abs/1965TrAlm...5...87E} {5, 87}

\bibitem[\protect\citeauthoryear{{Erkal} \& {Belokurov}}{{Erkal} \& {Belokurov}}{2015}]{stellar_streams3}
{Erkal} D.,  {Belokurov} V.,  2015, \mn@doi [\mnras] {10.1093/mnras/stv2122}, \href {https://ui.adsabs.harvard.edu/abs/2015MNRAS.454.3542E} {454, 3542}

\bibitem[\protect\citeauthoryear{{Errani} \& {Navarro}}{{Errani} \& {Navarro}}{2021}]{exp_truncation}
{Errani} R.,  {Navarro} J.~F.,  2021, \mn@doi [\mnras] {10.1093/mnras/stab1215}, \href {https://ui.adsabs.harvard.edu/abs/2021MNRAS.505...18E} {505, 18}

\bibitem[\protect\citeauthoryear{{Famaey} \& {McGaugh}}{{Famaey} \& {McGaugh}}{2012}]{mond2}
{Famaey} B.,  {McGaugh} S.~S.,  2012, \mn@doi [Living Reviews in Relativity] {10.12942/lrr-2012-10}, \href {https://ui.adsabs.harvard.edu/abs/2012LRR....15...10F} {15, 10}

\bibitem[\protect\citeauthoryear{{Flores} \& {Primack}}{{Flores} \& {Primack}}{1994}]{cusp_core1}
{Flores} R.~A.,  {Primack} J.~R.,  1994, \mn@doi [\apjl] {10.1086/187350}, \href {https://ui.adsabs.harvard.edu/abs/1994ApJ...427L...1F} {427, L1}

\bibitem[\protect\citeauthoryear{{Franx}, {Illingworth}  \& {de Zeeuw}}{{Franx} et~al.}{1991}]{triaxiality0}
{Franx} M.,  {Illingworth} G.,   {de Zeeuw} T.,  1991, \mn@doi [\apj] {10.1086/170769}, \href {https://ui.adsabs.harvard.edu/abs/1991ApJ...383..112F} {383, 112}

\bibitem[\protect\citeauthoryear{{Garrison-Kimmel} et~al.,}{{Garrison-Kimmel} et~al.}{2017}]{baryon_effects4}
{Garrison-Kimmel} S.,  et~al., 2017, \mn@doi [\mnras] {10.1093/mnras/stx1710}, \href {https://ui.adsabs.harvard.edu/abs/2017MNRAS.471.1709G} {471, 1709}

\bibitem[\protect\citeauthoryear{{Ghigna}, {Moore}, {Governato}, {Lake}, {Quinn}  \& {Stadel}}{{Ghigna} et~al.}{2000}]{subhalo_conc1}
{Ghigna} S.,  {Moore} B.,  {Governato} F.,  {Lake} G.,  {Quinn} T.,   {Stadel} J.,  2000, \mn@doi [\apj] {10.1086/317221}, \href {https://ui.adsabs.harvard.edu/abs/2000ApJ...544..616G} {544, 616}

\bibitem[\protect\citeauthoryear{{Gilman}, {Birrer}, {Treu}, {Nierenberg}  \& {Benson}}{{Gilman} et~al.}{2019}]{flux_ratios2}
{Gilman} D.,  {Birrer} S.,  {Treu} T.,  {Nierenberg} A.,   {Benson} A.,  2019, \mn@doi [\mnras] {10.1093/mnras/stz1593}, \href {https://ui.adsabs.harvard.edu/abs/2019MNRAS.487.5721G} {487, 5721}

\bibitem[\protect\citeauthoryear{{Gnedin}, {Kravtsov}, {Klypin}  \& {Nagai}}{{Gnedin} et~al.}{2004}]{adiabatic_contraction2}
{Gnedin} O.~Y.,  {Kravtsov} A.~V.,  {Klypin} A.~A.,   {Nagai} D.,  2004, \mn@doi [\apj] {10.1086/424914}, \href {https://ui.adsabs.harvard.edu/abs/2004ApJ...616...16G} {616, 16}

\bibitem[\protect\citeauthoryear{{Green} \& {van den Bosch}}{{Green} \& {van den Bosch}}{2019}]{subhalo_profiles3}
{Green} S.~B.,  {van den Bosch} F.~C.,  2019, \mn@doi [\mnras] {10.1093/mnras/stz2767}, \href {https://ui.adsabs.harvard.edu/abs/2019MNRAS.490.2091G} {490, 2091}

\bibitem[\protect\citeauthoryear{{Green}, {van den Bosch}  \& {Jiang}}{{Green} et~al.}{2021}]{artificial_disruption1}
{Green} S.~B.,  {van den Bosch} F.~C.,   {Jiang} F.,  2021, \mn@doi [\mnras] {10.1093/mnras/stab696}, \href {https://ui.adsabs.harvard.edu/abs/2021MNRAS.503.4075G} {503, 4075}

\bibitem[\protect\citeauthoryear{{Harris} et~al.,}{{Harris} et~al.}{2020}]{numpy}
{Harris} C.~R.,  et~al., 2020, \mn@doi [\nat] {10.1038/s41586-020-2649-2}, \href {https://ui.adsabs.harvard.edu/abs/2020Natur.585..357H} {585, 357}

\bibitem[\protect\citeauthoryear{{Hayashi}, {Navarro}, {Taylor}, {Stadel}  \& {Quinn}}{{Hayashi} et~al.}{2003}]{tidal_stripping}
{Hayashi} E.,  {Navarro} J.,  {Taylor} J.,  {Stadel} J.,   {Quinn} T.,  2003, \mn@doi [\apj] {10.1086/345788}, \href {https://ui.adsabs.harvard.edu/abs/2003ApJ...584..541H} {584, 541}

\bibitem[\protect\citeauthoryear{{Hernquist}}{{Hernquist}}{1990}]{hernquist_profile}
{Hernquist} L.,  1990, \mn@doi [\apj] {10.1086/168845}, \href {https://ui.adsabs.harvard.edu/abs/1990ApJ...356..359H} {356, 359}

\bibitem[\protect\citeauthoryear{{Hezaveh} et~al.,}{{Hezaveh} et~al.}{2016}]{hezaveh2016}
{Hezaveh} Y.~D.,  et~al., 2016, \mn@doi [\apj] {10.3847/0004-637X/823/1/37}, \href {https://ui.adsabs.harvard.edu/abs/2016ApJ...823...37H} {823, 37}

\bibitem[\protect\citeauthoryear{{Hsueh}, {Enzi}, {Vegetti}, {Auger}, {Fassnacht}, {Despali}, {Koopmans}  \& {McKean}}{{Hsueh} et~al.}{2020}]{flux_ratios1}
{Hsueh} J.~W.,  {Enzi} W.,  {Vegetti} S.,  {Auger} M.~W.,  {Fassnacht} C.~D.,  {Despali} G.,  {Koopmans} L.~V.~E.,   {McKean} J.~P.,  2020, \mn@doi [\mnras] {10.1093/mnras/stz3177}, \href {https://ui.adsabs.harvard.edu/abs/2020MNRAS.492.3047H} {492, 3047}

\bibitem[\protect\citeauthoryear{{Huchra} \& {Geller}}{{Huchra} \& {Geller}}{1982}]{fof}
{Huchra} J.~P.,  {Geller} M.~J.,  1982, \mn@doi [\apj] {10.1086/160000}, \href {https://ui.adsabs.harvard.edu/abs/1982ApJ...257..423H} {257, 423}

\bibitem[\protect\citeauthoryear{{Hunter}}{{Hunter}}{2007}]{matplotlib}
{Hunter} J.~D.,  2007, \mn@doi [Computing in Science and Engineering] {10.1109/MCSE.2007.55}, \href {https://ui.adsabs.harvard.edu/abs/2007CSE.....9...90H} {9, 90}

\bibitem[\protect\citeauthoryear{{Jaffe}}{{Jaffe}}{1983}]{jaffe_profile}
{Jaffe} W.,  1983, \mn@doi [\mnras] {10.1093/mnras/202.4.995}, \href {https://ui.adsabs.harvard.edu/abs/1983MNRAS.202..995J} {202, 995}

\bibitem[\protect\citeauthoryear{Jones, Oliphant  \& Peterson}{Jones et~al.}{2001}]{jones_scipy:_2001}
Jones E.,  Oliphant T.,   Peterson P.,  2001, {SciPy:} Open Source Scientific Tools for {Python}, \url {http://www.scipy.org}

\bibitem[\protect\citeauthoryear{{Kazantzidis}, {Mayer}, {Mastropietro}, {Diemand}, {Stadel}  \& {Moore}}{{Kazantzidis} et~al.}{2004}]{subhalo_trunc1}
{Kazantzidis} S.,  {Mayer} L.,  {Mastropietro} C.,  {Diemand} J.,  {Stadel} J.,   {Moore} B.,  2004, \mn@doi [\apj] {10.1086/420840}, \href {https://ui.adsabs.harvard.edu/abs/2004ApJ...608..663K} {608, 663}

\bibitem[\protect\citeauthoryear{{Klypin}, {Gottl{\"o}ber}, {Kravtsov}  \& {Khokhlov}}{{Klypin} et~al.}{1999a}]{klypin1999}
{Klypin} A.,  {Gottl{\"o}ber} S.,  {Kravtsov} A.,   {Khokhlov} A.,  1999a, \mn@doi [\apj] {10.1086/307122}, \href {https://ui.adsabs.harvard.edu/abs/1999ApJ...516..530K} {516, 530}

\bibitem[\protect\citeauthoryear{{Klypin}, {Kravtsov}, {Valenzuela}  \& {Prada}}{{Klypin} et~al.}{1999b}]{missing_satellites1}
{Klypin} A.,  {Kravtsov} A.~V.,  {Valenzuela} O.,   {Prada} F.,  1999b, \mn@doi [\apj] {10.1086/307643}, \href {https://ui.adsabs.harvard.edu/abs/1999ApJ...522...82K} {522, 82}

\bibitem[\protect\citeauthoryear{{Kroupa}}{{Kroupa}}{2012}]{mond1}
{Kroupa} P.,  2012, \mn@doi [Publications of the Astronomical Society of Australia] {10.1071/AS12005}, \href {https://ui.adsabs.harvard.edu/abs/2012PASA...29..395K} {29, 395}

\bibitem[\protect\citeauthoryear{{Lovell} \& {Zavala}}{{Lovell} \& {Zavala}}{2023}]{lovell2023}
{Lovell} M.~R.,  {Zavala} J.,  2023, \mn@doi [\mnras] {10.1093/mnras/stad216}, \href {https://ui.adsabs.harvard.edu/abs/2023MNRAS.520.1567L} {520, 1567}

\bibitem[\protect\citeauthoryear{{Lovell}, {Frenk}, {Eke}, {Jenkins}, {Gao}  \& {Theuns}}{{Lovell} et~al.}{2014}]{wdm_haloes2}
{Lovell} M.~R.,  {Frenk} C.~S.,  {Eke} V.~R.,  {Jenkins} A.,  {Gao} L.,   {Theuns} T.,  2014, \mn@doi [\mnras] {10.1093/mnras/stt2431}, \href {https://ui.adsabs.harvard.edu/abs/2014MNRAS.439..300L} {439, 300}

\bibitem[\protect\citeauthoryear{{Mashchenko}, {Wadsley}  \& {Couchman}}{{Mashchenko} et~al.}{2008}]{feedback8}
{Mashchenko} S.,  {Wadsley} J.,   {Couchman} H.~M.~P.,  2008, \mn@doi [Science] {10.1126/science.1148666}, \href {https://ui.adsabs.harvard.edu/abs/2008Sci...319..174M} {319, 174 pp.}

\bibitem[\protect\citeauthoryear{{Mastromarino}, {Despali}, {Moscardini}, {Robertson}, {Meneghetti}  \& {Maturi}}{{Mastromarino} et~al.}{2023}]{sidm5}
{Mastromarino} C.,  {Despali} G.,  {Moscardini} L.,  {Robertson} A.,  {Meneghetti} M.,   {Maturi} M.,  2023, \mn@doi [\mnras] {10.1093/mnras/stad1853}, \href {https://ui.adsabs.harvard.edu/abs/2023MNRAS.524.1515M} {524, 1515}

\bibitem[\protect\citeauthoryear{{Mateo}}{{Mateo}}{1998}]{dwarf_galaxies1}
{Mateo} M.~L.,  1998, \mn@doi [\araa] {10.1146/annurev.astro.36.1.435}, \href {https://ui.adsabs.harvard.edu/abs/1998ARA&A..36..435M} {36, 435}

\bibitem[\protect\citeauthoryear{{Merritt}, {Graham}, {Moore}, {Diemand}  \& {Terzi{\'c}}}{{Merritt} et~al.}{2006}]{better_einasto2}
{Merritt} D.,  {Graham} A.~W.,  {Moore} B.,  {Diemand} J.,   {Terzi{\'c}} B.,  2006, \mn@doi [\aj] {10.1086/508988}, \href {https://ui.adsabs.harvard.edu/abs/2006AJ....132.2685M} {132, 2685}

\bibitem[\protect\citeauthoryear{{Milgrom}}{{Milgrom}}{1983}]{mond5}
{Milgrom} M.,  1983, \mn@doi [\apj] {10.1086/161130}, \href {https://ui.adsabs.harvard.edu/abs/1983ApJ...270..365M} {270, 365}

\bibitem[\protect\citeauthoryear{{Milgrom}}{{Milgrom}}{2002}]{mond3}
{Milgrom} M.,  2002, \mn@doi [\nar] {10.1016/S1387-6473(02)00243-9}, \href {https://ui.adsabs.harvard.edu/abs/2002NewAR..46..741M} {46, 741}

\bibitem[\protect\citeauthoryear{{Minor}, {Gad-Nasr}, {Kaplinghat}  \& {Vegetti}}{{Minor} et~al.}{2021}]{minor2021}
{Minor} Q.,  {Gad-Nasr} S.,  {Kaplinghat} M.,   {Vegetti} S.,  2021, \mn@doi [\mnras] {10.1093/mnras/stab2247}, \href {https://ui.adsabs.harvard.edu/abs/2021MNRAS.507.1662M} {507, 1662}

\bibitem[\protect\citeauthoryear{{Molin{\'e}}, {S{\'a}nchez-Conde}, {Palomares-Ruiz}  \& {Prada}}{{Molin{\'e}} et~al.}{2017}]{moline2017}
{Molin{\'e}} {\'A}.,  {S{\'a}nchez-Conde} M.~A.,  {Palomares-Ruiz} S.,   {Prada} F.,  2017, \mn@doi [\mnras] {10.1093/mnras/stx026}, \href {https://ui.adsabs.harvard.edu/abs/2017MNRAS.466.4974M} {466, 4974}

\bibitem[\protect\citeauthoryear{{Molin{\'e}} et~al.,}{{Molin{\'e}} et~al.}{2023}]{moline2023}
{Molin{\'e}} {\'A}.,  et~al., 2023, \mn@doi [\mnras] {10.1093/mnras/stac2930}, \href {https://ui.adsabs.harvard.edu/abs/2023MNRAS.518..157M} {518, 157}

\bibitem[\protect\citeauthoryear{{Mollitor}, {Nezri}  \& {Teyssier}}{{Mollitor} et~al.}{2015}]{baryon_effects1}
{Mollitor} P.,  {Nezri} E.,   {Teyssier} R.,  2015, \mn@doi [\mnras] {10.1093/mnras/stu2466}, \href {https://ui.adsabs.harvard.edu/abs/2015MNRAS.447.1353M} {447, 1353}

\bibitem[\protect\citeauthoryear{{Moore}}{{Moore}}{1994}]{cusp_core2}
{Moore} B.,  1994, \mn@doi [\nat] {10.1038/370629a0}, \href {https://ui.adsabs.harvard.edu/abs/1994Natur.370..629M} {370, 629}

\bibitem[\protect\citeauthoryear{{Moore}, {Quinn}, {Governato}, {Stadel}  \& {Lake}}{{Moore} et~al.}{1999a}]{moore_profile}
{Moore} B.,  {Quinn} T.,  {Governato} F.,  {Stadel} J.,   {Lake} G.,  1999a, \mn@doi [\mnras] {10.1046/j.1365-8711.1999.03039.x}, \href {https://ui.adsabs.harvard.edu/abs/1999MNRAS.310.1147M} {310, 1147}

\bibitem[\protect\citeauthoryear{{Moore}, {Ghigna}, {Governato}, {Lake}, {Quinn}, {Stadel}  \& {Tozzi}}{{Moore} et~al.}{1999b}]{moore1999}
{Moore} B.,  {Ghigna} S.,  {Governato} F.,  {Lake} G.,  {Quinn} T.,  {Stadel} J.,   {Tozzi} P.,  1999b, \mn@doi [\apjl] {10.1086/312287}, \href {https://ui.adsabs.harvard.edu/abs/1999ApJ...524L..19M} {524, L19}

\bibitem[\protect\citeauthoryear{{Mu{\~n}oz}, {Kochanek}  \& {Keeton}}{{Mu{\~n}oz} et~al.}{2001}]{corecusp_profile2}
{Mu{\~n}oz} J.~A.,  {Kochanek} C.~S.,   {Keeton} C.~R.,  2001, \mn@doi [\apj] {10.1086/322314}, \href {https://ui.adsabs.harvard.edu/abs/2001ApJ...558..657M} {558, 657}

\bibitem[\protect\citeauthoryear{{Nadler}, {Mao}, {Wechsler}, {Garrison-Kimmel}  \& {Wetzel}}{{Nadler} et~al.}{2018}]{baryon_effects5}
{Nadler} E.~O.,  {Mao} Y.-Y.,  {Wechsler} R.~H.,  {Garrison-Kimmel} S.,   {Wetzel} A.,  2018, \mn@doi [\apj] {10.3847/1538-4357/aac266}, \href {https://ui.adsabs.harvard.edu/abs/2018ApJ...859..129N} {859, 129}

\bibitem[\protect\citeauthoryear{{Navarro}, {Frenk}  \& {White}}{{Navarro} et~al.}{1996}]{nfw2}
{Navarro} J.~F.,  {Frenk} C.~S.,   {White} S. D.~M.,  1996, \mn@doi [\apj] {10.1086/177173}, \href {https://ui.adsabs.harvard.edu/abs/1996ApJ...462..563N} {462, 563}

\bibitem[\protect\citeauthoryear{{Navarro}, {Frenk}  \& {White}}{{Navarro} et~al.}{1997}]{nfw1}
{Navarro} J.~F.,  {Frenk} C.~S.,   {White} S. D.~M.,  1997, \mn@doi [\apj] {10.1086/304888}, \href {https://ui.adsabs.harvard.edu/abs/1997ApJ...490..493N} {490, 493}

\bibitem[\protect\citeauthoryear{{Navarro} et~al.,}{{Navarro} et~al.}{2010}]{better_einasto1}
{Navarro} J.~F.,  et~al., 2010, \mn@doi [\mnras] {10.1111/j.1365-2966.2009.15878.x}, \href {https://ui.adsabs.harvard.edu/abs/2010MNRAS.402...21N} {402, 21}

\bibitem[\protect\citeauthoryear{{Nelson} et~al.,}{{Nelson} et~al.}{2019a}]{tng50data}
{Nelson} D.,  et~al., 2019a, \mn@doi [Computational Astrophysics and Cosmology] {10.1186/s40668-019-0028-x}, \href {https://ui.adsabs.harvard.edu/abs/2019ComAC...6....2N} {6, 2}

\bibitem[\protect\citeauthoryear{{Nelson} et~al.,}{{Nelson} et~al.}{2019b}]{nelson2019}
{Nelson} D.,  et~al., 2019b, \mn@doi [\mnras] {10.1093/mnras/stz2306}, \href {https://ui.adsabs.harvard.edu/abs/2019MNRAS.490.3234N} {490, 3234}

\bibitem[\protect\citeauthoryear{{Onions} et~al.,}{{Onions} et~al.}{2012}]{subfind_comparison1}
{Onions} J.,  et~al., 2012, \mn@doi [\mnras] {10.1111/j.1365-2966.2012.20947.x}, \href {https://ui.adsabs.harvard.edu/abs/2012MNRAS.423.1200O} {423, 1200}

\bibitem[\protect\citeauthoryear{{Pillepich} et~al.,}{{Pillepich} et~al.}{2018}]{tng_model2}
{Pillepich} A.,  et~al., 2018, \mn@doi [\mnras] {10.1093/mnras/stx2656}, \href {https://ui.adsabs.harvard.edu/abs/2018MNRAS.473.4077P} {473, 4077}

\bibitem[\protect\citeauthoryear{{Pillepich} et~al.,}{{Pillepich} et~al.}{2019}]{pillepich2019}
{Pillepich} A.,  et~al., 2019, \mn@doi [\mnras] {10.1093/mnras/stz2338}, \href {https://ui.adsabs.harvard.edu/abs/2019MNRAS.490.3196P} {490, 3196}

\bibitem[\protect\citeauthoryear{{Planck Collaboration} et~al.,}{{Planck Collaboration} et~al.}{2016}]{planck2015}
{Planck Collaboration} et~al., 2016, \mn@doi [\aap] {10.1051/0004-6361/201525830}, \href {https://ui.adsabs.harvard.edu/abs/2016A&A...594A..13P} {594, A13}

\bibitem[\protect\citeauthoryear{{Planck Collaboration} et~al.,}{{Planck Collaboration} et~al.}{2020}]{planck_main}
{Planck Collaboration} et~al., 2020, \mn@doi [\aap] {10.1051/0004-6361/201833880}, \href {https://ui.adsabs.harvard.edu/abs/2020A&A...641A...1P} {641, A1}

\bibitem[\protect\citeauthoryear{{Pontzen} \& {Governato}}{{Pontzen} \& {Governato}}{2012}]{feedback7}
{Pontzen} A.,  {Governato} F.,  2012, \mn@doi [Monthly Notices of the Royal Astronomical Society] {10.1111/j.1365-2966.2012.20571.x}, \href {https://ui.adsabs.harvard.edu/abs/2012MNRAS.421.3464P} {421, 3464}

\bibitem[\protect\citeauthoryear{{Read}, {Agertz}  \& {Collins}}{{Read} et~al.}{2016}]{feedback3}
{Read} J.~I.,  {Agertz} O.,   {Collins} M.~L.~M.,  2016, \mn@doi [Monthly Notices of the Royal Astronomical Society] {10.1093/mnras/stw713}, \href {https://ui.adsabs.harvard.edu/abs/2016MNRAS.459.2573R} {459, 2573}

\bibitem[\protect\citeauthoryear{{Robles}, {Kelley}, {Bullock}  \& {Kaplinghat}}{{Robles} et~al.}{2019}]{sidm0}
{Robles} V.~H.,  {Kelley} T.,  {Bullock} J.~S.,   {Kaplinghat} M.,  2019, \mn@doi [\mnras] {10.1093/mnras/stz2345}, \href {https://ui.adsabs.harvard.edu/abs/2019MNRAS.490.2117R} {490, 2117}

\bibitem[\protect\citeauthoryear{{Rocha}, {Peter}, {Bullock}, {Kaplinghat}, {Garrison-Kimmel}, {O{\~n}orbe}  \& {Moustakas}}{{Rocha} et~al.}{2013}]{sidm4}
{Rocha} M.,  {Peter} A. H.~G.,  {Bullock} J.~S.,  {Kaplinghat} M.,  {Garrison-Kimmel} S.,  {O{\~n}orbe} J.,   {Moustakas} L.~A.,  2013, \mn@doi [\mnras] {10.1093/mnras/sts514}, \href {https://ui.adsabs.harvard.edu/abs/2013MNRAS.430...81R} {430, 81}

\bibitem[\protect\citeauthoryear{{Rodriguez-Gomez} et~al.,}{{Rodriguez-Gomez} et~al.}{2015}]{sublink}
{Rodriguez-Gomez} V.,  et~al., 2015, \mn@doi [\mnras] {10.1093/mnras/stv264}, \href {https://ui.adsabs.harvard.edu/abs/2015MNRAS.449...49R} {449, 49}

\bibitem[\protect\citeauthoryear{{Sanders} \& {McGaugh}}{{Sanders} \& {McGaugh}}{2002}]{mond4}
{Sanders} R.~H.,  {McGaugh} S.~S.,  2002, \mn@doi [\araa] {10.1146/annurev.astro.40.060401.093923}, \href {https://ui.adsabs.harvard.edu/abs/2002ARA&A..40..263S} {40, 263}

\bibitem[\protect\citeauthoryear{{Sawala}, {Frenk}  \& {Fattahi et al.}}{{Sawala} et~al.}{2016}]{feedback2}
{Sawala} T.,  {Frenk} C.~S.,   {Fattahi et al.} A.,  2016, \mn@doi [Monthly Notices of the Royal Astronomical Society] {10.1093/mnras/stw145}, \href {https://ui.adsabs.harvard.edu/abs/2016MNRAS.457.1931S} {457, 1931}

\bibitem[\protect\citeauthoryear{{Schechter}}{{Schechter}}{1976}]{schechter1976}
{Schechter} P.,  1976, \mn@doi [\apj] {10.1086/154079}, \href {https://ui.adsabs.harvard.edu/abs/1976ApJ...203..297S} {203, 297}

\bibitem[\protect\citeauthoryear{{Springel}}{{Springel}}{2010}]{arepo2010}
{Springel} V.,  2010, \mn@doi [\mnras] {10.1111/j.1365-2966.2009.15715.x}, \href {https://ui.adsabs.harvard.edu/abs/2010MNRAS.401..791S} {401, 791}

\bibitem[\protect\citeauthoryear{{Springel}, {White}, {Tormen}  \& {Kauffmann}}{{Springel} et~al.}{2001}]{subfind1}
{Springel} V.,  {White} S. D.~M.,  {Tormen} G.,   {Kauffmann} G.,  2001, \mn@doi [\mnras] {10.1046/j.1365-8711.2001.04912.x}, \href {https://ui.adsabs.harvard.edu/abs/2001MNRAS.328..726S} {328, 726}

\bibitem[\protect\citeauthoryear{{Springel} et~al.,}{{Springel} et~al.}{2005}]{lhalotree}
{Springel} V.,  et~al., 2005, \mn@doi [\nat] {10.1038/nature03597}, \href {https://ui.adsabs.harvard.edu/abs/2005Natur.435..629S} {435, 629}

\bibitem[\protect\citeauthoryear{{Springel} et~al.,}{{Springel} et~al.}{2008}]{springel2008}
{Springel} V.,  et~al., 2008, \mn@doi [\mnras] {10.1111/j.1365-2966.2008.14066.x}, \href {https://ui.adsabs.harvard.edu/abs/2008MNRAS.391.1685S} {391, 1685}

\bibitem[\protect\citeauthoryear{{Tolstoy}, {Hill}  \& {Tosi}}{{Tolstoy} et~al.}{2009}]{dwarf_galaxies2}
{Tolstoy} E.,  {Hill} V.,   {Tosi} M.,  2009, \mn@doi [\araa] {10.1146/annurev-astro-082708-101650}, \href {https://ui.adsabs.harvard.edu/abs/2009ARA&A..47..371T} {47, 371}

\bibitem[\protect\citeauthoryear{{Vegetti}, {Czoske}  \& {Koopmans}}{{Vegetti} et~al.}{2010a}]{vegetti2010a}
{Vegetti} S.,  {Czoske} O.,   {Koopmans} L. V.~E.,  2010a, \mn@doi [\mnras] {10.1111/j.1365-2966.2010.16952.x}, \href {https://ui.adsabs.harvard.edu/abs/2010MNRAS.407..225V} {407, 225}

\bibitem[\protect\citeauthoryear{{Vegetti}, {Koopmans}, {Bolton}, {Treu}  \& {Gavazzi}}{{Vegetti} et~al.}{2010b}]{vegetti2010}
{Vegetti} S.,  {Koopmans} L.~V.~E.,  {Bolton} A.,  {Treu} T.,   {Gavazzi} R.,  2010b, \mn@doi [\mnras] {10.1111/j.1365-2966.2010.16865.x}, \href {https://ui.adsabs.harvard.edu/abs/2010MNRAS.408.1969V} {408, 1969}

\bibitem[\protect\citeauthoryear{{Vegetti}, {Lagattuta}, {McKean}, {Auger}, {Fassnacht}  \& {Koopmans}}{{Vegetti} et~al.}{2012}]{vegetti2012}
{Vegetti} S.,  {Lagattuta} D.~J.,  {McKean} J.~P.,  {Auger} M.~W.,  {Fassnacht} C.~D.,   {Koopmans} L.~V.~E.,  2012, \mn@doi [\nat] {10.1038/nature10669}, \href {https://ui.adsabs.harvard.edu/abs/2012Natur.481..341V} {481, 341}

\bibitem[\protect\citeauthoryear{{Vogelsberger}, {Zavala}  \& {Loeb}}{{Vogelsberger} et~al.}{2012}]{sidm3}
{Vogelsberger} M.,  {Zavala} J.,   {Loeb} A.,  2012, \mn@doi [\mnras] {10.1111/j.1365-2966.2012.21182.x10.1002/asna.19141991009}, \href {https://ui.adsabs.harvard.edu/abs/2012MNRAS.423.3740V} {423, 3740}

\bibitem[\protect\citeauthoryear{{Wechsler}, {Bullock}, {Primack}, {Kravtsov}  \& {Dekel}}{{Wechsler} et~al.}{2002}]{c_m_rel1}
{Wechsler} R.~H.,  {Bullock} J.~S.,  {Primack} J.~R.,  {Kravtsov} A.~V.,   {Dekel} A.,  2002, \mn@doi [\apj] {10.1086/338765}, \href {https://ui.adsabs.harvard.edu/abs/2002ApJ...568...52W} {568, 52}

\bibitem[\protect\citeauthoryear{{Weinberger} et~al.,}{{Weinberger} et~al.}{2017}]{tng_model1}
{Weinberger} R.,  et~al., 2017, \mn@doi [\mnras] {10.1093/mnras/stw2944}, \href {https://ui.adsabs.harvard.edu/abs/2017MNRAS.465.3291W} {465, 3291}

\bibitem[\protect\citeauthoryear{{Wetzel}, {Hopkins}, {Kim}, {Faucher-Gigu{\`e}re}, {Kere{\v{s}}}  \& {Quataert}}{{Wetzel} et~al.}{2016}]{feedback1}
{Wetzel} A.~R.,  {Hopkins} P.~F.,  {Kim} J.-h.,  {Faucher-Gigu{\`e}re} C.-A.,  {Kere{\v{s}}} D.,   {Quataert} E.,  2016, \mn@doi [The Astrophysical Journal Letters] {10.3847/2041-8205/827/2/L23}, \href {https://ui.adsabs.harvard.edu/abs/2016ApJ...827L..23W} {827, 6 pp.}

\bibitem[\protect\citeauthoryear{{Zavala} \& {Frenk}}{{Zavala} \& {Frenk}}{2019}]{halo_review}
{Zavala} J.,  {Frenk} C.~S.,  2019, \mn@doi [Galaxies] {10.3390/galaxies7040081}, \href {https://ui.adsabs.harvard.edu/abs/2019Galax...7...81Z} {7, 81}

\bibitem[\protect\citeauthoryear{{Zemp}, {Gnedin}, {Gnedin}  \& {Kravtsov}}{{Zemp} et~al.}{2011}]{triaxiality2}
{Zemp} M.,  {Gnedin} O.~Y.,  {Gnedin} N.~Y.,   {Kravtsov} A.~V.,  2011, \mn@doi [\apjs] {10.1088/0067-0049/197/2/30}, \href {https://ui.adsabs.harvard.edu/abs/2011ApJS..197...30Z} {197, 30}

\bibitem[\protect\citeauthoryear{{Zhao}}{{Zhao}}{1996}]{mnfw}
{Zhao} H.,  1996, \mn@doi [\mnras] {10.1093/mnras/278.2.488}, \href {https://ui.adsabs.harvard.edu/abs/1996MNRAS.278..488Z} {278, 488}

\bibitem[\protect\citeauthoryear{{Zhao}, {Jing}, {Mo}  \& {B{\"o}rner}}{{Zhao} et~al.}{2003}]{c_m_rel2}
{Zhao} D.~H.,  {Jing} Y.~P.,  {Mo} H.~J.,   {B{\"o}rner} G.,  2003, \mn@doi [\apjl] {10.1086/379734}, \href {https://ui.adsabs.harvard.edu/abs/2003ApJ...597L...9Z} {597, L9}

\bibitem[\protect\citeauthoryear{{Zhao}, {Jing}, {Mo}  \& {B{\"o}rner}}{{Zhao} et~al.}{2009}]{c_m_rel3}
{Zhao} D.~H.,  {Jing} Y.~P.,  {Mo} H.~J.,   {B{\"o}rner} G.,  2009, \mn@doi [\apj] {10.1088/0004-637X/707/1/354}, \href {https://ui.adsabs.harvard.edu/abs/2009ApJ...707..354Z} {707, 354}

\bibitem[\protect\citeauthoryear{{Zhu}, {Hernquist}, {Marinacci}, {Springel}  \& {Li}}{{Zhu} et~al.}{2017}]{baryon_effects2}
{Zhu} Q.,  {Hernquist} L.,  {Marinacci} F.,  {Springel} V.,   {Li} Y.,  2017, \mn@doi [\mnras] {10.1093/mnras/stw3387}, \href {https://ui.adsabs.harvard.edu/abs/2017MNRAS.466.3876Z} {466, 3876}

\bibitem[\protect\citeauthoryear{{{\c{S}}eng{\"u}l} \& {Dvorkin}}{{{\c{S}}eng{\"u}l} \& {Dvorkin}}{2022}]{sengul2022}
{{\c{S}}eng{\"u}l} A.~{\c{C}}.,  {Dvorkin} C.,  2022, \mn@doi [\mnras] {10.1093/mnras/stac2256}, \href {https://ui.adsabs.harvard.edu/abs/2022MNRAS.516..336S} {516, 336}

\bibitem[\protect\citeauthoryear{{van den Bosch} \& {Ogiya}}{{van den Bosch} \& {Ogiya}}{2018}]{artificial_disruption2}
{van den Bosch} F.~C.,  {Ogiya} G.,  2018, \mn@doi [\mnras] {10.1093/mnras/sty084}, \href {https://ui.adsabs.harvard.edu/abs/2018MNRAS.475.4066V} {475, 4066}

\bibitem[\protect\citeauthoryear{{van den Bosch}, {Ogiya}, {Hahn}  \& {Burkert}}{{van den Bosch} et~al.}{2018}]{artificial_disruption3}
{van den Bosch} F.~C.,  {Ogiya} G.,  {Hahn} O.,   {Burkert} A.,  2018, \mn@doi [\mnras] {10.1093/mnras/stx2956}, \href {https://ui.adsabs.harvard.edu/abs/2018MNRAS.474.3043V} {474, 3043}

\makeatother
\end{thebibliography}




\appendix

\section{Extra Material}
\label{sec:appendix}

\begin{figure}
  \includegraphics[width=\columnwidth]{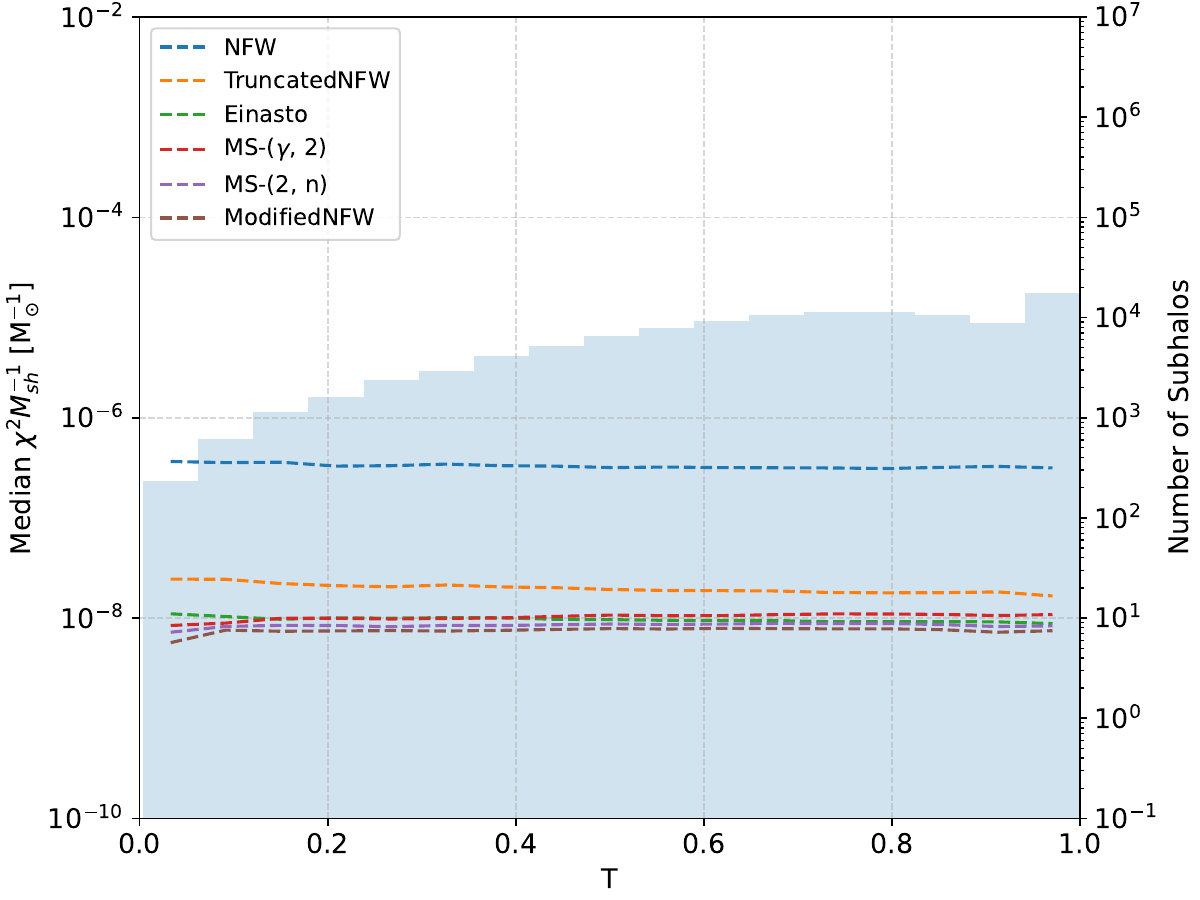}
  \caption{Median values of the goodness of fit per unit subhalo mass for subhaloes with different triaxialities. Additionally, the distribution of triaxialities is shown in the histogram.}
  \label{fig:performance_triaxiality}
\end{figure}

\begin{figure}
  \includegraphics[width=\columnwidth]{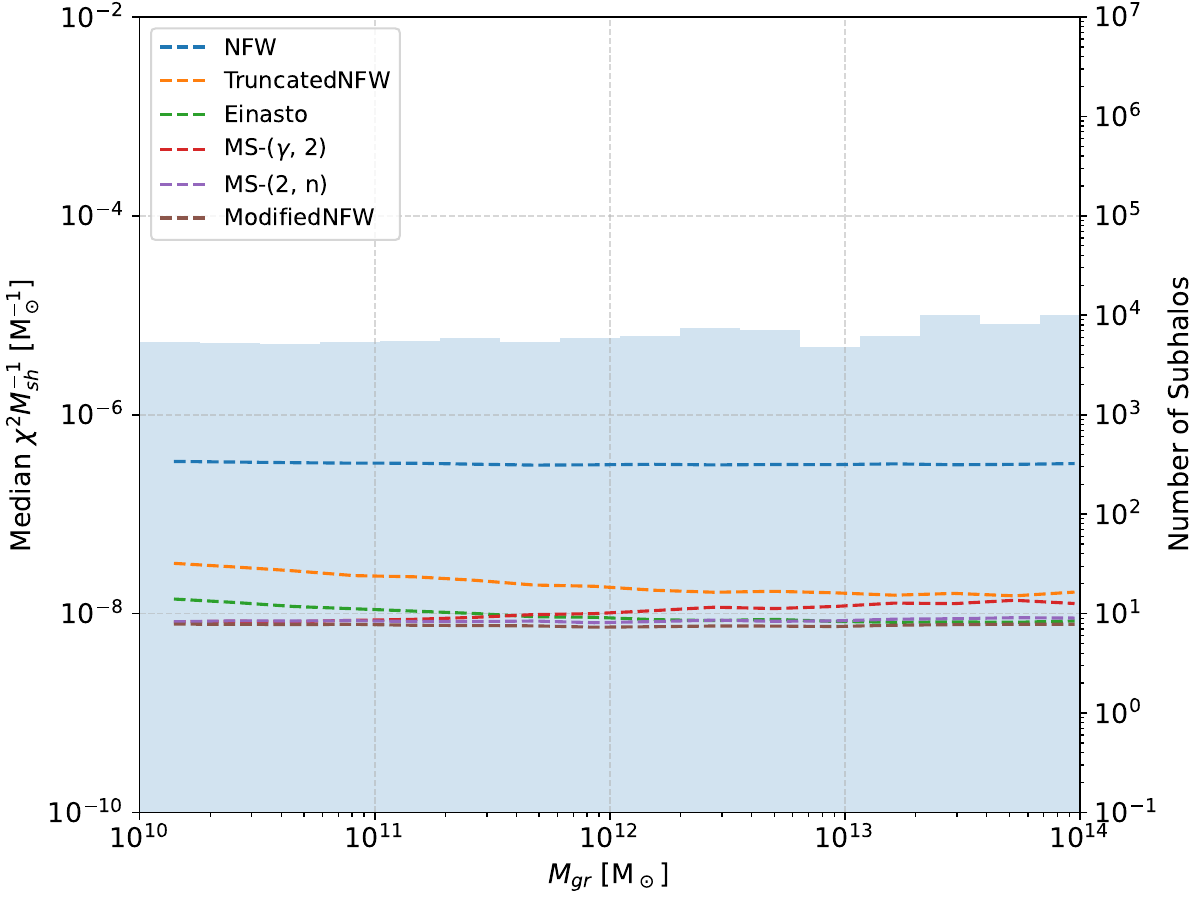}
  \caption{Median values of the goodness of fit per unit subhalo mass for subhaloes with different group masses. Additionally, the distribution of group masses is shown in the histogram.}
  \label{fig:performance_groupmass}
\end{figure}

\begin{figure}
  \includegraphics[width=\columnwidth]{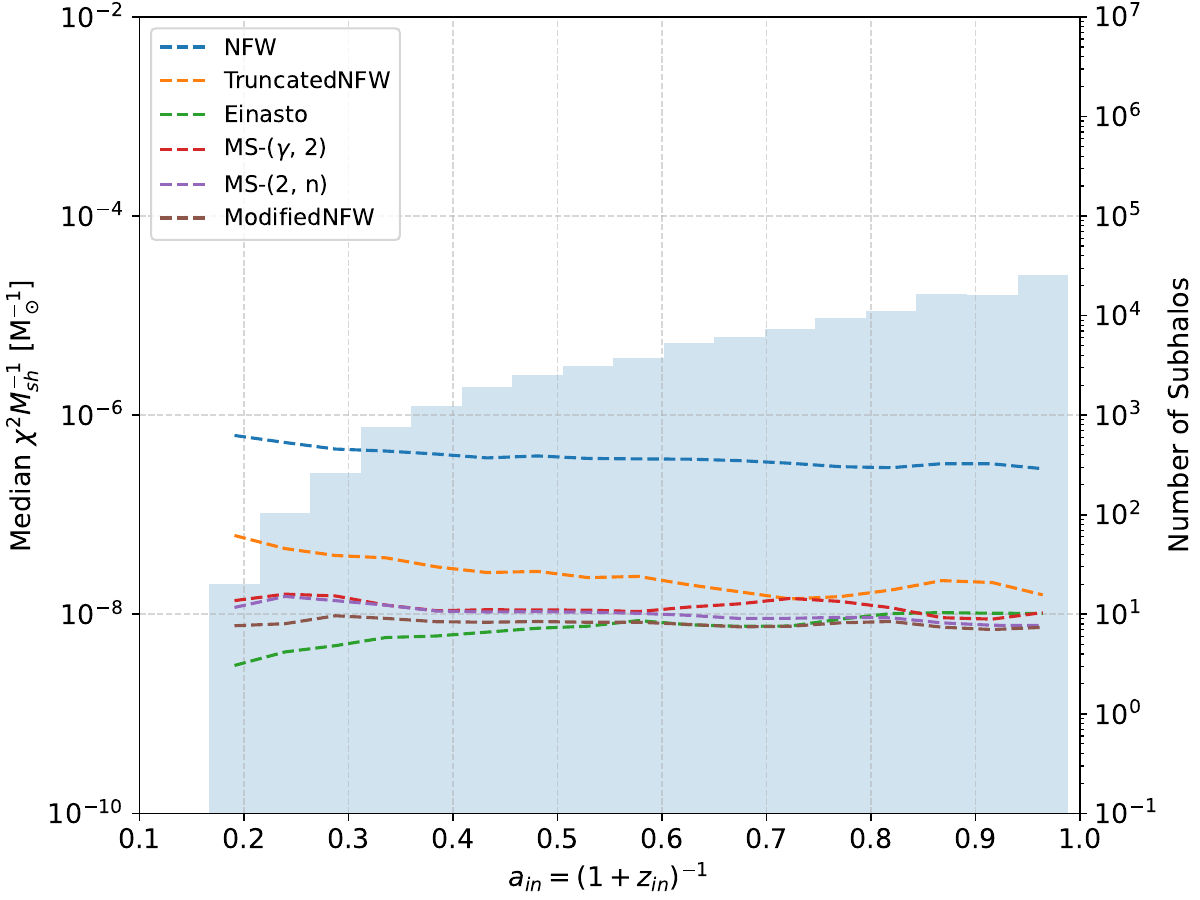}
  \caption{Median values of the goodness of fit per unit subhalo mass for subhaloes with different infall times. Additionally, the distribution of infall times is shown in the histogram.}
  \label{fig:performance_infall_time}
\end{figure}

\begin{figure*}
\includegraphics[width=\columnwidth]{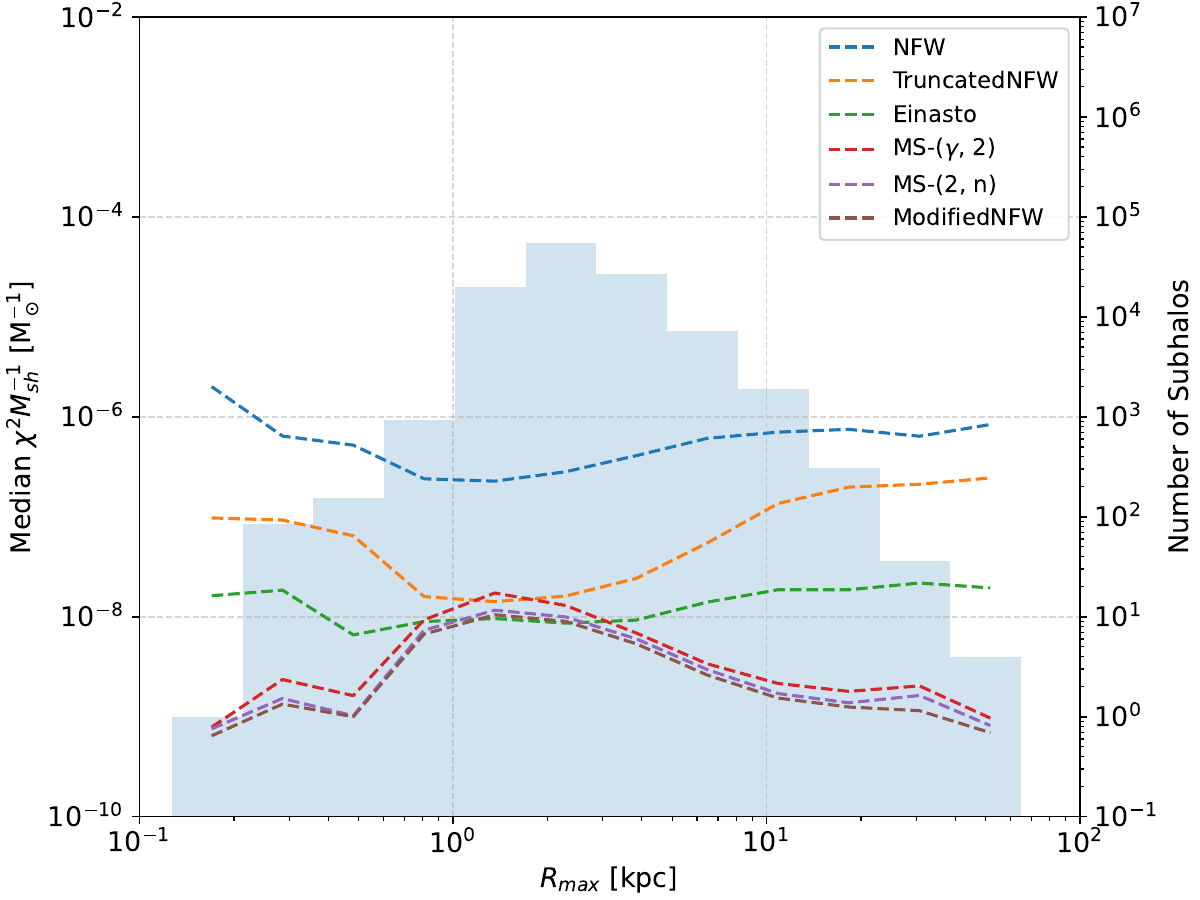}
\includegraphics[width=\columnwidth]{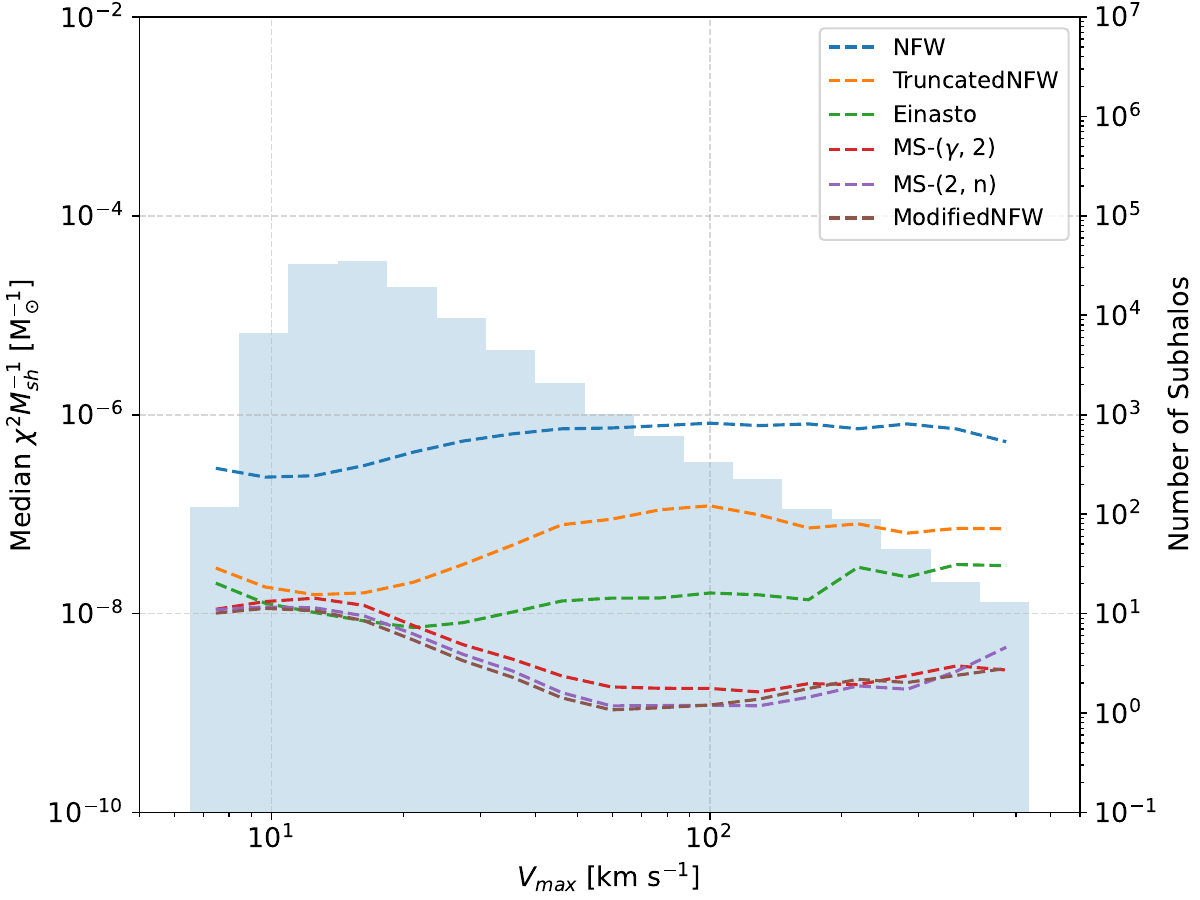}
\caption{Median values of the goodness of fit per unit subhalo mass for subhaloes with different $R_\mathrm{max}$ (left) and $V_\mathrm{max}$ (right). Additionally, the distributions of $R_\mathrm{max}$ and $V_\mathrm{max}$ are shown in the histograms.}
\label{fig:performace_vmax_rmax}
\end{figure*}

Figures \ref{fig:performance_triaxiality}, \ref{fig:performance_groupmass}, \ref{fig:performance_infall_time} and \ref{fig:performace_vmax_rmax} depict how the density profile models perform for different subhalo triaxialities, group masses, infall times, $R_\mathrm{max}$ and $V_\mathrm{max}$. \\

\begin{figure*}
\includegraphics[width=\columnwidth]{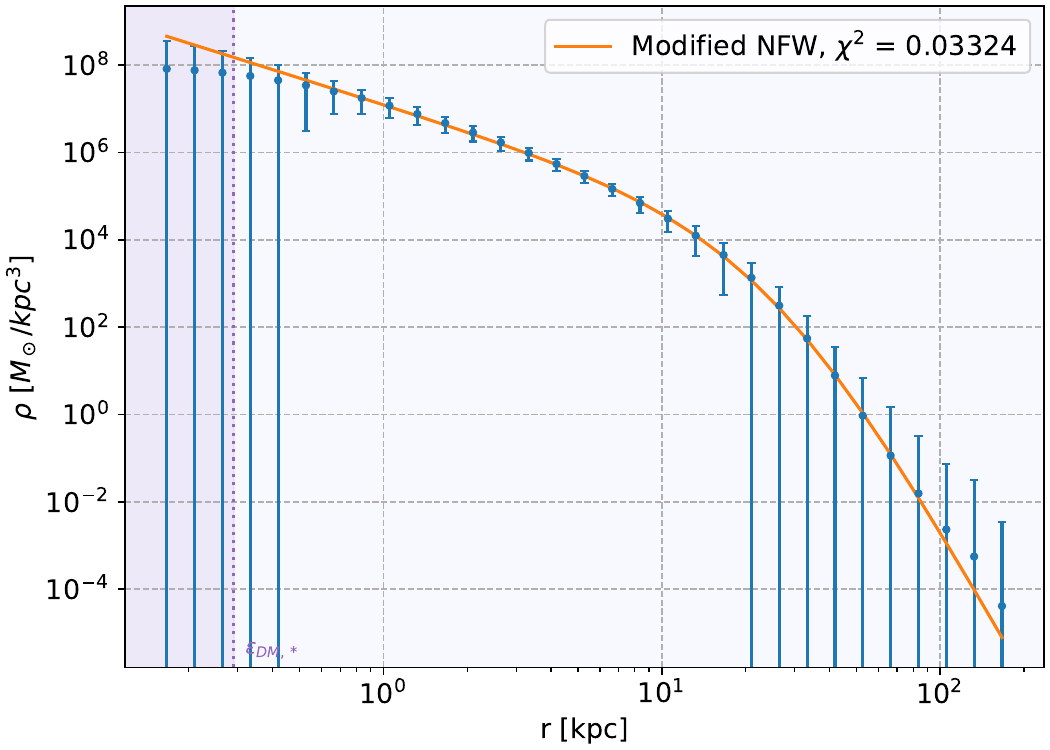}
\includegraphics[width=\columnwidth]{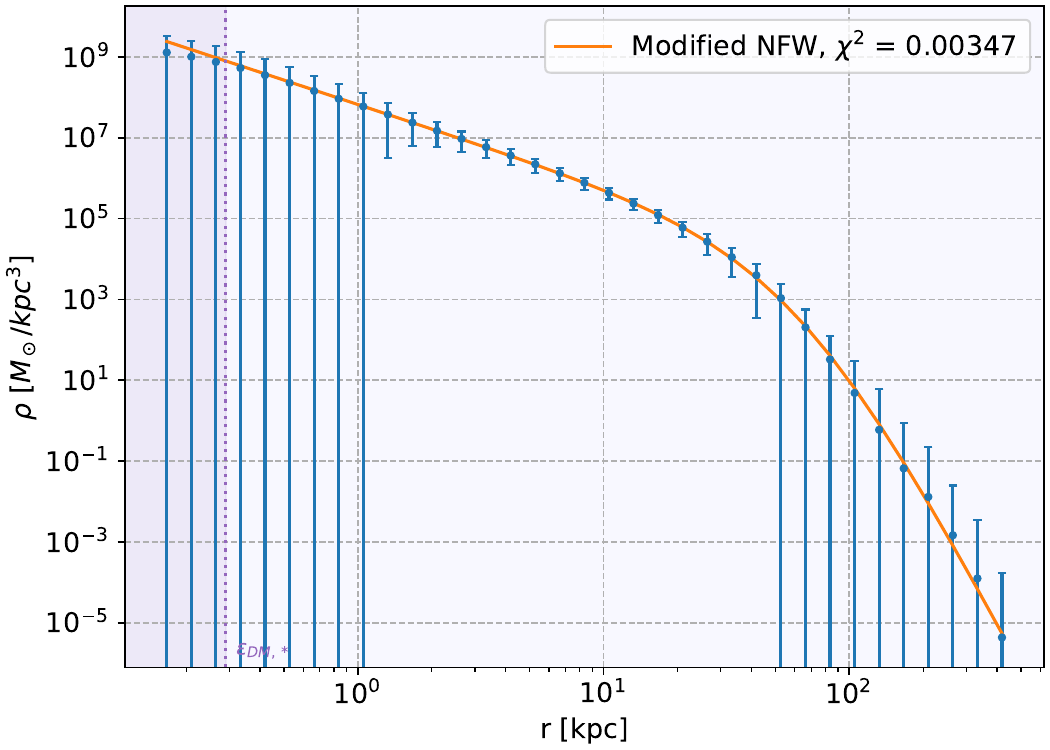}
\caption{Average density profiles with standard deviations for subhaloes with masses between $5.0 \times 10^8$ M$_\odot$ and $9.6 \times 10^8$ M$_\odot$ (left) as well as between $6.7 \times 10^9$ M$_\odot$ and $1.3 \times 10^{10}$ M$_\odot$ (right), together with the best modified NFW profile fits. The purple area indicates the regime below the softening length.}
\label{fig:average_profile_example}
\end{figure*}

Figure \ref{fig:average_profile_example} shows two example average density profiles for given mass bins, together with the best modified NFW profile fit. \\

\begin{figure*}
\includegraphics[width=\textwidth]{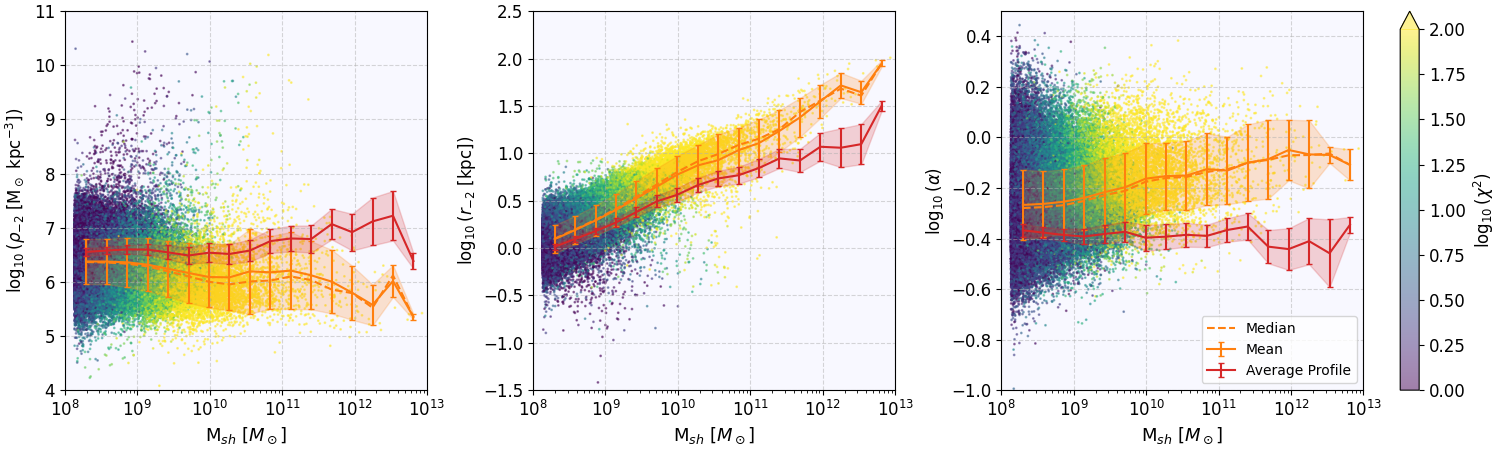}
\caption{Subhalo mass dependence of the parameters of the Einasto profile with colour-coded goodness of fit. The solid and dashed orange lines show the mean and median values of the distribution for each mass bin, together with the standard deviation. The red solid line shows the parameter values of the average density profiles for every mass bin, together with their uncertainties.}
\label{fig:parameter_mass_scatter_einasto}
\end{figure*}

\begin{figure*}
\includegraphics[width=\textwidth]{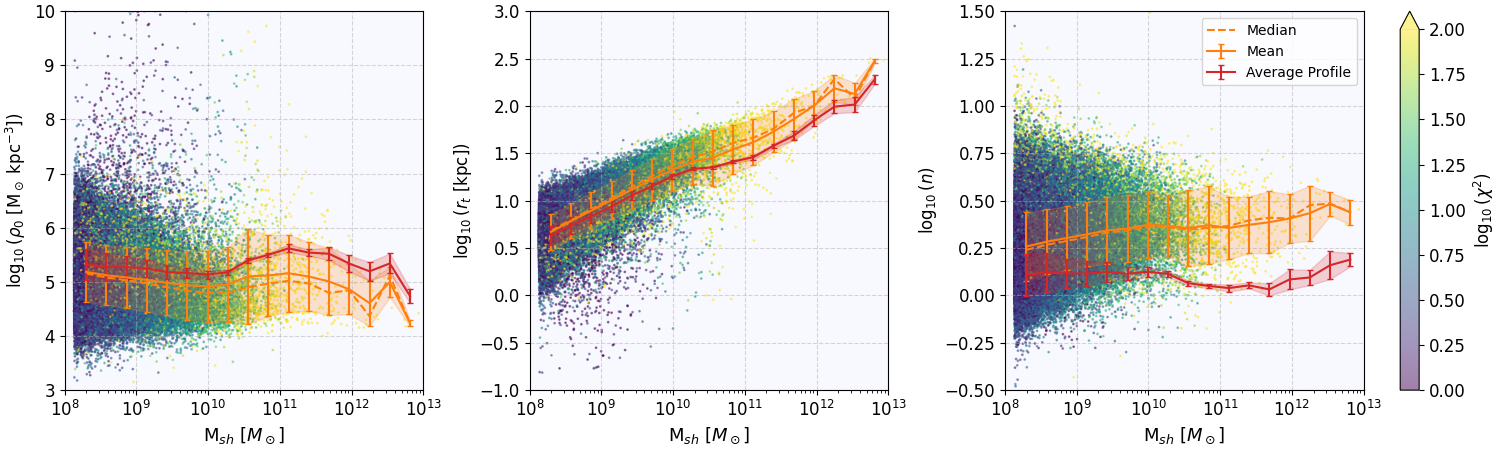}
\caption{Subhalo mass dependence of the parameters of the MS-(2, $n$) profile with colour-coded goodness of fit. The solid and dashed orange lines show the mean and median values of the distribution for each mass bin, together with the standard deviation. The red solid line shows the parameter values of the average density profiles for every mass bin, together with their uncertainties.}
\label{fig:parameter_mass_scatter_tpl}
\end{figure*}

\begin{figure*}
\includegraphics[width=\textwidth]{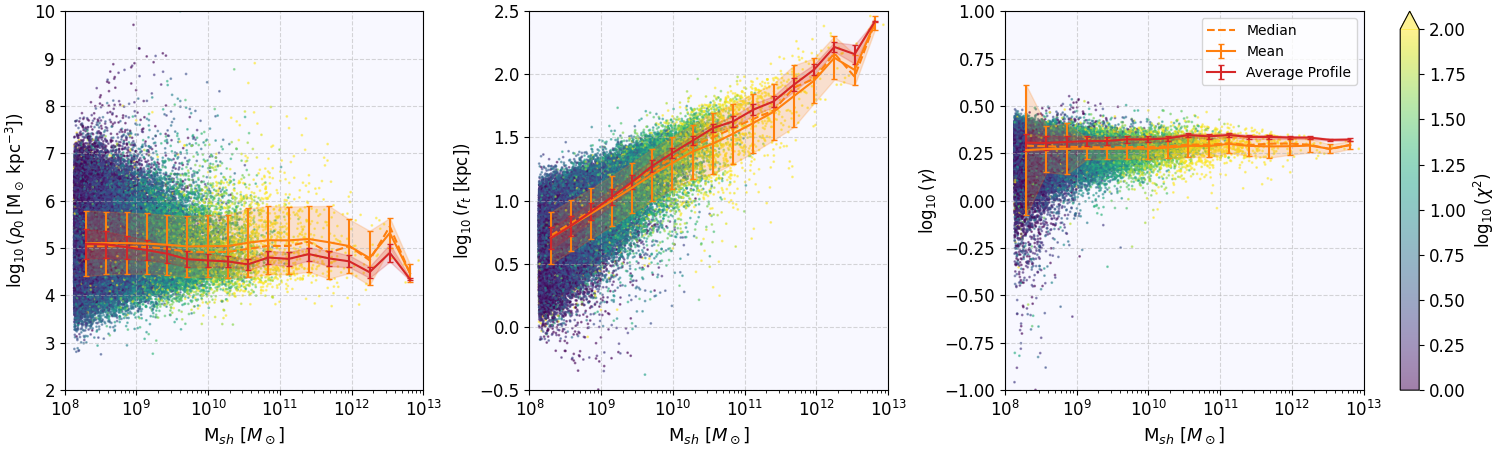}
\caption{Subhalo mass dependence of the parameters of the MS-($\gamma$, 2) profile with colour-coded goodness of fit. The solid and dashed orange lines show the mean and median values of the distribution for each mass bin, together with the standard deviation. The red solid line shows the parameter values of the average density profiles for every mass bin, together with their uncertainties.}
\label{fig:parameter_mass_scatter_schechter}
\end{figure*}

\begin{figure*}
\includegraphics[width=\textwidth]{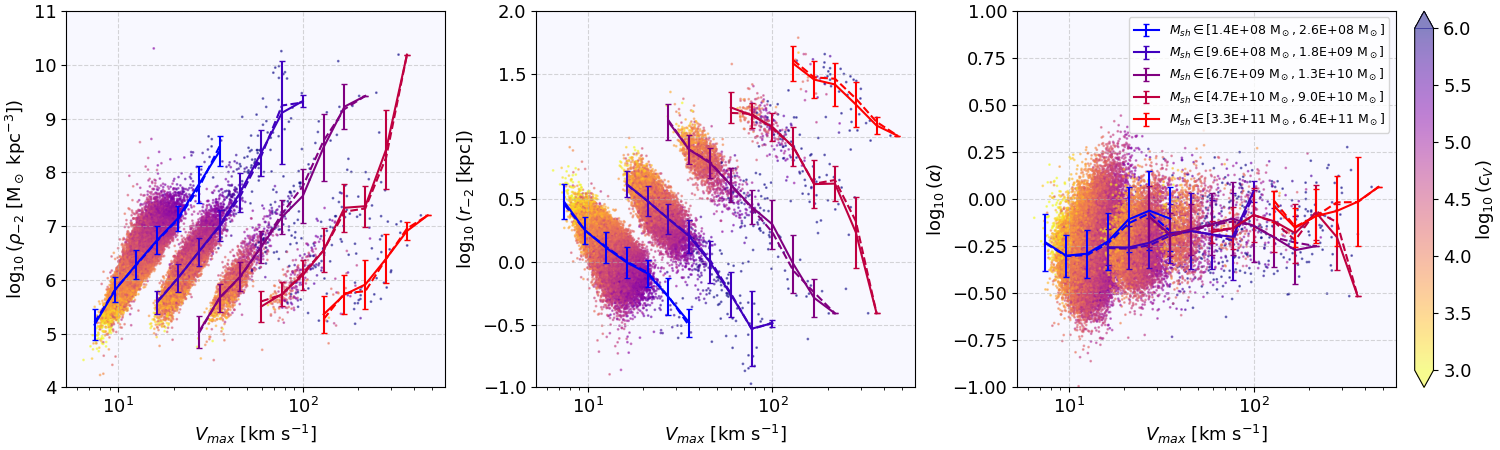}
\caption{Dependence of the Einasto profile parameter values on $V_\mathrm{max}$ for five different mass bins with colour-coded subhalo concentration. The solid and dashed lines indicate the mean and median of each distribution for several $V_\mathrm{max}$ bins, together with the standard deviations.}
\label{fig:parameter_vmax_scatter_einasto}
\vspace{10pt}
\end{figure*}

\begin{figure*}
\includegraphics[width=\textwidth]{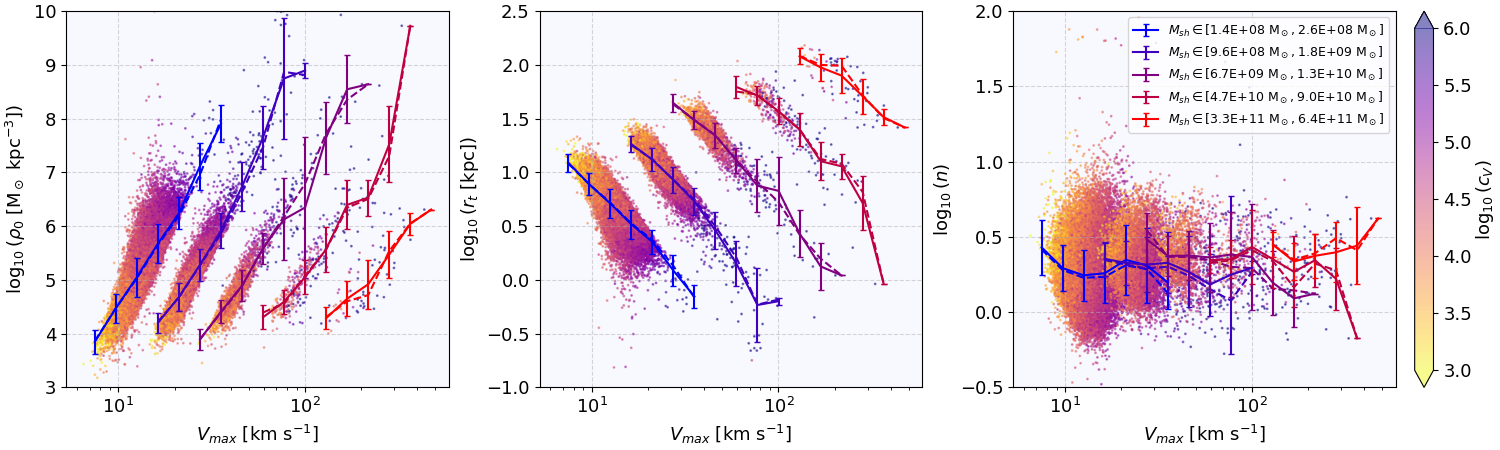}
\caption{Dependence of the MS-(2, $n$) profile parameter values on $V_\mathrm{max}$ for five different mass bins with colour-coded subhalo concentration. The solid and dashed lines indicate the mean and median of each distribution for several $V_\mathrm{max}$ bins, together with the standard deviations.}
\label{fig:parameter_vmax_scatter_tpl}
\vspace{10pt}
\end{figure*}

\begin{figure*}
\includegraphics[width=\textwidth]{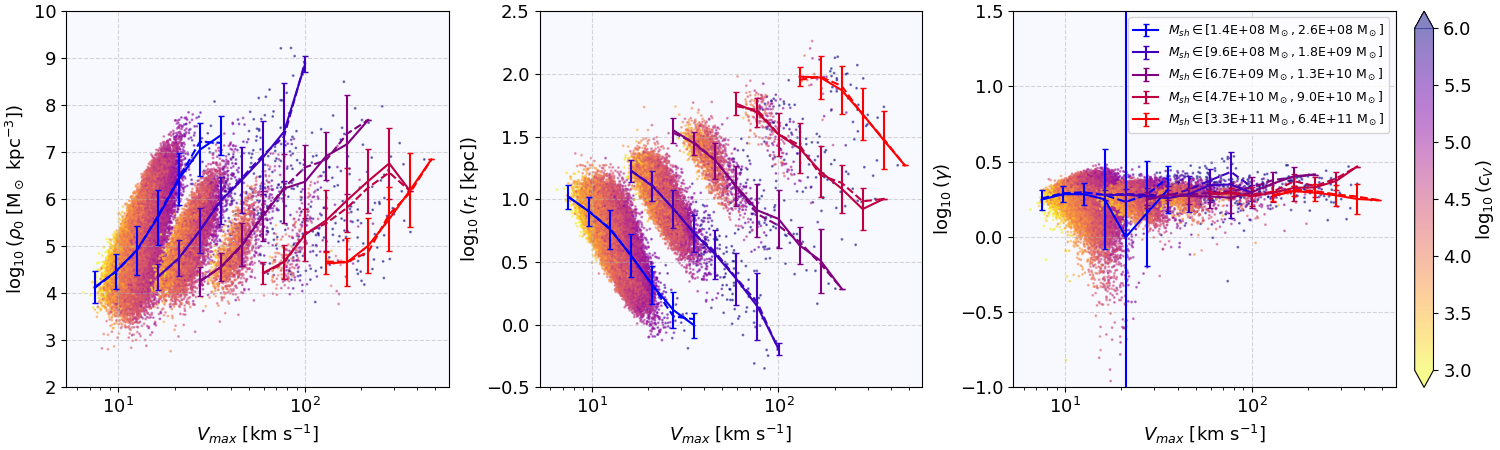}
\caption{Dependence of the MS-($\gamma$, 2) parameter values on $V_\mathrm{max}$ for five different mass bins with colour-coded subhalo concentration. The solid and dashed lines indicate the mean and median of each distribution for several $V_\mathrm{max}$ bins, together with the standard deviations.}
\label{fig:parameter_vmax_scatter_schechter}
\end{figure*}

Figures \ref{fig:parameter_mass_scatter_einasto}, \ref{fig:parameter_mass_scatter_tpl} and \ref{fig:parameter_mass_scatter_schechter} show the mass dependence of the parameter values for the Einasto profile, the truncated power law and the modified Schechter profile. Figures \ref{fig:parameter_vmax_scatter_einasto}, \ref{fig:parameter_vmax_scatter_tpl} and \ref{fig:parameter_vmax_scatter_schechter} show the dependence of the parameter values on $V_\mathrm{max}$ for different mass bins for the Einasto profile, the truncated power law and the modified Schechter profile. One can see that for all of the diagrams the overlap and scatter of the distributions is significantly larger than for the modified NFW profile (Figures \ref{fig:parameter_mass_scatter} and \ref{fig:parameter_vmax_scatter}), which is another thing that makes it the best model of the ones we have analysed. We didn't show the distributions for the NFW profile and its smoothly truncated version, since the parameters take extreme values and do not show nice scaling relations. \\

\begin{figure*}
\includegraphics[width=\textwidth]{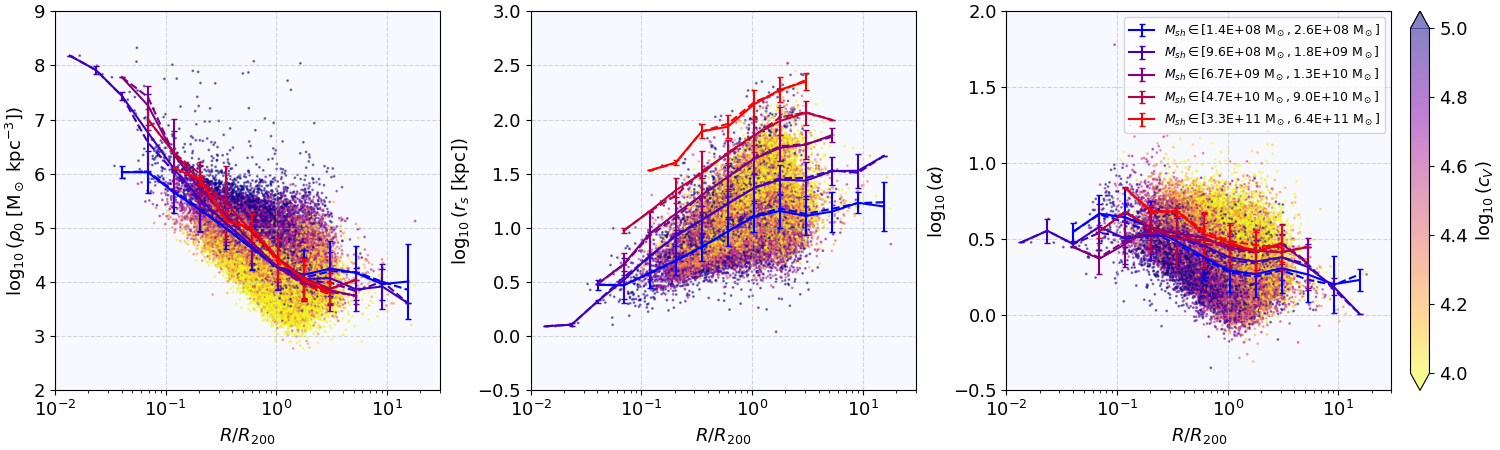}
\caption{Complete version of Figure \ref{fig:parameter_distance_scatter}, which shows the dependence of the modified NFW parameter values on the subhalo distance from the host halo centre for five different mass bins with colour-coded concentration $c_V$. The solid and dashed lines indicate the mean and median of each distribution for different distance bins together with the standard deviation.}
\label{fig:parameter_distance_scatter_complete}
\end{figure*}

Figure \ref{fig:parameter_distance_scatter_complete} is the complete version of Figure \ref{fig:parameter_distance_scatter} and shows how the parameter values of the modified NFW profile depend on the distance from the host halo centre. \\

\begin{figure*}
\includegraphics[width=\columnwidth]{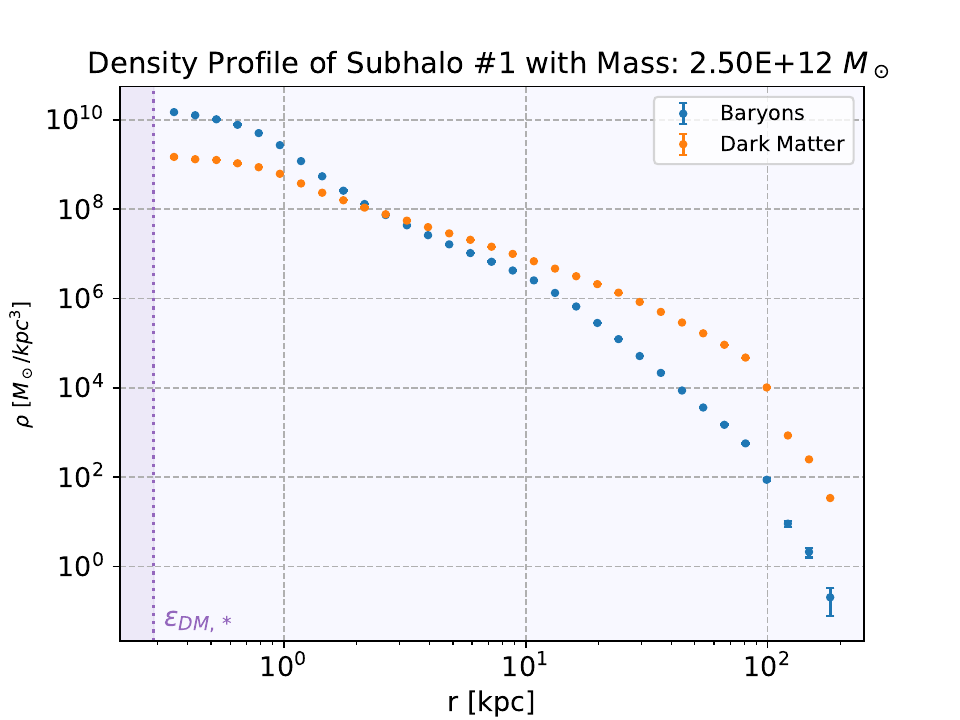}
\includegraphics[width=\columnwidth]{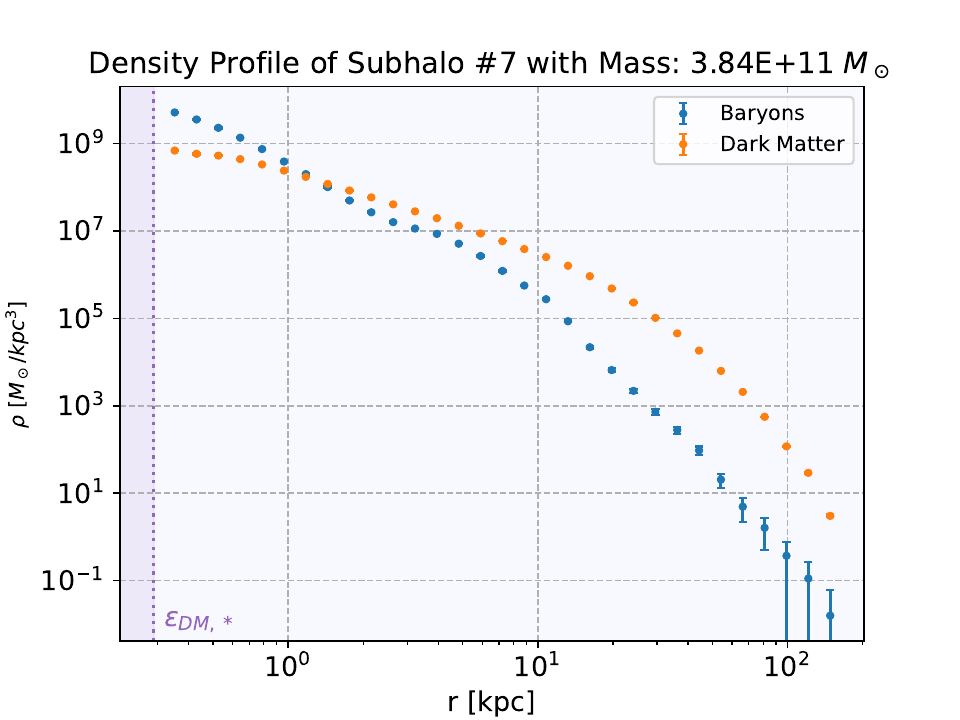}
\caption{The density profiles of the first two subhaloes from Figure \ref{fig:example_profiles} decomposed into their dark matter (orange) and baryon components (blue). The purple area indicates the regime below the softening length. For these density profiles we did not compute data points below the softening length.}
\label{fig:density_profile_decomposition}
\end{figure*}

Figure \ref{fig:density_profile_decomposition} shows the density profiles of the first two subhaloes in Figure \ref{fig:example_profiles} decomposed into their dark matter and baryon components. \\

Table \ref{tab:scaling_rel} lists the equations for the scaling relations for the parameter values of the Einasto profile, the truncated power law and the modified Schechter profile. \\

\newpage

\begin{table*}
\caption{Scaling relations for the parameter values of the Einasto profile and the modified Schechter profiles.}
\label{tab:scaling_rel}
 \begin{tabular}{lccc}
  \hline
  Density Profile Model & Normalisation Parameter & Radius Parameter & Shape Parameter\\
   & [$\unit{M_{\sun} \ kpc^{-3}}$] & [kpc] & \\
  \hline
  Einasto Profile 
  & $\rho_{-2} = 930 \ \left( \frac{M}{10^{13} \unit{M_\odot}} \right)^{-1.6} \left( \frac{V_{\mathrm{max}}}{10^2 \unit{km \ s^{-1}}} \right)^{4.8}$
  & $r_{-2} = 1.6 \ \left( \frac{M}{10^{10} \unit{M_\odot}} \right)^{0.9} \left( \frac{V_{\mathrm{max}}}{10^2 \unit{km \ s^{-1}}} \right)^{-1.6}$ 
  & $\alpha = 0.64 \ \left( \frac{M}{10^{10} \unit{M_\odot}} \right)^{0.047}$\\\\
  MS-(2, $n$)
  & $\rho_0 = 1.2 \ \left( \frac{M}{10^{13} \unit{M_\odot}} \right)^{-2.2} \left( \frac{V_{\mathrm{max}}}{10^2 \unit{km \ s^{-1}}} \right)^{5.6}$
  & $r_\mathrm{t} = 4.4 \ \left( \frac{M}{10^{10} \unit{M_\odot}} \right)^{0.9} \left( \frac{V_{\mathrm{max}}}{10^2 \unit{km \ s^{-1}}} \right)^{-1.8}$ 
  & $n = 2.2 \ \left( \frac{M}{10^{10} \unit{M_\odot}} \right)^{0.046}$\\\\
  MS-($\gamma$, 2) 
  & $\rho_0 = 13 \ \left( \frac{M}{10^{13} \unit{M_\odot}} \right)^{-2.0} \left( \frac{V_{\mathrm{max}}}{10^2 \unit{km \ s^{-1}}} \right)^{5.9}$
  & $r_\mathrm{t} = 5.5 \ \left( \frac{M}{10^{10} \unit{M_\odot}} \right)^{0.9} \left( \frac{V_{\mathrm{max}}}{10^2 \unit{km \ s^{-1}}} \right)^{-1.5}$ 
  & $\gamma = 1.94 \ \left( \frac{M}{10^{10} \unit{M_\odot}} \right)^{0.001}$\\
  \hline
 \end{tabular}
\end{table*}

\newpage

\bsp	
\label{lastpage}
\end{document}